\documentclass[11pt]{article}
\usepackage{etoolbox}
\usepackage{tikz}
\newrobustcmd*{\mysquare}[1]{\tikz{\filldraw[draw=#1,fill=#1] (0,0)
rectangle (0.2cm,0.2cm);}}
\newrobustcmd*{\mycircle}[1]{\tikz{\filldraw[draw=#1,fill=#1] (0,0) circle [radius=0.1cm];}}
\newrobustcmd*{\mytriangle}[1]{\tikz{\filldraw[draw=#1,fill=#1] (0,0) --
(0.2cm,0) -- (0.1cm,0.2cm);}}
\usepackage{amssymb}
\usepackage{amsmath}
\usepackage{tcolorbox}
\usepackage{graphics}
\usepackage[final]{pdfpages}
\usepackage{slashed}
\usepackage{amssymb}
\usepackage{epstopdf}
\catcode`@11
\def\seceqaa{\@addtoreset{equation}{section}
\def\theequation{A\arabic{equation}}}
\def\seceqbb{\@addtoreset{equation}{section}
\def\theequation{B\arabic{equation}}}
\def\seceqcc{\@addtoreset{equation}{section}
\def\theequation{C\arabic{equation}}}
\def\seceqdd{\@addtoreset{equation}{section}
\def\theequation{D\arabic{equation}}}
\def\seceqee{\@addtoreset{equation}{section}
\def\theequation{E\arabic{equation}}}
\def\seceqff{\@addtoreset{equation}{section}
\def\theequation{F\arabic{equation}}}
\def\seceqgg{\@addtoreset{equation}{section}
\def\theequation{G\arabic{equation}}}
\def\seceqhh{\@addtoreset{equation}{section}
\def\theequation{H\arabic{equation}}}
\catcode`@11

\date{\today}

\begin{document}

\begin{titlepage}
\begin{center}
{\large \bf Top-Down Holographic $G$-Structure Glueball Spectroscopy at (N)LO in $N$ and Finite Coupling}\\
 Karunava Sil\footnote{e-mail: krusldph@iitr.ac.in},
 Vikas Yadav\footnote{email: viitr.dph2015@iitr.ac.in}
  and
 Aalok Misra\footnote{e-mail: aalokfph@iitr.ac.in}
 \\
Department of Physics, Indian Institute of Technology,
Roorkee - 247 667, Uttaranchal, India

 \date{\today}
\end{center}
\thispagestyle{empty}
\begin{abstract}

\end{abstract}
The top-down type IIB holographic dual of large-$N$ thermal QCD as constructed in \cite{metrics} involving a fluxed resolved warped deformed conifold, its delocalized type IIA SYZ mirror as well as its  M-theory uplift constructed in \cite{MQGP} - both  in the finite coupling ($g_s\stackrel{<}{\sim}1$)/`MQGP' limit  of \cite{MQGP} - were shown explicitly to possess a local $SU(3)/G_2$-structure in \cite{NPB}. Glueballs spectra 
 in the {\it  finite-gauge-coupling limit} (and not just large-t'Hooft coupling limit) - a limit expected to be directly relevant to strongly coupled systems at finite temperature such as QGP \cite{Natsuume} - has thus far been missing in the literature. In this paper, we fill this gap by calculating the masses of the $0^{++},  0^{-+},0^{--}, 1^{++}, 2^{++}$ (`glueball') states (which correspond to fluctuations in the dilaton or complexified two-forms or appropriate metric components) in the aforementioned backgrounds of $G$-structure in the `MQGP' limit of \cite{MQGP}. We use WKB quantization conditions on one hand and impose Neumann/Dirichlet boundary conditions at an IR cut-off (`$r_0$')/horizon radius (`$r_h$') on the solutions to the equations of motion on the other. We find that the former technique produces results closer to the lattice results. We also discuss $r_h=0$-limits of all calculations. In this context we also calculate the  $0^{++}, 0^{--},1^{++}, 2^{++}$ glueball masses up to NLO in $N$ and find a $\frac{g_sM^2}{N}(g_sN_f)$-suppression similar to and further validating a similar semi-universality of NLO corrections to transport coefficients, observed in \cite{EPJC-2}.
\end{titlepage}

\section{Introduction}
The AdS/CFT correspondence \cite{maldacena} remarkably establishes an equivalence between the partition functions of a five dimensional gravitational theory (bulk theory) and a four dimensional supersymmetric and scale invariant gauge theory (boundary theory). A generalization of the AdS/CFT correspondence is necessary to explore more realistic gauge theories (less supersymmetric and non-conformal) such as QCD with $SU(3)$ gauge group. The top-down model that we have considered in this work is motivated to capture QCD-like gauge theories from a suitable gravitational background. QCD is a strongly coupled theory at low energies. The low energy dynamics of QCD involves the color neutral bound states of gluons, known as glueballs. Hence, the non-perturbative aspects of QCD can be largely understood from the glueball sector of the theory. Moreover, the plasma phase of QCD (QGP) occurs at high temperatures $T>T_c$.
 In QGP medium the quarks and the gluons stay in a deconfined state due to Debye screening. However, the recent RHIC experiments indicate strongly that non-perturbative effects of QCD are present in the plasma phase.
   In fact the lattice results of \cite{lattice result} conclude that QGP must be non-perturbative in the temperature regime $T_{c}\leq T\leq 5T_{c}$. This is precisely the reason why we concentrate on the glueball spectra in the {\it  finite-gauge-coupling limit} (and not just large-t'Hooft coupling limit) - a limit expected to be directly relevant to strongly coupled systems at finite temperature such as QGP \cite{Natsuume}.

QCD is a non-abelian gauge theory, in which gauge fields play the role of dynamical degrees of freedom.  Non-abelian nature of QCD  allows the gauge bosons to form
color-neutral bound states of gluons known as glueballs (gg, ggg, etc.). Therefore, the study of glueballs and their spectra enables us to gain
a better understanding of the non-perturbative regime of QCD.  Glueball state is represented
by quantum numbers  $\text{J}^{\text{PC}}$, where $J, P$ and $C$ correspond  to total angular momentum,
parity and charge conjugation respectively


Different generalized versions of the AdS/CFT correspondence has thus far been proposed to study non-supersymmetric field theories with a running gauge coupling constant. The original proposal was given by Witten to obtain a gravity dual for non-supersymmetric field theories. As per Witten's formalism, non-supersymmetric Yang-Mills theory can be obtained by compactifying one of the spatial direction on a circle
and imposing antiperiodic boundary conditions on the fermions around this circle. This makes the fermions and scalars massive and they get decoupled leaving only gauge fields as
degrees of freedom. The gravity dual of this compactified theory was asymptotically AdS. In a particular case of ${\cal N}=4$ SU(N) super-Yang-Mills theory dual to type IIB string theory on $AdS_{5}\times S^{5}$, the procedure described above gives an effective model of three dimensional Yang-Mills theory, i.e., $\text{QCD}_{3}$.

 The gravity dual of non-supersymmetric theories in the low energy limit is typically given by supergravity backgrounds involving $\text{AdS}_{p}\times\text{M}_{q}$ where AdS is the anti de Sitter space with dimension $p$ and $\text{M}_{q}$ is the internal manifold with dimension $q$. In the supergravity theory the Klauza-Klein modes on $\text{M}_{q}$ can be classified according to the spherical harmonics of the $\text{M}_{q}$, which forms representations of the isometry group of $\text{M}_{q}$. The states carrying the non-trivial isometry group quantum numbers are heavier and do not couple to the pure gluonic operators on the boundary. Thus the glueballs are identified with singlet states of the isometry group.

In the past decade, glueballs have been studied extensively to gain new insight into the non-perturbative regime of QCD. Various holographic setups such as soft-wall model, hard wall model, modified soft wall model, etc. have been used to obtain the glueball's spectra. In \cite {Colangelo,Nicotri} a soft wall holographic model was used to study the glueball spectrum. In \cite {Colangelo} glueballs and scalar mesons were studied at finite temperature.
It was found that the masses of the hadronic states decreases and the widths become broader as T increases. But for temperature range of the order of 40-60 MeV, states disappear from the
glueball and meson spectral function. Both hard wall and soft wall holographic models were considered \cite {Forkel,ForkelStructure} to obtain the glueball correlation functions to study the dynamics of QCD. Decay rates were obtained for glueballs in both models. Dynamical content of the
correlators was investigated\cite {ForkelStructure} by obtaining their spectral density and relating it with various other quantities to obtain the estimates for three lowest dimensional gluon
condensate. In \cite {Li} 
a two-flavor quenched dynamical holographic QCD model was considered with two different forms for the dilaton field 
given as: $\Phi=\mu_{G}^{2}z^{2}$ and $\Phi=\mu_{G}^{2}z^{2}\tanh (\mu_{G}^{4}z^{2}/\mu_{G}^{2}),$ ($z$ being a radial coordinate).
In \cite {FolcoCapossoli} an AdS$_5$ mass renormalization was implemented in a modified holographic softwall model to obtain the spectrum of scalar and higher even glueball spin states with $P=C=+1$.
In \cite {Jugeau,Colangelolight} a bottom-up approach 
was used to obtain the mass spectra of the scalar and vector glueballs. In this case, 
the Vector glueball masses were found 
to be heavier than that
of the scalar glueballs while higher values for both were reported
in other approaches.
In \cite {Huang} holographic description was used for supersymmetric and non-supersymmetric, non-commutative dipole gauge theory in 4D. WKB approximation was used to obtain the 
mass by solving the dilaton and antisymmetric tensor field equations in the bulk. For the supersymmetric theory, dipole length plays the role of an intrinsic scale while for non-supersymmetric theory the same role is played by the temperature 
Two different phases for baryons were found, a big baryon dual to the static string and a small baryon dual to a moving string.
In \cite {Chen} spectrum for scalar, vector and tensor two-gluon and trigluon glueballs were obtained in 5-D holographic QCD model with a metric structure deformed by the dilaton field. The spectrum
was compared with the results obtained from both soft-wall and hard-wall holograpic QCD models. Here, the spectra of the two-gluon glueball was found to be in agreement with the lattice data. For
trigluon glueballs, the masses for $1^{\pm -}$,$2^{--}$ were matched while masses for $0^{--},0^{+-}$ and $2^{+-}$ were lighter than lattice data which indicates that the latter glueballs are dominated by three-gluon
condensate. In \cite {Gordeli} holographic glueball spectrum was obtained in the singlet sector of ${\cal N}$=1
supersymmetric Klebanov-Strassler model.  States containing the bifundamental $A_i$ and $B_i$ fields were not considered. Comparison with the lattice data showed a nice agreement for $1^{+-}$ and $1^{--}$ states
while $0^{+-}$ results were different because of its fermionic component.

Glueballs appear in the meson spectra of QCD and difficulty in their
identification in the meson spectra is largely due to lack of information about their coupling with mesons in strongly coupled QCD. Lattice QCD gives an estimate for the masses but it
does not give any information about the glueball couplings and their decay widths both of which are required for identification of glueballs. Holographic approach gives a better understanding of  glueball decay rates than lattice QCD. Various holographic models such as Witten-Sakai-Sugimoto model, Soft wall model and supersymmetric Klebanov-Strassler model, etc. have been used to obtain the coupling between mesons and glueballs to obtain expressions for the glueball decay widths.

In \cite {Hashimoto,Brunner0,Brunner1,ParganlijaScalar,Brunnerdecay,Brunner2} top-down Witten-Sakai-Sugimoto model was used as a holographic setup for low energy QCD to obtain the
coupling of scalar glueballs to mesons and subsequently obtain their decay widths.  Results obtained were compared with the experimental data available for lattice counterparts f$_0$(1500) and
f$_0$(1710) of scalar glueballs. In \cite {Hashimoto} results obtained for decay widths and branching ratios for scalar glueball decays were found to be consistent with experimental data
for f$_0$(1500) state in \cite {Hashimoto} while in \cite {Brunner1} results favored f$_0$(1710) as scalar glueball candidate instead of f$_0$(1500). Decay patterns were
obtained for scalar glueball candidate f$_0$(1710) in top-down holographic Witten-Sakai-Sugimoto model for low energy QCD in \cite {Brunner0}. It was shown that there exists a narrow
pseudoscalar glueball heavier than the scalar glueball whose decay pattern involves $\eta$ and $\eta\prime$  mesons. In \cite {ParganlijaScalar,Brunnerdecay,Brunner2} Witten-Sakai-Sugimoto model
was used to study the phenomenology of scalar glueball states. A dilaton and an exotic mode were obtained as two sets of scalar glueball states in \cite {ParganlijaScalar,Brunnerdecay}. Calculation of mass spectra showed that out of two modes, dilaton mass is quite close to both f$_0$(1710) and f$_0$(1500) scalar glueball candidates while calculation  of decay width showed that f$_0$(1710) is the favored glueball candidate corresponding to dilaton mode. In \cite {Parganlija} the holographic top-down Witten-Sakai-Sugimoto model was used to study the tensor $2^{++}$ glueball mass spectrum and decay width . Decay width was found to be above 1 GeV for glueball mass
M$_T$=2400 MeV while for M$_T$=2000 MeV it was reduced to 640 MeV. In \cite {Capossoli} modified holographic soft-wall model was used to calculate the mass spectrum and Regge trajectories of lightest scalar glueball and higher spin glueball states. Results were obtained for both even and odd spins glueball states.

In this paper, we use a large-$N$ top-down holographic dual of QCD  to obtain the spin $2^{++}, 1^{++},$ $0^{++}, 0^{--}, 0^{-+}$ glueball spectrum explicitly for QCD$_{3}$  from type IIB, type IIA and M theory perspectives. Now for the computation of the glueball masses, we need to introduce a scale in our theory. In other words, the conformal invariance has to be broken. This can be done in two different ways. The first approach, after Witten \cite{witten}, corresponds to the compactification of the time direction on a circle of finite radius, forming a black hole in the background. In this case the masses are determined in units of the horizon radius $r_h$ of the black hole. The other approach is to consider a cut-off at $r=r_0$ in the gravitational background ($r$ being the non-compact radial direction)\cite{polchinski}. This forbids the arbitrary low energy excitations of the boundary field theory and hence breaks the conformal invariance. So, in this case the required scale to address strong interaction is introduced by the IR cut-off  $r_0$. From a top-down perspective this IR cut-off will in fact be proportional to two-third power of the Ouyang embedding parameter obtained from the minimum radial distance (corresponding to the lightest quarks) requiring one to be at the South Poles in the $\theta_{1,2}$ coordinates, in the holomorphic Ouyang embedding of flavor $D7$-branes. In the spirit of \cite{witten}, the time direction for both cases will be compact with fermions obeying anti-periodic boundary conditions along this compact direction, and hence we will be evaluating three-dimensional glueball masses.

Glueball masses can be obtained by evaluating the correlation functions of gauge invariant local operator. The first step to obtain the glueball spectrum in QCD$_3$ is to identify the operators in the gauge theory that have quantum number corresponding to the glueballs of interest. According to
the gauge/gravity duality each supergravity mode corresponds to a gauge theory operator. This operator couples to the supergravity mode at the boundary of the AdS space, for example,
the lowest dimension operator with quantum numbers $J^{PC}=0^{++}$ is Tr$F^2=$Tr$F_{\mu\nu}F^{\mu\nu}$ and this operator couples to the dilaton mode on the boundary. To calculate $0^{++}$ glueball mass we need to evaluate the correlator $\left< TrF^{2}(x)TrF^2(y)\right>$ =$\Sigma_{i} c_{i} e^{-m{i}|x-y|}$, where
$m_{i}$ give the value for glueball mass. However the masses can also be obtained by solving the wave equations for supergravity modes which couples to the gauge theory operators on the
boundary. The latter approach is used in this paper.

The 11D metric obtained as the uplift of the delocalized SYZ type IIA metric, up to LO in $N$, can be interpreted as a black $M5$-brane wrapping a two-cycle, i.e. a black $M3$-brane \cite{transport-coefficients,NPB}. Taking this as the starting point, compactifying again along the M-theory circle, we land up at the type IIA metric and then compactifying again along the periodic temporal circle (with the radius given by the reciprocal of the temperature), one obtains ${\rm QCD}_3$ corresponding to the three non-compact directions of the black $M3$-brane world volume. The Type IIB background of \cite{metrics}, in principle, involves $M_4\times$ RWDC($\equiv$ Resolved Warped Deformed Conifold); asymptotically the same becomes $AdS_5\times T^{1,1}$. To determine the gauge theory fields that would couple to appropriate supergravity fields a la gauge-gravity duality, ideally one should work the same out for the $M_4\times$ RWDC background (which would also involve solving for the Laplace equation for the internal RWDC). We do not attempt to do the same here. Motivated however by, e.g.,

\noindent {\bf (a)} asymptotically the type IIB background of \cite{metrics} and its delocalized type IIA mirror of \cite{MQGP} consist of $AdS_5$

\noindent and

\noindent {\bf (b)} terms of the type $Tr(F^2(AB)^k), (F^4(AB)^k)$ where $F^2 = F_{\mu\nu}F^{\mu\nu}, F^4 = F_{\mu_1}^{\ \mu_2}F_{\mu_2}^{\ \mu_3}F_{\mu_3}^{\ \mu_4}F_{\mu_4}^{\ \mu_1} - \frac{1}{4}\left(F_{\mu_1}^{\ \mu_2}F_{\mu_2}^{\ \mu_1}\right)^2$, $A, B$ being the bifundamental fields that appear in the gauge theory superpotential corresponding to $AdS_5\times T^{1,1}$ in \cite{KW}, form part of the gauge theory operators corresponding to the solution to the Laplace equation on $T^{1,1}$ \cite{Gubser-Tpq} (the operator $Tr F^2$ which shares the quantum numbers of the $0^{++}$ glueball couples to the dilaton and $Tr F^4$ which also shares the quantum numbers of the $0^{++}$ glueball couples to trace of metric fluctuations and the four-form potential, both in the internal angular directions),

\noindent we calculate in this paper:\\
$\bullet$ type IIB dilaton fluctuations, which we refer to as $0^{++}$ glueball\\
 $\bullet$ type IIB complexified two-form fluctuations that couple to
 $d^{abc}Tr(F_{\mu\rho}^aF^{b\ \rho}_{\lambda}F^{c\ \lambda}_{\ \ \ \ \ \nu})$, which we refer to as $0^{--}$ glueball \\
$\bullet$ type IIA one-form fluctuations that couple to $Tr(F\wedge F)$, which we refer to as $0^{-+}$ glueball \\
$\bullet$  M-theory metric's scalar fluctuations which we refer to as another (lighter) $0^{++}$ glueball \\
$\bullet$ M-theory metric's vector fluctuations which we refer to as $1^{++}$ glueball, \\
and \\
$\bullet$ M-theory metric's tensor fluctuations which we refer to as $2^{++}$ glueball.

All holographic glueball spectra calculations done thus far, have only considered a large t'Hooft coupling limit: $g_{\rm YM}^2N\gg1,\ N\gg1$. However, holographic duals of thermal QCD laboratories like sQGP also require a finite gauge coupling \cite{Natsuume}. This was addressed as part of the `MQGP limit' in \cite{MQGP}. It is in this regard that results of this paper - which discusses supergravity glueball spectra at finite string coupling - are particularly significant. Also, the recent observation - see, e.g., \cite{glueball-QCD-instantons} -  that the non-perturbative properties of quark gluon plasma can be related to the change of properties of scalar and pseudoscalar glueballs, makes the study of glueballs quite important.

The rest of the paper is organized as follows. In Sec. {\bf 2}, via five sub-sections, we summarize the top-down type IIB holographic dual of large-$N$ thermal QCD of \cite{metrics}, its delocalized SYZ type IIA mirror and its M theory uplift of \cite{MQGP,NPB}. In Sec. {\bf 3}, we discuss a supergravity calculation of the spectrum of $0^{++}$ glueball at finite horizon radius $r_h$ ({\bf 3.1}) and setting $r_h=0$ ({\bf 3.2}). The $r_h\neq0$ computations are given in {\bf 3.1}, corresponding to use of WKB quantization conditions using coordinate/field redefinitions of  \cite{Minahan}. The $r_h=0$ calculations are subdivided into sub-section {\bf 3.2.1} corresponding to solving the $0^{++}$ equation of motion up to LO in $N$ and imposing Neumann/Dirichlet boundary condition at the horizon, and sub-section {\bf 3.2.2} corresponding to WKB quantization conditions inclusive of non-conformal/NLO-in-$N$ corrections using the redefinitions of \cite{Minahan}. Sec. {\bf 4} has to do with the $0^{-+}$ glueball spectrum. Further therein, {\bf 4.1.1} and {\bf 4.1.2} respectively are on obtaining the $r_h\neq0$ spectrum and its $r_h=0$ limit using Neumann/Dirichlet boundary conditions on the solutions up to LO in $N$ respectively at the horizon and the IR cutoff. Then subsections {\bf 4.2.1} and {\bf 4.2.2} respectively are on WKB quantization at finite and zero $r_h$ up to LO in $N$, using redefinitions of \cite{Minahan}. Sec. {\bf 5} is on $0^{--}$ glueball spectrum. Therein, subsections {\bf 5.1} and {\bf 5.2} are on getting the spectrum by imposing Neumann/Dirichlet boundary condition on the solutions up to LO in $N$ to the EOM respectively at the horizon and the IR cut-off. Subsection {\bf 5.3} has to do with obtaining the spectrum using WKB quantization up to LO in $N$ at $r_h\neq0$ using the redefinitions of \cite{Minahan}; {\bf 5.4} has to do with a similar calculation in the $r_h=0$ limit at LO in $N$ in {\bf 5.4.1} and up to NLO in $N$ in {\b 5.4.2}. Section {\bf 6} has to do with M-theory calculations of $0^{++}, 1^{++}, 2^{++}$ glueballs arising from appropriate metric fluctuations. Subsection {\bf 6.1} is on such a $0^{++}$ glueball spectrum  (imposing Neumann/Dirichlet boundary conditions at the horizon/IR cut-off in {\bf 6.1.1}, and using WKB quantization conditions and the redefinitions of \cite{Minahan} at finite/zero horizon radius in {\bf 6.1.2}). Subsection {\bf 6.2} is on such a $2^{++}$ glueball spectrum (via imposing Neumann/Dirichlet boundary conditions at the horizon/IR cut-off in {\bf 6.2.1}/{\bf 6.2.2}, and via WKB quantization conditions using redefinitions of \cite{Minahan} at LO in $N$ at $r_h\neq0$ in {\bf 6.2.3} and zero $r_h$ in {\bf 6.2.4} as well as up to NLO in $N$ in the $r_h=0$-limit, in {\bf 6.2.5}). Subsection {\bf 6.3} is on such a $1^{++}$ glueball spectrum (via WKB quantization conditions using the redefinitions of \cite{Minahan} up to LO in $N$ in the finite/zero horizon radius limit in {\bf 6.3.1}/{\bf 6.3.2}, and up to NLO in $N$ in the zero-horizon limit in {\bf 6.3.3}). From the point of view of comparing string theory and M theory glueball spectrum calculations, we obtain the $2^{++}$ glueball spectrum arising from tensor mode of metric fluctuations in the type IIB background of \cite{metrics} in Sec. {\bf 7}. Subsection {\bf 7.1} has to do with a supergravity calculation (via Neumann/Dirichlet boundary conditions at the horizon in {\bf 7.1.1} and WKB quantization condition using redefinitions of \cite{Minahan} in {\bf 7.1.2}), and {\bf 7.2} has to do with zero-horizon radius limit calculation (WKB quantization condition using redefinitions of \cite{Minahan} at LO in $N$ in {\bf 7.2.1} and up to NLO in $N$ in {\bf 7.2.2}). Section {\bf 8} contains a summary of and discussion on the results obtained in this paper. There a appendix {\bf A} on the square of different fluxes that appear in EOM relevant to spin-two perturbations of the type IIB metric.

\section{Background: A Top-Down Type IIB Holographic Large-$N$ Thermal QCD and its M-Theory Uplift in the `MQGP' Limit}

In this section, via five sub-sections we will:

\begin{itemize}
\item
provide a short review of the type IIB background of \cite{metrics} which is supposed to provide a UV complete holographic dual of large-$N$ thermal QCD, as well as their precursors in subsection {\bf 2.1},

\item
discuss the 'MQGP' limit of \cite{MQGP} and the motivation for considering the same in subsection {\bf 2.2},

\item
briefly review issues as discussed in \cite{MQGP} pertaining to construction of delocalized S(trominger) Y(au) Z(aslow) mirror and approximate supersymmetry, in subsection {\bf 2.3},

\item
briefly review the new results of \cite{NPB} and \cite{EPJC-2} pertaining to construction of explicit $SU(3)$ and $G_2$ structures respectively of type IIB/IIA and M-theory uplift,

\item
briefly discuss the new Physics-related results of \cite{NPB} and \cite{EPJC-2}, in subsection {\bf 2.4}

\end{itemize}

\subsection{Type IIB Dual of Large-$N$ Thermal QCD}

In this subsection, we will discuss a UV complete holographic dual of large-$N$ thermal QCD as given in  Dasgupta-Mia et al \cite{metrics}. As partly mentioned in Sec. {\bf 1}, this was inspired by the zero-temperature Klebanov-Witten model \cite{KW}, the non-conformal Klebanov-Tseytlin model \cite{KT}, its IR completion as given in the Klebanov-Strassler model \cite{KS} and Ouyang's inclusion \cite{ouyang} of flavor in the same \footnote{See \cite{Leo-i} for earlier attempts at studying back-reacted $D3/D7$ geometry at zero temperature; we thank L. Zayas for bringing \cite{Leo-i,Leo-ii} to our attention.}, as well as the non-zero temperature/non-extremal version of \cite{Buchel} (the solution however was not regular as the non-extremality/black hole function and the ten-dimensional warp factor vanished simultaneously at the horizon radius), \cite{Gubser-et-al-finitetemp} (valid only at large temperatures) of the Klebanov-Tseytlin model  and \cite{Leo-ii} (addressing the IR), in the absence of flavors.  \\
 \noindent (a) \underline{Brane construction}

 In order to include fundamental quarks at non-zero temperature in the context of type IIB string theory, to the best of our knowledge, the following model proposed in \cite{metrics} is the closest to a UV complete holographic dual of large-$N$ thermal QCD. The KS model (after a duality cascade) and QCD have similar IR behavior: $SU(M)$ gauge group and IR confinement. However, they differ drastically in the UV as the former yields a logarithmically divergent gauge coupling (in the UV) - Landau pole. This necessitates modification of the UV sector of the KS model apart from inclusion of non-extremality factors. With this in mind and building up on all of the above, the type IIB holographic dual of
 \cite{metrics} was constructed. The setup of \cite{metrics} is summarized below.

 \begin{itemize}
 \item
  From a gauge-theory perspective, the authors of \cite{metrics} considered  $N$ black $D3$-branes placed at the tip of six-dimensional conifold, $M\ D5$-branes wrapping the vanishing two-cycle and $M\ \overline{D5}$-branes  distributed along the resolved two-cycle and placed at the outer boundary  of the IR-UV interpolating region/inner boundary of the UV region.

  \item
  More specifically, the $M\ \overline{D5}$ are distributed around the antipodal point relative to the location of $M\ D5$ branes on the blown-up $S^2$. If  the $D5/\overline{D5}$ separation is given by ${\cal R}_{D5/\overline{D5}}$, then this provides the boundary common to the outer UV-IR interpolating region and the inner UV region. The region $r>{\cal R}_{D5/\overline{D5}}$ is the UV.  In other words, the radial space, in \cite{metrics} is divided into the IR, the IR-UV interpolating region and the UV. To summarize the above:
  \begin{itemize}
  \item
  $r_0/r_h<r<|\mu_{\rm Ouyang}|^{\frac{2}{3}}(r_h=0)/{\cal R}_{D5/\overline{D5}}(r_h\neq0)$: the IR/IR-UV interpolating regions with $r\sim\Lambda$: deep IR where the $SU(M)$ gauge theory confines

  \item
  $r>|\mu_{\rm Ouyang}|^{\frac{2}{3}} (r_h=0)/ {\cal R}_{D5/\overline{D5}}(r_h\neq0)$: the UV region.

  \end{itemize}

\item
$N_f\ D7$-branes, via Ouyang embedding,  are holomorphically embedded in the UV (asymptotically $AdS_5\times T^{1,1}$), the IR-UV interpolating region and dipping into the (confining) IR (up to a certain minimum value of $r$ corresponding to the lightest quark)  and $N_f\ \overline{D7}$-branes present in the UV and the UV-IR interpolating (not the confining IR). This is to ensure turning off of three-form fluxes, constancy of the axion-dilaton modulus and hence conformality and absence of Landau poles in the UV.

\item
The resultant ten-dimensional geometry hence involves a resolved warped deformed conifold. Back-reactions are included, e.g., in the ten-dimensional warp factor. Of course, the gravity dual, as in the Klebanov-Strassler construct, at the end of the Seiberg-duality cascade will have no $D3$-branes and the $D5$-branes are smeared/dissolved over the blown-up $S^3$ and thus replaced by fluxes in the IR.
\end{itemize}

The delocalized S(trominger) Y(au) Z(aslow) type IIA mirror of the aforementioned type IIB background of \cite{metrics} and its M-theory uplift had been obtained in \cite{MQGP,NPB,EPJC-2}.

\noindent (b) \underline{Seiberg duality cascade, IR confining $SU(M)$ gauge theory at finite temperature and}\\ \underline{$N_c = N_{\rm eff}(r) + M_{\rm eff}(r)$}

\begin{enumerate}
\item
{\bf IR Confinement after Seiberg Duality Cascade}: Footnote numbered 3 shows that one effectively adds on to the number of $D3$-branes in the UV and hence, one has $SU(N+M)\times SU(N+M)$ color gauge group (implying an asymptotic $AdS_5$) and $SU(N_f)\times SU(N_f)$ flavor gauge group, in the UV: $r\geq {\cal R}_{D5/\overline{D5}}$. It is expected that there will be a partial Higgsing of $SU(N+M)\times SU(N+M)$ to $SU(N+M)\times SU(N)$ at $r={\cal R}_{D5/\overline{D5}}$  \cite{K. Dasgupta et al [2012]}. The two gauge couplings, $g_{SU(N+M)}$ and $g_{SU(N)}$ flow  logarithmically  and oppositely in the IR:
\begin{equation}
\label{RG}
4\pi^2\left(\frac{1}{g_{SU(N+M)}^2} + \frac{1}{g_{SU(N)}^2}\right)e^\phi \sim \pi;\
 4\pi^2\left(\frac{1}{g_{SU(N+M)}^2} - \frac{1}{g_{SU(N)}^2}\right)e^\phi \sim \frac{1}{2\pi\alpha^\prime}\int_{S^2}B_2.
\end{equation}
  Had it not been for $\int_{S^2}B_2$, in the UV, one could have set $g_{SU(M+N)}^2=g_{SU(N)}^2=g_{YM}^2\sim g_s\equiv$ constant (implying conformality) which is the reason for inclusion of $M$ $\overline{D5}$-branes at the common boundary of the UV-IR interpolating and the UV regions, to annul this contribution. In fact, the running also receives a contribution from the $N_f$ flavor $D7$-branes which needs to be annulled via $N_f\ \overline{D7}$-branes. The gauge coupling $g_{SU(N+M)}$ flows towards strong coupling and the $SU(N)$ gauge coupling flows towards weak coupling. Upon application of Seiberg duality, $SU(N+M)_{\rm strong}\stackrel{\rm Seiberg\ Dual}{\longrightarrow}SU(N-(M - N_f))_{\rm weak}$ in the IR;  assuming after repeated Seiberg dualities or duality cascade, $N$ decreases to 0 and there is a finite $M$, {one will be left with $SU(M)$ gauge theory with $N_f$ flavors that confines in the IR - the finite temperature version of the same is what was looked at by \cite{metrics}}.

 \item
{\bf Obtaining $N_c=3$, and Color-Flavor Enhancement of Length Scale in the IR}:  So, in the IR, at the end of the duality cascade, what gets identified with the number of colors $N_c$ is $M$, which in the `MQGP limit' to be discussed below, can be tuned to equal 3. One can identify $N_c$ with $N_{\rm eff}(r) + M_{\rm eff}(r)$, where $N_{\rm eff}(r) = \int_{\rm Base\ of\ Resolved\ Warped\ Deformed\ Conifold}F_5$ and $M_{\rm eff} = \int_{S^3}\tilde{F}_3$ (the $S^3$ being dual to $\ e_\psi\wedge\left(\sin\theta_1 d\theta_1\wedge d\phi_1 - B_1\sin\theta_2\wedge d\phi_2\right)$, wherein $B_1$ is an asymmetry factor defined in \cite{metrics}, and $e_\psi\equiv d\psi + {\rm cos}~\theta_1~d\phi_1 + {\rm cos}~\theta_2~d\phi_2$) where $\tilde{F}_3 (\equiv F_3 - \tau H_3)\propto M(r)\equiv 1 - \frac{e^{\alpha(r-{\cal R}_{D5/\overline{D5}})}}{1 + e^{\alpha(r-{\cal R}_{D5/\overline{D5}})}}, \alpha\gg1$  \cite{IR-UV-desc_Dasgupta_etal}. The effective number $N_{\rm eff}$ of $D3$-branes varies between $N\gg1$ in the UV and 0 in the deep IR, and the effective number $M_{\rm eff}$ of $D5$-branes varies between 0 in the UV and $M$ in the deep IR (i.e., at the end of the duality cacade in the IR). Hence, the number of colors $N_c$ varies between $M$ in the deep IR and a large value [even in the MQGP limit of (\ref{limits_Dasguptaetal-ii}) (for a large value of $N$)] in the UV.  {Hence, at very low energies, the number of colors $N_c$ can be approximated by $M$, which in the MQGP limit is taken to be finite and can hence be taken to be equal to three. } However, in this discussion, the low energy or the IR is relative to the string scale. But these energies which are much less than the string scale, can still be much larger than $T_c$. Therefore, for all practical purposes, as regard the energy scales relevant to QCD, the number of colors can be tuned to three.

 In the IR in the MQGP limit, with the inclusion of terms higher order in $g_s N_f$  in the RR and NS-NS three-form fluxes and the NLO terms in the angular part of the metric, there occurs an IR color-flavor enhancement of the length scale as compared to a Planckian length scale in KS for ${\cal O}(1)$ $M$, thereby showing that quantum corrections will be suppressed. Using \cite{metrics}:
\begin{eqnarray}
\label{NeffMeffNfeff}
& & N_{\rm eff}(r) = N\left[ 1 + \frac{3 g_s M_{\rm eff}^2}{2\pi N}\left(\log r + \frac{3 g_s N_f^{\rm eff}}{2\pi}\left(\log r\right)^2\right)\right],\nonumber\\
& & M_{\rm eff}(r) = M + \frac{3g_s N_f M}{2\pi}\log r + \sum_{m\geq1}\sum_{n\geq1} N_f^m M^n f_{mn}(r),\nonumber\\
& & N^{\rm eff}_f(r) = N_f + \sum_{m\geq1}\sum_{n\geq0} N_f^m M^n g_{mn}(r).
\end{eqnarray}
it was argued in \cite{NPB} that  the length scale of the OKS-BH metric in the IR will be given by:
\begin{eqnarray}
\label{length-IR}
& & L_{\rm OKS-BH}\sim\sqrt{M}N_f^{\frac{3}{4}}\sqrt{\left(\sum_{m\geq0}\sum_{n\geq0}N_f^mM^nf_{mn}(\Lambda)\right)}\left(\sum_{l\geq0}\sum_{p\geq0}N_f^lM^p g_{lp}(\Lambda)\right)^{\frac{1}{4}}g_s^{\frac{1}{4}}\sqrt{\alpha^\prime}\nonumber\\
& & \equiv N_f^{\frac{3}{4}}\left.\sqrt{\left(\sum_{m\geq0}\sum_{n\geq0}N_f^mM^nf_{mn}(\Lambda)\right)}\left(\sum_{l\geq0}\sum_{p\geq0}N_f^lM^p g_{lp}(\Lambda)\right)^{\frac{1}{4}} L_{\rm KS}\right|_{\Lambda:\log \Lambda{<}{\frac{2\pi}{3g_sN_f}}},
\end{eqnarray}
which implies that  {in the IR, relative to KS, there is a color-flavor enhancement of the length scale in the OKS-BH metric}. Hence,  in the IR, even for $N_c^{\rm IR}=M=3$ and $N_f=2$ (light flavors) upon inclusion of of $n,m>1$  terms in
$M_{\rm eff}$ and $N_f^{\rm eff}$ in (\ref{NeffMeffNfeff}), $L_{\rm OKS-BH}\gg L_{\rm KS}(\sim L_{\rm Planck})$ in the MQGP limit involving $g_s\stackrel{\sim}{<}1$, implying that {the stringy corrections are suppressed and one can trust supergravity calculations}. As a reminder one will generate higher powers of $M$ and $N_f$ in the double summation in $M_{\rm eff}$ in (\ref{NeffMeffNfeff}), e.g., from the terms higher order in $g_s N_f$ in the RR and NS-NS three-form fluxes that become relevant for the aforementioned values of $g_s, N_f$.

 \item
  Further, the global  flavor group in the UV-IR interpolating and UV regions, due to presence of $N_f$ $D7$ and $N_f\ \overline{D7}$-branes, is $SU(N_f)\times SU(N_f)$, which is broken in the IR to $SU(N_f)$ as the IR has only $N_f$ $D7$-branes.

\end{enumerate}

Hence, the following features of the type IIB model of \cite{metrics} make it an ideal holographic dual of thermal QCD:

\begin{itemize}
\item
the theory having quarks transforming in the fundamental representation, is UV conformal and IR confining with the required chiral symmetry breaking in the IR and restoration at high temperatures

\item
the theory is UV complete with the gauge coupling remaining finite in the UV (absence of Landau poles)

\item
the theory is not just defined for large temperatures but for low and high temperatures

\item
(as will become evident in Sec. {\bf 3}) with the inclusion of a finite baryon chemical potential, the theory provides a lattice-compatible QCD confinement-deconfinement temperature $T_c$ for the right number of light quark flavors and masses, and is also thermodynamically stable; given the IR proximity of the value of the lattice-compatible $T_c$,  after the end of the Seiberg duality cascade, the number of quark flavors approximately equals $M$ which in the `MQGP' limit of (\ref{limits_Dasguptaetal-ii}) can be tuned to equal 3

\item
in the MQGP limit (\ref{limits_Dasguptaetal-ii}) which requires considering a finite gauge coupling and hence string coupling, the theory was shown in \cite{MQGP} to be holographically renormalizable from an M-theory perspective with the M-theory uplift also being thermodynamically stable.

\end{itemize}


\noindent (d) \underline{Supergravity solution on resolved warped deformed conifold}

The working metric is given by :
\begin{equation}
\label{metric}
ds^2 = \frac{1}{\sqrt{h}}
\left(-g_1 dt^2+dx_1^2+dx_2^2+dx_3^2\right)+\sqrt{h}\biggl[g_2^{-1}dr^2+r^2 d{\cal M}_5^2\biggr].
\end{equation}
 $g_i$'s are black hole functions in modified OKS(Ouyang-Klebanov-Strassler)-BH (Black Hole) background and are assumed to be:
$ g_{1,2}(r,\theta_1,\theta_2)= 1-\frac{r_h^4}{r^4} + {\cal O}\left(\frac{g_sM^2}{N}\right)$
where $r_h$ is the horizon, and the ($\theta_1, \theta_2$) dependence come from the
${\cal O}\left(\frac{g_sM^2}{N}\right)$ corrections. The  $h_i$'s are expected to receive corrections of
${\cal O}\left(\frac{g_sM^2}{N}\right)$ \cite{K. Dasgupta  et al [2012]}. We assume the same to also be true of the `black hole functions' $g_{1,2}$.  The compact five dimensional metric in (\ref{metric}), is given as:
\begin{eqnarray}
\label{RWDC}
& & d{\cal M}_5^2 =  h_1 (d\psi + {\rm cos}~\theta_1~d\phi_1 + {\rm cos}~\theta_2~d\phi_2)^2 +
h_2 (d\theta_1^2 + {\rm sin}^2 \theta_1 ~d\phi_1^2) +   \nonumber\\
&&  + h_4 (h_3 d\theta_2^2 + {\rm sin}^2 \theta_2 ~d\phi_2^2) + h_5~{\rm cos}~\psi \left(d\theta_1 d\theta_2 -
{\rm sin}~\theta_1 {\rm sin}~\theta_2 d\phi_1 d\phi_2\right) + \nonumber\\
&&  + h_5 ~{\rm sin}~\psi \left({\rm sin}~\theta_1~d\theta_2 d\phi_1 +
{\rm sin}~\theta_2~d\theta_1 d\phi_2\right),
\end{eqnarray}
$r\gg a, h_5\sim\frac{({\rm deformation\ parameter})^2}{r^3}\ll  1$ for $r \gg({\rm deformation\ parameter})^{\frac{2}{3}}$, i.e. in the UV/IR-UV interpolating region.  The $h_i$'s appearing in internal metric as well as $M, N_f$ are not constant and up to linear order depend on $g_s, M, N_f$ are given as below:
\begin{eqnarray}
\label{h_i}
& & \hskip -0.45in h_1 = \frac{1}{9} + {\cal O}\left(\frac{g_sM^2}{N}\right),\  h_2 = \frac{1}{6} + {\cal O}\left(\frac{g_sM^2}{N}\right),\ h_4 = h_2 + \frac{a^2}{r^2},\nonumber\\
& & h_3 = 1 + {\cal O}\left(\frac{g_sM^2}{N}\right),\ h_5\neq0,\
 L=\left(4\pi g_s N\right)^{\frac{1}{4}}.
\end{eqnarray}
One sees from (\ref{RWDC}) and (\ref{h_i}) that one has a non-extremal resolved warped deformed conifold involving
an $S^2$-blowup (as $h_4 - h_2 = \frac{a^2}{r^2}$), an $S^3$-blowup (as $h_5\neq0$) and squashing of an $S^2$ (as $h_3$ is not strictly unity). The horizon (being at a finite $r=r_h$) is warped squashed $S^2\times S^3$. In the deep IR, in principle one ends up with a warped squashed $S^2(a)\times S^3(\epsilon),\ \epsilon$ being the deformation parameter. Assuming $\epsilon^{\frac{2}{3}}>a$ and given that $a={\cal O}\left(\frac{g_s M^2}{N}\right)r_h$ \cite{K. Dasgupta  et al [2012]}, in the IR and in the MQGP limit, $N_{\rm eff}(r\in{\rm IR})=\int_{{\rm warped\ squashed}\ S^2(a)\times S^3(\epsilon)}F_5(r\in{\rm IR})\ll   M = \int_{S^3(\epsilon)}F_3(r\in{\rm IR})$; we have a confining $SU(M)$ gauge theory in the IR.

 The warp factor that includes the back-reaction, in the IR is given as:
\begin{eqnarray}
\label{eq:h}
&& \hskip -0.45in h =\frac{L^4}{r^4}\Bigg[1+\frac{3g_sM_{\rm eff}^2}{2\pi N}{\rm log}r\left\{1+\frac{3g_sN^{\rm eff}_f}{2\pi}\left({\rm
log}r+\frac{1}{2}\right)+\frac{g_sN^{\rm eff}_f}{4\pi}{\rm log}\left({\rm sin}\frac{\theta_1}{2}
{\rm sin}\frac{\theta_2}{2}\right)\right\}\Biggr],
\end{eqnarray}
where, in principle, $M_{\rm eff}/N_f^{\rm eff}$ are not necessarily the same as $M/N_f$; we however will assume that up to ${\cal O}\left(\frac{g_sM^2}{N}\right)$, they are. Proper UV behavior requires \cite{K. Dasgupta et al [2012]}:
\begin{eqnarray}
\label{h-large-small-r}
& & h = \frac{L^4}{r^4}\left[1 + \sum_{i=1}\frac{{\cal H}_i\left(\phi_{1,2},\theta_{1,2},\psi\right)}{r^i}\right],\ {\rm large}\ r;
\nonumber\\
& & h = \frac{L^4}{r^4}\left[1 + \sum_{i,j; (i,j)\neq(0,0)}\frac{h_{ij}\left(\phi_{1,2},\theta_{1,2},\psi\right)\log^ir}{r^j}\right],\ {\rm small}\ r.
\end{eqnarray}


  In the IR, up to ${\cal O}(g_s N_f)$ and setting $h_5=0$, the three-forms are as given in \cite{metrics}:
\begin{eqnarray}
\label{three-form-fluxes}
& & \hskip -0.4in (a) {\widetilde F}_3  =  2M { A_1} \left(1 + \frac{3g_sN_f}{2\pi}~{\rm log}~r\right) ~e_\psi \wedge
\frac{1}{2}\left({\rm sin}~\theta_1~ d\theta_1 \wedge d\phi_1-{ B_1}~{\rm sin}~\theta_2~ d\theta_2 \wedge
d\phi_2\right)\nonumber\\
&& \hskip -0.3in -\frac{3g_s MN_f}{4\pi} { A_2}~\frac{dr}{r}\wedge e_\psi \wedge \left({\rm cot}~\frac{\theta_2}{2}~{\rm sin}~\theta_2 ~d\phi_2
- { B_2}~ {\rm cot}~\frac{\theta_1}{2}~{\rm sin}~\theta_1 ~d\phi_1\right)\nonumber \\
&& \hskip -0.3in -\frac{3g_s MN_f}{8\pi}{ A_3} ~{\rm sin}~\theta_1 ~{\rm sin}~\theta_2 \left(
{\rm cot}~\frac{\theta_2}{2}~d\theta_1 +
{ B_3}~ {\rm cot}~\frac{\theta_1}{2}~d\theta_2\right)\wedge d\phi_1 \wedge d\phi_2, \nonumber\\
& & \hskip -0.4in (b) H_3 =  {6g_s { A_4} M}\Biggl(1+\frac{9g_s N_f}{4\pi}~{\rm log}~r+\frac{g_s N_f}{2\pi}
~{\rm log}~{\rm sin}\frac{\theta_1}{2}~
{\rm sin}\frac{\theta_2}{2}\Biggr)\frac{dr}{r}\nonumber \\
&& \hskip -0.3in \wedge \frac{1}{2}\Biggl({\rm sin}~\theta_1~ d\theta_1 \wedge d\phi_1
- { B_4}~{\rm sin}~\theta_2~ d\theta_2 \wedge d\phi_2\Biggr)
+ \frac{3g^2_s M N_f}{8\pi} { A_5} \Biggl(\frac{dr}{r}\wedge e_\psi -\frac{1}{2}de_\psi \Biggr)\nonumber  \\
&&  \wedge \Biggl({\rm cot}~\frac{\theta_2}{2}~d\theta_2
-{ B_5}~{\rm cot}~\frac{\theta_1}{2} ~d\theta_1\Biggr). \nonumber\\
\end{eqnarray}
The asymmetry factors in (\ref{three-form-fluxes}) are given by: $ A_i=1 +{\cal O}\left(\frac{a^2}{r^2}\ {\rm or}\ \frac{a^2\log r}{r}\ {\rm or}\ \frac{a^2\log r}{r^2}\right) + {\cal O}\left(\frac{{\rm deformation\ parameter }^2}{r^3}\right),$ $  B_i = 1 + {\cal O}\left(\frac{a^2\log r}{r}\ {\rm or}\ \frac{a^2\log r}{r^2}\ {\rm or}\ \frac{a^2\log r}{r^3}\right)+{\cal O}\left(\frac{({\rm deformation\ parameter})^2}{r^3}\right)$.    As in the UV, $\frac{({\rm deformation\ parameter})^2}{r^3}\ll  \frac{({\rm resolution\ parameter})^2}{r^2}$, we will assume the same three-form fluxes for $h_5\neq0$.

Further, to ensure UV conformality, it is important to ensure that the axion-dilaton modulus approaches a constant implying a vanishing beta function in the UV. This was discussed in detail in  appendix B of \cite{NPB}, wherein in particular, assuming an F-theory uplift involving, locally, an elliptically fibered $K3$, it was shown that UV conformality and the Ouyang embedding are mutually consistent.

\subsection{The `MQGP Limit'}

In \cite{MQGP}, we had considered the following two limits:
\begin{eqnarray}
\label{limits_Dasguptaetal-i}
&   & \hskip -0.17in (i) {\rm weak}(g_s){\rm coupling-large\ t'Hooft\ coupling\ limit}:\nonumber\\
& & \hskip -0.17in g_s\ll  1, g_sN_f\ll  1, \frac{g_sM^2}{N}\ll  1, g_sM\gg1, g_sN\gg1\nonumber\\
& & \hskip -0.17in {\rm effected\ by}: g_s\sim\epsilon^{d}, M\sim\left({\cal O}(1)\epsilon\right)^{-\frac{3d}{2}}, N\sim\left({\cal O}(1)\epsilon\right)^{-19d}, \epsilon\ll  1, d>0
 \end{eqnarray}
(the limit in the first line  though not its realization in the second line, considered in \cite{metrics});
\begin{eqnarray}
\label{limits_Dasguptaetal-ii}
& & \hskip -0.17in (ii) {\rm MQGP\ limit}: \frac{g_sM^2}{N}\ll  1, g_sN\gg1, {\rm finite}\
 g_s, M\ \nonumber\\
& & \hskip -0.17in {\rm effected\ by}:  g_s\sim\epsilon^d, M\sim\left({\cal O}(1)\epsilon\right)^{-\frac{3d}{2}}, N\sim\left({\cal O}(1)\epsilon\right)^{-39d}, \epsilon\lesssim 1, d>0.
\end{eqnarray}

Let us enumerate the motivation for considering the MQGP limit which was discussed in detail in \cite{NPB}. There are principally two.
\begin{enumerate}
\item
Unlike the AdS/CFT limit wherein $g_{\rm YM}\rightarrow0, N\rightarrow\infty$ such that $g_{\rm YM}^2N$ is large, for strongly coupled thermal systems like sQGP, what is relevant is $g_{\rm YM}\sim{\cal O}(1)$ and $N_c=3$. From the discussion in the previous paragraphs specially the one in point (c) of sub-section {\bf 2.1}, one sees that in the IR after the Seiberg duality cascade, effectively $N_c=M$ which in the MQGP limit of (\ref{limits_Dasguptaetal-ii})  can be tuned to 3. Further, in the same limit, the string coupling $g_s\stackrel{<}{\sim}1$. The finiteness of the string coupling necessitates addressing the same from an M theory perspective. This is the reason for coining the name: `MQGP limit'. In fact this is the reason why one is required to first construct a type IIA mirror, which was done in \cite{MQGP} a la delocalized Strominger-Yau-Zaslow mirror symmetry, and then take its M-theory uplift.

\item
From the perspective of calculational simplification in supergravity, the following are examples of the same and constitute therefore the second set of reasons for looking at the MQGP limit of (\ref{limits_Dasguptaetal-ii}):
\begin{itemize}
\item
In the UV-IR interpolating region and the UV,
$(M_{\rm eff}, N_{\rm eff}, N_f^{\rm eff})\stackrel{\rm MQGP}{\approx}(M, N, N_f)$
\item
Asymmetry Factors $A_i, B_j$(in three-form fluxes)$\stackrel{MQGP}{\rightarrow}1$  in the UV-IR interpolating region and the UV.

\item
Simplification of ten-dimensional warp factor and non-extremality function in MQGP limit
\end{itemize}
\end{enumerate}

 With ${\cal R}_{D5/\overline{D5}}$ denoting the boundary common to the UV-IR interpolating region and the UV region, $\tilde{F}_{lmn}, H_{lmn}=0$ for $r\geq {\cal R}_{D5/\overline{D5}}$ is required to ensure conformality in the UV.  Near the $\theta_1=\theta_2=0$-branch, assuming: $\theta_{1,2}\rightarrow0$ as $\epsilon^{\gamma_\theta>0}$ and $r\rightarrow {\cal R}_{\rm UV}\rightarrow\infty$ as $\epsilon^{-\gamma_r <0}, \lim_{r\rightarrow\infty}\tilde{F}_{lmn}=0$ and  $\lim_{r\rightarrow\infty}H_{lmn}=0$ for all components except $H_{\theta_1\theta_2\phi_{1,2}}$; in the MQGP limit and near $\theta_{1,2}=\pi/0$-branch, $H_{\theta_1\theta_2\phi_{1,2}}=0/\left.\frac{3 g_s^2MN_f}{8\pi}\right|_{N_f=2,g_s=0.6, M=\left({\cal O}(1)g_s\right)^{-\frac{3}{2}}}\ll  1.$ So, the UV nature too is captured near $\theta_{1,2}=0$-branch in the MQGP limit. This mimics addition of $\overline{D5}$-branes in \cite{metrics} to ensure cancellation of $\tilde{F}_3$.

\subsection{Approximate Supersymmetry, Construction of  the Delocalized SYZ IIA Mirror and Its M-Theory Uplift in the MQGP Limit}

A central issue to \cite{MQGP,transport-coefficients} has been implementation of delocalized mirror symmetry via the Strominger Yau Zaslow prescription according to which the mirror of a Calabi-Yau can be constructed via three T dualities along a special Lagrangian $T^3$ fibered over a large base in the Calabi-Yau. This sub-section is a quick review of precisely this.

{ To implement the quantum mirror symmetry a la S(trominger)Y(au)Z(aslow) \cite{syz}, one needs a special Lagrangian (sLag) $T^3$ fibered over a large base (to nullify contributions from open-string disc instantons with boundaries as non-contractible one-cycles in the sLag). Defining delocalized T-duality coordinates, $(\phi_1,\phi_2,\psi)\rightarrow(x,y,z)$ valued in $T^3(x,y,z)$ \cite{MQGP}:
\begin{equation}
\label{xyz defs}
x = \sqrt{h_2}h^{\frac{1}{4}}sin\langle\theta_1\rangle\langle r\rangle \phi_1,\ y = \sqrt{h_4}h^{\frac{1}{4}}sin\langle\theta_2\rangle\langle r\rangle \phi_2,\ z=\sqrt{h_1}\langle r\rangle h^{\frac{1}{4}}\psi,
\end{equation}
using the results of \cite{M.Ionel and M.Min-OO (2008)} it was shown in \cite{transport-coefficients,EPJC-2} that the following conditions are satisfied:
\begin{eqnarray}
\label{sLag-conditions}
& & \left.i^* J\right|_{\rm RC/DC} \approx 0,\nonumber\\
& & \left.\Im m\left( i^*\Omega\right)\right|_{\rm RC/DC} \approx 0,\nonumber\\
& & \left.\Re e\left(i^*\Omega\right)\right|_{\rm RC/DC}\sim{\rm volume \ form}\left(T^3(x,y,z)\right),
\end{eqnarray}
for the $T^2$-invariant sLag of \cite{M.Ionel and M.Min-OO (2008)} for a deformed conifold.  Hence, if the resolved warped deformed conifold is predominantly either resolved or deformed, the local $T^3$ of (\ref{xyz defs}) is the required sLag to effect SYZ mirror construction.}

{Interestingly, in the `delocalized limit' \cite{M. Becker et al [2004]}  $\psi=\langle\psi\rangle$, under the coordinate transformation:
\begin{equation}
\label{transformation_psi}
\left(\begin{array}{c} sin\theta_2 d\phi_2 \\ d\theta_2\end{array} \right)\rightarrow \left(\begin{array}{cc} cos\langle\psi\rangle & sin\langle\psi\rangle \\
- sin\langle\psi\rangle & cos\langle\psi\rangle
\end{array}\right)\left(\begin{array}{c}
sin\theta_2 d\phi_2\\
d\theta_2
\end{array}
\right),
\end{equation}
and $\psi\rightarrow\psi - \cos\langle{\bar\theta}_2\rangle\phi_2 + \cos\langle\theta_2\rangle\phi_2 - \tan\langle\psi\rangle ln\sin{\bar\theta}_2$, the $h_5$ term becomes $h_5\left[d\theta_1 d\theta_2 - sin\theta_1 sin\theta_2 d\phi_1d\phi_2\right]$, $e_\psi\rightarrow e_\psi$, i.e.,  one introduces an isometry along $\psi$ in addition to the isometries along $\phi_{1,2}$. This clearly is not valid globally - the deformed conifold does not possess a third global isometry}.

{ To enable use of SYZ-mirror duality via three T dualities, one also needs to ensure a large base (implying large complex structures of the aforementioned two two-tori) of the $T^3(x,y,z)$ fibration. This is effected via
\cite{F. Chen et al [2010]}:
\begin{eqnarray}
\label{SYZ-large base}
& & d\psi\rightarrow d\psi + f_1(\theta_1)\cos\theta_1 d\theta_1 + f_2(\theta_2)\cos\theta_2d\theta_2,\nonumber\\
& & d\phi_{1,2}\rightarrow d\phi_{1,2} - f_{1,2}(\theta_{1,2})d\theta_{1,2},
\end{eqnarray}
for appropriately chosen large values of $f_{1,2}(\theta_{1,2})$. The three-form fluxes
 remain invariant. The fact that one can choose such large values of $f_{1,2}(\theta_{1,2})$, was justified in \cite{MQGP}.  The guiding principle is that one requires the metric obtained after SYZ-mirror transformation applied to the non-K\"{a}hler  resolved warped deformed conifold is like a non-K\"{a}hler warped resolved conifold at least locally. Then $G^{IIA}_{\theta_1\theta_2}$ needs to vanish \cite{MQGP}. This was explicitly shown in \cite{NPB}.


As in Klebanov-Strassler construction, a single T-duality along a direction orthogonal to the $D3$-brane world volume, e.g., $z$ of (\ref{xyz defs}), yields $D4$ branes straddling a pair of $NS5$-branes consisting of world-volume coordinates $(\theta_1,x)$ and $(\theta_2,y)$. Further, T-dualizing along $x$ and then $y$ would yield a Taub-NUT space  from each of the two $NS5$-branes \cite{T-dual-NS5-Taub-NUT-Tong}. The $D7$-branes yield $D6$-branes which get uplifted to Kaluza-Klein monopoles in M-theory \cite{KK-monopoles-A-Sen} which too involve Taub-NUT spaces. Globally, probably the eleven-dimensional uplift would involve a seven-fold of $G_2$-structure, analogous to the uplift of $D5$-branes wrapping a two-cycle in a resolved warped conifold \cite{Dasguptaetal_G2_structure}.

\subsection{$G$-Structures}

The mirror type IIA metric after performing three T-dualities, first along $x$, then along $y$ and finally along $z$, utilizing the results of \cite{M. Becker et al [2004]} was worked out in \cite{MQGP}. The type IIA metric components  were worked out in \cite{MQGP}.

Now, any metric-compatible connection can be written in terms of the Levi-Civita connection and the contorsion tensor $\kappa$. Metric compatibility requires $\kappa\in\Lambda^1\otimes\Lambda^2$, $\Lambda^n$ being the space of $n$-forms. Alternatively, in $d$ complex dimensions, since $\Lambda^2\cong so(d)$, $\kappa$  also be thought of as $\Lambda^1\otimes so(d)$. Given the existence of a $G$-structure, we can decompose $so(d)$
into a part in the Lie algebra $g$ of $G \subset SO(d)$ and its orthogonal complement $g^\perp = so(d)/g$.  The contorsion $\kappa$ splits accordingly into
$\kappa = \kappa^0 + \kappa^g$, where $\kappa^0$ - the intrinsinc torsion - is the part in $\Lambda^1\otimes g^\perp$. One can decompose $\kappa^0$ into irreducible $G$ representations providing a classification of $G$-structures in terms of which representations appear in the decomposition. Let us consider the decomposition of $T^0$ in the case of $SU(3)$-structure. The relevant
representations are
$\Lambda^1\sim 3\oplus\bar{3}, g \sim 8, g^\perp\sim  1 \oplus 3 \oplus \bar{3}.$
Thus the intrinsic torsion, an element of $\Lambda^1\oplus su(3)^\perp$, can be decomposed into the following $SU(3)$ representations:
\begin{eqnarray}
& & \Lambda^1 \otimes su(3)^\perp = (3 \oplus \bar{3}) \otimes (1 \oplus 3 \oplus \bar{3)}
\nonumber\\
& & = (1 \oplus 1) \oplus (8 \oplus 8) \oplus (6 \oplus \bar{6}) \oplus (3 \oplus \bar{3}) \oplus (3 \oplus \bar{3})^\prime\equiv W_1\oplus W_2\oplus W_3\oplus W_4\oplus W_5.
\end{eqnarray}
The $SU(3)$ structure torsion classes \cite{torsion} can be defined in terms of J, $ \Omega $, dJ, $ d{\Omega}$ and
the contraction operator  $\lrcorner : {\Lambda}^k T^{\star} \otimes {\Lambda}^n
T^{\star} \rightarrow {\Lambda}^{n-k} T^{\star}$,  $J$ being given by:
$$ J  =  e^1 \wedge e^2 + e^3 \wedge e^4 + e^5 \wedge e^6, $$
and
the (3,0)-form $ \Omega $ being given by
$$ \Omega  =  ( e^1 + ie^2) \wedge (e^3 +
ie^4) \wedge (e^5 + ie^6). $$
The torsion classes are defined in the following way:
\begin{itemize}
\item
$W_1 \leftrightarrow [dJ]^{(3,0)}$, given by real numbers
$W_1=W_1^+ + W_1^-$
with $ d {\Omega}_+ \wedge J = {\Omega}_+ \wedge dJ = W_1^+ J\wedge J\wedge J$
and $ d {\Omega}_- \wedge J = {\Omega}_- \wedge dJ = W_1^- J \wedge J \wedge J$;

\item
$W_2 \leftrightarrow [d \Omega]_0^{(2,2)}$ :
$(d{\Omega}_+)^{(2,2)}=W_1^+ J \wedge J + W_2^+ \wedge J$
and $(d{\Omega}_-)^{(2,2)}=W_1^- J \wedge J + W_2^- \wedge J$;

\item
 $W_3 \leftrightarrow [dJ]_0^{(2,1)}$ is defined
as $W_3=dJ^{(2,1)} -[J \wedge W_4]^{(2,1)}$;

\item
 $W_4 \leftrightarrow J \wedge dJ$ : $W_4 =\frac{1}{2} J\lrcorner dJ$;

 \item
 $W_5 \leftrightarrow
[d \Omega]_0^{(3,1)}$: $W_5 = \frac{1}{2} {\Omega}_+\lrcorner d{\Omega}_+$
(the subscript 0 indicative of the primitivity of the respective forms).
\end{itemize}
 In \cite{transport-coefficients}, we saw that the five $SU(3)$ structure torsion classes, in the MQGP limit, satisfied (schematically):
\begin{eqnarray}
\label{T-IIB-i}
& & T_{SU(3)}^{\rm IIB}\in W_1 \oplus W_2 \oplus W_3 \oplus W_4 \oplus W_5 \sim \frac{e^{-3\tau}}{\sqrt{g_s N}} \oplus \left(g_s N\right)^{\frac{1}{4}} e^{-3\tau}\oplus \sqrt{g_s N}e^{-3\tau}\oplus -\frac{2}{3} \oplus -\frac{1}{2}
\end{eqnarray}
$(r\sim e^{\frac{\tau}{3}})$, such that
 \begin{equation}
 \label{T-IIB-ii}
 \frac{2}{3}W^{\bar{3}}_5=W^{\bar{3}}_4
 \end{equation}
  in the UV-IR interpolating region/UV, implying a Klebanov-Strassler-like supersymmetry
  \cite{Butti et al [2004]}. Locally around $\theta_1\sim\frac{1}{N^{\frac{1}{5}}}, \theta_2\sim\frac{1}{N^{\frac{3}{10}}}$, the type IIA torsion classes of the delocalized SYZ type IIA mirror metric (\ref{Type_IIA_metric}), were shown in \cite{NPB} to be:
 \begin{eqnarray}
 \label{T-IIA}
 T_{SU(3)}^{\rm IIA} &\in& W_2 \oplus W_3 \oplus W_4 \oplus W_5 \sim \gamma_2g_s^{-\frac{1}{4}} N^{\frac{3}{10}} \oplus g_s^{-\frac{1}{4}}N^{-\frac{1}{20}} \oplus g_s^{-\frac{1}{4}} N^{\frac{3}{10}} \oplus g_s^{-\frac{1}{4}} N^{\frac{3}{10}}\approx \gamma W_2\oplus W_4\oplus W_5\nonumber\\
 & & \stackrel{\scriptsize\rm fine\ tuning:\gamma\approx0}{\longrightarrow}\approx W_4\oplus W_5.
 \end{eqnarray}
   Further,
   \begin{equation}
   W_4\sim \Re e W_5
   \end{equation}
    indicative of supersymmetry after constructing the delocalized SYZ mirror.

Apart from quantifying the departure from $SU(3)$ holonomy due to intrinsic contorsion supplied by the NS-NS three-form $H$, via the evaluation of the $SU(3)$ structure torsion classes, to our knowledge for the first time in the context of holographic thermal QCD {\bf at finite gauge coupling} in \cite{NPB}: \\
(i) the existence of approximate supersymmetry of the type IIB holographic dual of \cite{metrics} in the MQGP limit near the coordinate branch $\theta_1=\theta_2=0$ was demonstrated, which apart from the existence of a special Lagrangian three-cycle (as shown in \cite{transport-coefficients,NPB}) is essential for construction of the local SYZ type IIA mirror;\\
 (ii)   it was demonstrated that the large-$N$ suppression of the deviation of the type IIB resolved warped deformed conifold from being a complex manifold, is lost on being duality-chased to type IIA - it was also shown that one further fine tuning  $\gamma_2=0$ in $W_2^{\rm IIA}$ can ensure that the local type IIA mirror is complex;\\
(iii)  for the local type IIA $SU(3)$ mirror,  the possibility of surviving approximate supersymmetry was demonstrated which is essential from the point of view of the end result of application of the SYZ mirror prescription.

We can get a one-form type IIA potential from the triple T-dual (along $x, y, z$) of the type IIB $F_{1,3,5}$ in \cite{MQGP} and using which the following $D=11$ metric was obtained in \cite{MQGP} ($u\equiv\frac{r_h}{r}$):
\begin{eqnarray}
\label{Mtheory met}
& &\hskip -0.6in   ds^2_{11} = e^{-\frac{2\phi^{IIA}}{3}} \left[g_{tt}dt^2 + g_{\mathbb{R}^3}\left(dx^2 + dy^2 + dz^2\right) +  g_{uu}du^2  +   ds^2_{IIA}({\theta_{1,2},\phi_{1,2},\psi})\right] \nonumber\\
& & \hskip -0.6in+ e^{\frac{4{\phi}^{IIA}}{3}}\Bigl(dx_{11} + A^{F_1}+A^{F_3}+A^{F_5}\Bigr)^2 \equiv\ {\rm Black}\ M3-{\rm Brane}+{\cal O}\left(\left[\frac{g_s M^2 \log N}{N}\right] \left(g_sM\right)N_f\right).
\end{eqnarray}

If $V$ is a seven-dimensional real vector space, then a three-form $\varphi $ is said to be positive if it lies in the $GL\left( 7,
\mathbb{R}\right) $ orbit of $\varphi _{0}$, where $\varphi_0$ is a three-form on $\mathbb{R}^7$ which is preserved by $G_2$-subgroup of $GL(7,\mathbb{R})$.  The pair $\left( \varphi ,g\right)$ for a positive $3$-form $\varphi $ and corresponding metric $g$ constitute a $G_{2}$-structure.
The space of $p$-forms decompose as following irreps of $G_{2}$ \cite{Grigorian}:
\begin{eqnarray}
\Lambda ^{1} &=&\Lambda _{7}^{1}  \label{l1decom} \nonumber\\
\Lambda ^{2} &=&\Lambda _{7}^{2}\oplus \Lambda _{14}^{2}  \label{l2decom} \nonumber\\
\Lambda ^{3} &=&\Lambda _{1}^{3}\oplus \Lambda _{7}^{3}\oplus \Lambda
_{27}^{3}  \label{l3decom} \nonumber\\
\Lambda ^{4} &=&\Lambda _{1}^{4}\oplus \Lambda _{7}^{4}\oplus \Lambda
_{27}^{4}  \label{l4decom} \nonumber\\
\Lambda ^{5} &=&\Lambda _{7}^{5}\oplus \Lambda _{14}^{5}  \label{l5decom} \nonumber\\
\Lambda ^{6} &=&\Lambda _{7}^{6}  \label{l6decom}
\end{eqnarray}
The subscripts denote the dimension of representation and components of same representation/dimensionality, are isomorphic to each other.
Let $M$ be a $7$-manifold with a $G_{2}$-structure $\left( \varphi ,g\right)
$.  Then the components of spaces of $2$-, $3$-, $4$-, and $5$-forms are given in \cite{Grigorian,Bryant}. The metric $g$ defines a reduction of the frame bundle F to a principal $SO\left( 7\right) $-sub-bundle $Q$, that is, a sub-bundle of oriented orthonormal frames. Now, $g$ also defines a Levi-Civita connection $\nabla $ on the tangent bundle
$TM$, and hence on $F$. However, the $G_{2}$-invariant $3$-form $\varphi $
reduces the orthonormal bundle further to a principal $G_{2}$-subbundle $Q$.
The Levi-Civita connection can be pulled back to $Q$. On $Q$,  $\nabla $ can be uniquely decomposed as
\begin{equation}
\nabla =\bar{\nabla}+\mathcal{T}  \label{tors}
\end{equation}
where $\bar{\nabla}$ is a $G_{2}$-compatible canonical connection on $P$, taking values in the sub-algebra $\mathfrak{g}_{2}\subset \mathfrak{so}%
\left( 7\right) $, while $\mathcal{T}$ is a $1$-form taking values in $%
\mathfrak{g}_{2}^{\perp }\subset \mathfrak{so}\left( 7\right) $; $\mathcal{T}$ is known as the intrinsic torsion of the
$G_{2}$-structure - the obstruction to the
Levi-Civita connection being $G_{2}$-compatible. Now $\mathfrak{so}%
\left( 7\right) $ splits under $G_{2}$ as
\begin{equation}
\mathfrak{so}\left( 7\right) \cong \Lambda ^{2}V\cong \Lambda _{7}^{2}\oplus
\Lambda _{14}^{2}.
\end{equation}
But $\Lambda _{14}^{2}\cong \mathfrak{g}_{2}$, so the orthogonal complement $\mathfrak{g%
}_{2}^{\perp }\cong \Lambda _{7}^{2}\cong V$. Hence $\mathcal{T}$ can be
represented by a tensor $T_{ab}$ which lies in $W\cong V\otimes V$. Now,
since $\varphi $ is $G_{2}$-invariant, it is $\bar{\nabla}$-parallel. So, the
torsion is determined by $\nabla \varphi $. from Lemma 2.24 of \cite{karigiannis-2007}:
\begin{equation}
\nabla \varphi \in \Lambda _{7}^{1}\otimes \Lambda _{7}^{3}\cong W.
\label{torsphiW}
\end{equation}%
Due to the isomorphism between the $\Lambda^{a=1,...,5}_7$s, $\nabla \varphi $ lies in the same space as $T_{AB}$ and thus
completely determines it. Equation (\ref{torsphiW}) is equivalent to:
\begin{equation}
\nabla _{A}\varphi _{BCD}=T_{A}^{\ \ E}\psi _{EBCD}  \label{fulltorsion}
\end{equation}%
where $T_{AB}$ is the full torsion tensor. Equation (\ref{fulltorsion}) can be inverted to yield:
\begin{equation}
T_{A}^{\ M}=\frac{1}{24}\left( \nabla _{A}\varphi _{BCD}\right) \psi ^{MBCD}.
\label{tamphipsi}
\end{equation}%
The tensor $T_A^{\ M}$, like the space W, possesses 49 components and hence fully defines $\nabla \varphi $. In general $T_{AB}$ cab be split into torsion components
as
\begin{equation}
T=T _{1}g+T _{7}\lrcorner \varphi +T _{14}+T _{27}
\label{torsioncomps}
\end{equation}
where $T _{1}$ is a function and gives the $\mathbf{1}$ component of $T$
. We also have $T _{7}$, which is a $1$-form and hence gives the $\mathbf{
7}$ component, and, $T _{14}\in \Lambda _{14}^{2}$ gives the $\mathbf{14}$
component. Further, $T _{27}$ is traceless symmetric, and gives the $\mathbf{27}$
component. Writing $T_i$ as $W_i$, we can split $W$ as
\begin{equation}
W=W_{1}\oplus W_{7}\oplus W_{14}\oplus W_{27}.  \label{Wsplit}
\end{equation}
From \cite{G2-Structure}, we see that a $G_2$ structure can be defined as:
\begin{equation}
\label{G_2_i}
\varphi_0 = \frac{1}{3!}{f}_{ABC}e^{ABC} = e^{-\phi^{IIA}}{f}_{abc}e^{abc} + e^{-\frac{2\phi^{IIA}}{3}}J\wedge e^{x_{10}},
\end{equation}
where $A,B,C=1,...,6,10; a,b,c,=1,...,6$ and ${f}_{ABC}$ are the structure constants of the imaginary octonions.
Using the same, the $G_2$-structure torsion classes were worked out around $\theta_1\sim\frac{1}{N^{\frac{1}{5}}}, \theta_2\sim\frac{1}{N^{\frac{3}{10}}}$ in \cite{NPB} to:
  \begin{equation}
  \label{G2}
  T_{G_2}\in W_2^{14} \oplus W_3^{27} \sim \frac{1}{\left(g_sN\right)^{\frac{1}{4}} }\oplus \frac{1}{\left(g_sN\right)^{\frac{1}{4}}}.
  \end{equation}
   Hence, the approach of the seven-fold, locally, to having a $G_2$ holonomy ($W_1^{G_2}=W_2^{G_2}=W_3^{G_2}=W_4^{G_2}=0$)  is accelerated in the MQGP limit.

As stated earlier, the global uplift to M-theory of the type $IIB$ background of \cite{metrics} is expected to involve a seven-fold of $G_2$ structure (not $G_2$-holonomy due to non-zero $G_4$). It is hence extremely important to be able to see this, at least locally. It is in this sense that the results of \cite{MQGP} are of great significance as one explicitly sees, for the first time, in the context of  holographic thermal QCD {\bf at finite gauge coupling}, though locally, the aforementioned $G_2$ structure having worked out the non-trivial $G_2$-structure torsion classes.

\section{$0^{++(***...}$ Glueball spectrum from type $IIB$ supergravity background}

In this section we discuss the $0^{++}$ glueball spectrum by solving the dilaton wave equation in the type $IIB$ background discussed in section $2$. The type $IIB$ metric as given in equation \ref{metric} with the warp factor $h$ given in \ref{eq:h} can be simplified by working around a particular value of $\theta_1$ and $\theta_2$: $\{\theta_1=\frac{1}{N^{1/5}}, \theta_2=\frac{1}{N^{3/10}}\}$, then keeping terms upto (N)ext to (L)eading (O)rder in $N$ in the large $N$ limit. The simplified type $IIB$ metric is given as:
 \begin{equation}
\label{IIB metric}
ds^2 =g_{tt}dt^2+g_{x_1x_1}\left(dx_1^2+dx_2^2+dx_3^2\right)+g_{rr}dr^2+\sqrt{h}r^2 d{\cal M}_5^2,
\end{equation}
with the components $g_{tt}, g_{x_1x_1}(=g_{x_2x_2}=g_{x_3x_3}), g_{rr}$ as given below:
\begin{equation}
\label{metriccomponent}
 g_{tt}=\frac{r^2 (1-B(r)) \left(\frac{r_h^4}{r^4}-1\right)}{2 \sqrt{\pi } \sqrt{N} \sqrt{g_s}}
 ~~~~g_{x_1x_1}=\frac{r^2 (1-B(r))}{2 \sqrt{\pi } \sqrt{N} \sqrt{g_s}}
~~~~g_{rr}=\frac{2 \sqrt{\pi } \sqrt{N} \left(r^2-3 a^2\right) (B(r)+1) \sqrt{g_s}}{r^4 \left(1-\frac{r_h^4}{r^4}\right)},
\end{equation}
where
 $B(r)=\frac{3 M^2 g_s \log (r) \left(12 N_f g_s \log (r)+6 N_f g_s+N_f g_s (-\log (N))-2 \log (4) N_f g_s+8 \pi \right)}{32 \pi ^2 N}$,
  and $a$ being the resolution parameter is proportional to the horizon radius $r_h$. Hence while computing the spectrum with a cut-off in the radial direction and no horizon, we must put both $r_h$ and $a$ to zero in the above equation.

Moreover the dilaton profiles with/without the black-hole are given below as:
\begin{eqnarray}
\label{dilaton}
& (a) & r_h\neq 0 :\nonumber\\
& & e^{-\Phi} = \frac{1}{g_s} - \frac{N_f}{8\pi}\log(r^6 + a^2 r^4) - \frac{N_f}{2\pi}\log\left(\sin\frac{\theta_1}{2}\sin\frac{\theta_2}{2}\right),\ r<{\cal R}_{D5/\overline{D5}},\nonumber\\
& & e^{-\Phi} = \frac{1}{g_s},\ r>{\cal R}_{D5/\overline{D5}};\nonumber\\
& (b) & r_h=0:\nonumber\\
& & e^{-\Phi} = \frac{1}{g_s} - \frac{3 N_f}{4\pi}\log r - \frac{N_f}{2\pi}\log\left(\sin\frac{\theta_1}{2}\sin\frac{\theta_2}{2}\right),\ r<\left|\mu_{\rm Ouyang}\right|^{\frac{2}{3}},\nonumber\\
& & e^{-\Phi} = \frac{1}{g_s},\ r>\left|\mu_{\rm Ouyang}\right|^{\frac{2}{3}}.
\end{eqnarray}

Again working around the particular choices of $\theta_1$ and $\theta_2$ the above profile can be simplified upto NLO in $N$.
The dilaton equation that has to be solved is given as:
\begin{equation}
\label{EOM-0++}
\begin{split}
\partial_{\mu}\left(e^{-2\Phi}\sqrt{g}g^{\mu\nu}\partial_{\nu}\phi\right)&=0.
\end{split}
\end{equation}

To solve the above dilaton equation we assume $\phi$ in (\ref{EOM-0++}) to be of the form $\phi=e^{i k.x}\tilde{\phi}(r)$. Now with this, we adopt the WKB method to get to the final result. The first step towards the WKB method is to convert the glueball equation of motion into a Schr\"{o}dinger-like equation. Then the WKB quantization condition can be applied on the potential term obtained from the Schr\"{o}dinger-like equation.  For $(0^{++})$ glueball spectrum with no horizon $(r_h=0)$, one of the solution was obtained by imposing Neumann boundary condition at the cut-off.

\subsection{$r_h\neq0$ using WKB quantization method}
In the following we discuss the spectrum of $0^{++}$ glueball in the type $IIB$ background with a black hole, implying a horizon of radius $r_h$ in the geometry. The results corresponding to the coordinate and field redefinitions of  \cite{Minahan},  are discussed below.

 Using the redefinitions of \cite{Minahan} with $r=\sqrt{y},r_h=\sqrt{y_h}$ and finally $y=y_h\left(1 + e^z\right)$, the $0^{++}$ EOM (\ref{EOM-0++}), with $k^2=-m^2$ can be written as:
\begin{equation}\label{ee}
\partial_{z}(E_{z}\partial_{z}\tilde{\phi})+y^2_{h}F_{z}m^2\tilde{\phi}=0,
\end{equation}
where at leading order in $N$, $E_z$ and $F_z$ are given with $L=(4\pi g_s N)^{1/4}$ as:
\begin{equation}
\begin{split}
E_{z}&=\frac{1}{128 \pi ^2 L^5 g_s^2}\left(e^z+2\right) y_h^2\Biggl(8 \pi+ g_s N_f \log {256}+2 N_f g_s \log {N}-3g_s N_f  \log \left[\left(e^z+1\right) y_h\right]\Biggr)
\\&\Biggl[2 \Biggl(3 a^2 \left\{4 \pi+ g_s N_f  (\log {16}-6) \right\}+\left(e^z+1\right) y_h \left\{8 \pi+g_s N_f\log {256}\right\}\Biggr)
\\&+2 g_s N_f \log {N} \left\{3 a^2+2 \left(e^z+1\right) y_h\right\}-3g_s N_f  \left\{3 a^2+2 \left(e^z+1\right) y_h\right\} \log \left[\left(e^z+1\right)
   y_h\right]\Biggr]
\end{split}
\end{equation}
\begin{equation}
\begin{split}
F_{z}&=\frac{1}{128 \pi ^2 g_s^2 L y_h \left(e^z+1\right)}e^z\left(4 \pi+ g_s N_f \log {16}+g_s N_f \log {N}-\frac{3}{2}g_s N_f  \log \left[\left(e^z+1\right) y_h\right]\right)
\\&\Biggl[\left(e^z+1\right) y_h \Biggl(8 \pi+2 g_s N_f \log {N}+g_s N_f \log {256} -3g_s N_f  \log \left[\left(e^z+1\right) y_h\right] \Biggr)
\\&-3 a^2 \left(4 \pi+6g_s N_f +g_s N_f  \log {N}+g_s N_f  \log {16} -\frac{3}{2}g_s N_f  \log \left[\left(e^z+1\right) y_h\right]\right)\Biggr]
\end{split}
\end{equation}
 Now, redefining the wave function $\tilde{\phi}$ as $\psi(z)=\sqrt{E_{z}}\tilde{\phi}(z)$ equation (\ref{ee}) reduces to a Schr\"{o}dinger-like equation
\begin{equation}\label{pote}
\left(\frac{d^2}{dy^2} + V(z)\right)\psi(z)=0
\end{equation}
 where the potential $V(z)$ is a rather cumbersome expression which we will not explicitly write out.
The WKB quantization condition becomes: $\int_{z_1}^{z_2}\sqrt{V(z)} = \left(n + \frac{1}{2}\right)\pi$ where $z_{1,2}$ are the turning points of $V(z)$. We will work below with a dimensionless glueball mass $\tilde{m}$ assumed to be large and defined via: $m = \tilde{m} \frac{r_h}{L^2}$. To determine the turning points of the potential $V(z)$, we consider two limits of the same - $r\in[r_h,\sqrt{3}a(r_h)\approx \sqrt{3} b r_h]\cup[\sqrt{3} b r_h,\infty)\equiv(IR,IR/UV\ {\rm interpolating}\ + UV)$. In the IR, we have to take the limit $z\rightarrow-\infty$.  Now in the large $\tilde{m}$ and large $\log{N}$ limit this potential at small $z$ can be shown to be given as:
\begin{eqnarray}
\label{Vsmall}
& & \hskip -1in V(z\ll0)= \frac{1}{8} \left(1-3 b^2\right) {e^z} {\tilde{m}}^2 + {\cal O}\left(e^{2z},\left(\frac{1}{{\tilde{m}}}\right)^2,\frac{1}{\log N}\right) <0\nonumber\\
   & &
\end{eqnarray}
Hence, there are no turning points in the IR.

Now, in the UV, apart from taking the large $z$ limit we also have to take $N_f=M=0$, to get:
\begin{eqnarray}
\label{Vlarger}
& & V(z\gg1) = -\frac{3 \left(b^2+1\right) \left({y_h}{\tilde{m}}^2+3\right)}{4 {y_h}{e^{2z}}}+\frac{3 b^2+{y_h}\tilde{m}^2+6}{4 {y_h}{e^z}}-1
   + {\cal O}(e^{-3z}).
\end{eqnarray}

The turning points of (\ref{Vlarger}) are:

$z_1=\frac{1}{8} \left(3 b^2-\sqrt{9 b^4-6 b^2 \left(7 {y_h}\tilde{m}^2+18\right)+{y_h}\tilde{m}^4-36 {y_h}\tilde{m}^2-108}+{y_h}\tilde{m}^2+6\right)$, \\$z_2=\frac{1}{8} \left(3 b^2+\sqrt{9 b^4-6 b^2 \left(7 {y_h}\tilde{m}^2+18\right)+{y_h}\tilde{m}^4-36 {y_h}\tilde{m}^2-108}+{y_h}\tilde{m}^2+6\right)$ which in the large $\tilde{m}$ limit is given as:
 $\Biggl\{z_1=(3 + 3 b^2) + {\cal O}\left(\frac{1}{\tilde{m}^2}\right), z_2=\frac{\tilde{m}^2}{4} - \frac{3(2 + 3 b^2)}{4} + {\cal O}\left(\frac{1}{\tilde{m}^2}\right)\Biggr\}$.

To obtain a real spectrum, one  first notes:
\begin{eqnarray}
\label{Vlargezlargemtilde}
& & \sqrt{V(z\gg1,N_f=M=0; b=0.6)} = \sqrt{0.25 e^{-z}-1.02 e^{-2 z}}\tilde{m} + {\cal O}\left(e^{-3z},\frac{1}{\tilde{m}^2}\right).
\end{eqnarray}
and
\begin{eqnarray}
\label{intsqrtVlarger}
& & \int\sqrt{V(z)}dz = \frac{\sqrt{e^{-2 z} \left(0.25 e^z-1.0023\right)} \tilde{m} \left(0.124856 e^z \tan ^{-1}\left(\frac{0.124856 e^z-1.00115}{\sqrt{0.25
   e^z-1.0023}}\right)-1. \sqrt{0.25 e^z-1.0023}\right)}{\sqrt{0.25 e^z-1.0023}}.\nonumber\\
   & & = \frac{\sqrt{e^{-2 z} \left(0.25 e^z-1.02\right)} \tilde{m} \left(0.123768 e^z \tan ^{-1}\left(\frac{0.123768 e^{1. z}-1.00995}{\sqrt{0.25
   e^z-1.02}}\right)-1. \sqrt{0.25 e^z-1.02}\right)}{\sqrt{0.25 e^z-1.02}}
\end{eqnarray}
Therefore
\begin{eqnarray}
\label{intsqrtVWKB}
& & \int_{4.08}^{0.25 \tilde{m}^2 - 2.31}\sqrt{V(z)} = 0.39 \tilde{m} - 2 = \left(n + \frac{1}{2}\right)\pi,
\end{eqnarray}
yielding:
\begin{equation}
\label{mn0++T>0}
m_n^{0^{++}} = 9.18 + 8.08 n.
\end{equation}

\subsection{Glueball Mass for $r_h=0$ and an IR cut-off $r_0$}

In this background the type $IIB$ metric and the dilaton profile has to be modified by the limit $r_h\rightarrow0$ and hence with $a\rightarrow0$. This time also we have provided two solutions to the dilaton equation. The first solutions was obtained by first following the redefinition of variables in \cite{WKB-i} and then imposing a neumann boundary condition at the radial cut-off. For the other solution we again consider the WKB method after a redefinition of variables as given in \cite{Minahan}.

\subsubsection{Neumann boundary condition at $r_0$}
Following \cite{WKB-i}, we redefine the radial coordinate as $z=\frac{1}{r}$. With this change of variable, the radial cut-off now maps to $z=z_0$, with $z_0=\frac{1}{r_0}$. The dilaton equation using the metric and the dilaton background in the limit $(r_h,a)\rightarrow0$ is given as:
\begin{equation}
\begin{split}
e^{2U}\partial_z\left(e^{-2U}\partial_z\tilde{\phi}\right)+\left(\frac{(4g_s N\pi)(B(z)+1)}{(1-B(z))}\right)(m^2)\tilde{\phi}=0,
\end{split}
\end{equation}
where upto NLO in $N$ we have,
\begin{equation}
\begin{split}
e^U & =\frac{8 \times 2^{1/4} g_s^{13/8}\pi^{13/8}N^{5/8} z^{3/2}}{4 \left(\pi  \log (4) N_f g_s+\pi \right)+2 \pi  N_f g_s \log (N)+3 N_f g_s \log (z)}
\\ & -\frac{15
   M^2 z^{3/2} g_s^{21/8} \log (z) \left(-12 N_f g_s \log (z)+6 N_f g_s+N_f g_s (-\log (N))-\log (16) N_f g_s+8 \pi \right)}{8\ 2^{3/4} \pi ^{3/8} N^{3/8}
   \left(3 N_f g_s \log (z)+2 \pi  N_f g_s \log (N)+4 \left(\pi  \log (4) N_f g_s+\pi \right)\right)}.
\end{split}
\end{equation}
Now to convert the above equation in a one-dimensional schrodinger like form we introduce a new field variable $\psi(z)$ as: $\psi(z)=e^{-U}\tilde{\phi}(z)$.

With this one can write the equation in the following schrodinger like form,
\begin{equation}
\begin{split}\label{pot}
\frac{\partial^2\psi(z)}{\partial z^2} & = V(z)\psi(z).
\end{split}
\end{equation}
     The potential $V(z)$, in the large-$N$ large-$\log{N}$ limit is given as:
\begin{eqnarray}
\label{VV}
& & \hskip -0.6in V(z)= 4 \pi  {g_s} m^2 N+\frac{6}{\pi  z^2 \log (N)}-\frac{15}{4 z^2} + {\cal O}\left(\frac{1}{(\log N)^2},\frac{g_sM^2}{N}\right).
\end{eqnarray}
Hence, the Schr\"{o}dinger equation becomes:
\begin{equation}
\label{Schroedinger}
\psi''(z)+\psi(z) \left(4 \pi  {g_s} m^2 N+\frac{6}{\pi  z^2 \log (N)}-\frac{15}{4 z^2}\right)=0,
\end{equation}
whose solution is given as under:
\begin{eqnarray}
\label{Schroedinger-solution}
& & \psi(z) = c_1 \sqrt{z} J_{\sqrt{\frac{2}{\pi }} \sqrt{\frac{2 \pi  \log (N)-3}{\log (N)}}}\left(2 \sqrt{{g_s}} m \sqrt{N} \sqrt{\pi } z\right)+c_2 \sqrt{z}
   Y_{\sqrt{\frac{2}{\pi }} \sqrt{\frac{2 \pi  \log (N)-3}{\log (N)}}}\left(2 \sqrt{{g_s}} m \sqrt{N} \sqrt{\pi } z\right).
\end{eqnarray}
Requiring finiteness of $\psi(z)$ at $z=0$ requires setting $c_2=0$. Then imposing Neumann boundary condition on $\tilde{\phi}(z)$ at $z=z_0$ implies:
\begin{eqnarray}
\label{Df}
& & \hskip -0.4in \tilde{\phi}'(z_0) = \frac{1}{\left(-2 \pi  {g_s} {N_f} \log \left(\frac{1}{N}\right)+3 {g_s}
   {N_f} \log (z_0)+4 (\pi  {g_s} {N_f} \log (4)+\pi )\right)^2}\nonumber\\
   & & \hskip -0.4in\times\Biggl\{4 \sqrt[4]{2} \pi ^{13/8} {g_s}^{13/8} N^{5/8} z_0 \Biggl[3 \Biggl(-2 \pi  {g_s} {N_f} \log \left(\frac{1}{N}\right)+3 {g_s} {N_f} \log (z_0)-2
   {g_s} {N_f}+4 \pi  {g_s} {N_f} \log (4)+4 \pi \Biggr)\nonumber\\
    & & \hskip -0.4in\times J_{\sqrt{4-\frac{6}{\pi  \log (N)}}}\left(2 \sqrt{{g_s}} m \sqrt{N} \sqrt{\pi }
   z_0\right)+\left(-2 \pi  {g_s} {N_f} \log \left(\frac{1}{N}\right)+3 {g_s} {N_f} \log (z_0)+4 (\pi  {g_s} {N_f} \log (4)+\pi )\right)\nonumber\\
   & &\hskip -0.4in \times \Biggl(2
   \sqrt{\pi } \sqrt{{g_s}} m \sqrt{N} z_0 J_{\sqrt{4-\frac{6}{\pi  \log (N)}}-1}\left(2 \sqrt{{g_s}} m \sqrt{N} \sqrt{\pi } z_0\right) -2 \sqrt{\pi }
   \sqrt{{g_s}} m \sqrt{N} z_0 J_{\sqrt{4-\frac{6}{\pi  \log (N)}}+1}\left(2 \sqrt{{g_s}} m \sqrt{N} \sqrt{\pi } z_0\right)\nonumber\\
   & &\hskip -0.4in +J_{\sqrt{4-\frac{6}{\pi  \log
   (N)}}}\left(2 \sqrt{{g_s}} m \sqrt{N} \sqrt{\pi } z_0\right)\Biggr)\Biggr]\Biggr\}=0,
\end{eqnarray}
implying in the large-$N$ large-$z$ (as the Neumann boundary condition will be implemented in the IR) limit:
\begin{eqnarray}
\label{Neumann_bc}
& & \frac{1}{2} x_0 J_{\sqrt{4-\frac{6}{\pi  \log (N)}}-1}(x_0)-\frac{1}{2} x_0 J_{\sqrt{4-\frac{6}{\pi  \log (N)}}+1}(x_0)+2 J_{\sqrt{4-\frac{6}{\pi  \log (N)}}}(x_0)=0,
\end{eqnarray}
where $x_0\equiv 2\sqrt{g_s N \pi}m z_0$. The graphical solution points out that the ground state has a zero mass and the lightest (first excited state) glueball mass is approximately given by $3.71 \frac{r_0}{L^2}$.

\begin{figure}
\begin{center}
 \includegraphics[scale=0.8]
 {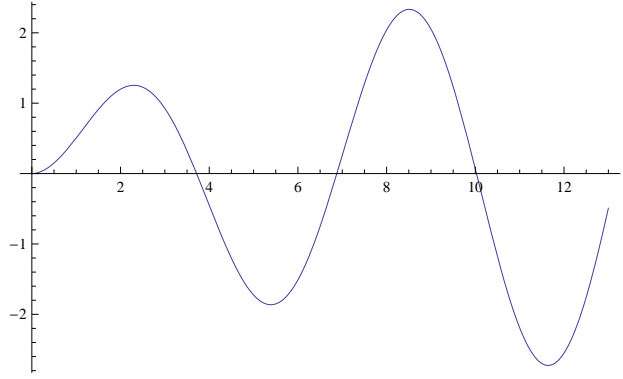}
 \end{center}
 \caption{$0^{++*,++**,++***}$ Masses obtained by graphical solution of the Neumann boundary condition in the $r_h=0$ Limit}
\end{figure}

\subsubsection{WKB method: Including the Non-Conformal/NLO (in $N$) Corrections}

Again following the redefinition of variables as given in \cite{Minahan}: $r=\sqrt{y}, y=y_0\left(1 + e^z\right)$, where $r_0=\sqrt{y_0}$ is the radial cut-off, and using the type $IIB$ metric as well as the dilaton profile in the limit $(rh,a)\rightarrow 0$,  the dilaton equation (\ref{EOM-0++}) can be written as:
\begin{equation}\label{ff}
\partial_{z}(C_{z}\partial_{z}\tilde{\phi})+y^2_{h}D_{z}m^2\tilde{\phi}=0,
\end{equation}
where $C_{z}$ and $D_{z}$ are given up to NLO in $N$ by,
\begin{equation}\begin{split}
C_{z}=\frac{1}{32768 \sqrt{2} ~\pi ^{21/4} N^{9/4} g_s^{13/4}} e^{-z} \left(e^z+1\right)^3~y_0^3\Biggl(8 \pi+2 g_s N_f \log {N}+4 g_s N_f  \log {4} -3 g_s N_f  \log \left[y_0 \left(e^z+1\right)\right]\Biggr)^2
\\\Biggl(128 \pi ^2 N+15 g_s  M^2 \left\{-8 \pi+g_s N_f
   (\log {16}-6) +g_s N_f  \log {N} \right\}\log \left[ \left(e^z+1\right)y_0\right]-90 M^2 g_s^2 N_f  \log \left[ \left(e^z+1\right)y_0\right]^2\Biggr)
\end{split}
\end{equation}
\begin{equation}\begin{split}
D_{z}&=\frac{1}{32768 \sqrt{2}~ \pi^{17/4} N^{5/4} g_s^{9/4}} e^{z}\Biggl(8 \pi+2 g_s N_f \log {N}+4 g_s N_f  \log {4} -3 g_s N_f  \log \left[y_0 \left(e^z+1\right)\right]\Biggr)^2
\\&\Biggl(128 \pi ^2 N+3 g_s  M^2 \left\{-8 \pi+g_s N_f
   (\log {16}-6) +g_s N_f  \log {N} \right\}\log \left[ \left(e^z+1\right)y_0\right]-18 M^2 g_s^2 N_f  \log \left[ \left(e^z+1\right)y_0\right]^2\Biggr)
\end{split}
\end{equation}

Defining a new variable $\psi(z)$ such that: $\psi(z)=\sqrt{C_z}\tilde{\phi}(z)$, the above equation can be converted into a Schr\"{o}dinger-like equation,
\begin{equation}\label{poten}
\left(\frac{d^2}{dy^2} + V(z)\right)\psi(z)=0,
\end{equation}
where the potential is given up to NLO in $N$ as:
\begin{eqnarray}
\label{V0++-IR}
& & \hskip -0.7in V(z) = -\frac{1}{4} -\frac{3 {e^{2z}} {g_s}^2 {\log N} M^2 \tilde{m}^2 {N_f} \log ({y_0})}{128 \pi ^2 N}\nonumber\\
 & & \hskip -0.7in +\frac{{e^{2z}} \left(2 \left({g_s} {N_f}
   \left({\log N} \left(\tilde{m}^2-3\right)+\tilde{m}^2 \log (16)+12-6 \log (4)\right)+4 \pi  \left(\tilde{m}^2-3\right)\right)-3 {g_s}
   \left(\tilde{m}^2-3\right) {N_f} \log ({y_0})\right)}{4 (2 {g_s} {N_f} ({\log N}+\log (16))-3 {g_s} {N_f} \log ({y_0})+8 \pi
   )}\nonumber\\
   & & \hskip -0.7in + {\cal O}\left(\frac{1}{N^2}\right).
\end{eqnarray}
The domain of integration over which $V(z)>0$ can be shown to be:\\ $\{\log\left(\frac{3 {g_s}^2 M^2 {N_f} \log (N) \log ({y_0})}{64 \pi ^2 \tilde{m} N}+\frac{1}{\tilde{m}}\right),\log\left(\delta^2-1\right)\}$, where $\mu^{\frac{2}{3}} = \delta \sqrt{y_0}$. Note that the IR cut-off $r_0$ or $y_0$ is not put in by hand but is proportional to the Ouyang embedding parameter (corresponding to the embedding of the flavor $D7$-branes
) raised to two-third power. The proportionality constant $\delta$ could be determined, as discussed in point number 5 in Section {\bf 8}, by matching with lattice calculations and turns out to be ${\cal O}(1)$.

Expanding $\sqrt{V}$ first in $N$ and then in $\tilde{m}$ and then integrating over the above mentioned domain, one gets the following quantization condition,
\begin{eqnarray}
\label{WKB-integral-ii}
& & \int_{z=\log \left(\frac{3 M^2 {g_s}^2  {N_f}{\log N}  \log {y_0}}{64 \pi ^2 \tilde{m} N}+\frac{1}{\tilde{m}}\right)}^{\log(\delta^2-1)}\sqrt{V(z)}\nonumber\\
 & & = \frac{1}{2} \left(\delta ^2-1\right) \tilde{m} \left(\frac{3 M^2{g_s}^2{N_f} {\log N}   \log {y_0}}{64 \pi ^2 N}+1\right) - 0.75 + {\cal O}\left(\frac{1}{\tilde{m}N},\frac{1}{\tilde{m}^2}\right) = \left(n + \frac{1}{2}\right)\pi,
\end{eqnarray}
 yielding:
\begin{equation}
\label{WKB-result0++_T=0}
m_n^{0^{++}}=\frac{(6.28319 n+4.64159) \left(1-\frac{0.00474943 {g_s}^2 {\log N} M^2 {N_f} \log ({y_0})}{N}\right)}{\delta ^2-1}.
\end{equation}

\section{Scalar Glueball ($0^{-+(***...)}$) Masses}

As $Tr(F\tilde{F})$ has $P=-,C=+$ and it couples to $A_1$ in the Wess-Zumino term for the type IIA $D4-$brane: $\int_{\Sigma_{4,1}}A\wedge F\wedge F$, one considers $A_1$'s EOM:
\begin{equation}
\label{0-+-i}
\partial_\nu\left(\sqrt{g^{\rm IIA}}g_{\rm IIA}^{\mu\sigma}g_{\rm IIA}^{\nu\rho}\left(\partial_{[\sigma}A_{\rho]}\right)\right)=0,
\end{equation}
where $\mu,\nu,...=a(\equiv 0,1,2,3),r,\alpha(\equiv 5,...,9)$. Like \cite{Czaki_et_al-0-+}, assume $A_\mu=\delta^{\theta_2}_\mu a_{\theta_2}(r)e^{i k\cdot x}, k^2=-m^2$ as the fluctuation about the type IIA $A_1$ that was worked out in \cite{MQGP}. The $0^{-+}$ EOM then reduces to:
\begin{equation}
\label{0-+-ii}
 \sqrt{g^{\rm IIA}}g_{\rm IIA}^{\theta_2\theta_2}g_{\rm IIA}^{rr}a_{\theta_2}^{\prime\prime}(r) + \partial_r\left(\sqrt{g^{\rm IIA}}g_{\rm IIA}^{\theta_2\theta_2}g_{\rm IIA}^{rr}\right)a_{\theta_2}^\prime(r) + \sqrt{h}m^2\sqrt{g^{\rm IIA}}g_{\rm IIA}^{\theta_2\theta_2}a_{\theta_2}(r)= 0.
\end{equation}

\subsection{Neumann/Dirichlet Boundary Conditions}

\subsubsection{$r_h\neq0$}

Then taking the large-N limit followed by a small-$\theta_1$ small-$\theta_2$ limit one can show that the equation of motion (\ref{0-+-ii}) yields::
\begin{eqnarray}
\label{0-+-vi}
& & -\frac{8 \pi  \left(r^4-{r_h}^4\right) a_{\theta_2}^{\prime\prime}(r)}{\sqrt{3}}+\frac{32 \pi  \left(r^4-2 {r_h}^4\right) a^{\prime}_{\theta_2}(r)}{\sqrt{3} r}-32
   \sqrt[6]{3} \pi ^2 {g_s} m^2 N {a_{\theta_2}}(r)=0.
\end{eqnarray}
Working near $r=r_h$, approximating (\ref{0-+-vii}) by:
\begin{eqnarray}
\label{0-+-vii}
& & -\frac{32 \pi  {r_h}^3 (r-{r_h}) {a_{\theta_2}}''(r)}{\sqrt{3}}+\left(\frac{160 \pi  {r_h}^2 (r-{r_h})}{\sqrt{3}}-\frac{32 \pi
   {r_h}^3}{\sqrt{3}}\right) {a_{\theta_2}}'(r)-32 \sqrt[6]{3} \pi ^2 {g_s} m^2 N {a_{\theta_2}}(r)=0,
\end{eqnarray}
whose solution is given as under:
\begin{eqnarray}
\label{0-+-viii}
a_{\theta_2}(r) = c_1 U\left(-\frac{3^{2/3} {g_s} m^2 N \pi }{5 {r_h}^2},1,\frac{5 r}{{r_h}}-5\right)+c_2 L_{\frac{3^{2/3} \pi  {g_s} m^2 N}{5
   {r_h}^2}}\left(\frac{5 r}{{r_h}}-5\right)
\end{eqnarray}
Imposing Neumann boundary condition: $a_{\theta_2}^\prime(r=r_h)=0$, utilizing:
\begin{eqnarray}
\label{0-+-ix}
& & U\left(1-\frac{3^{2/3} \pi  {g_s} m^2 N}{5 |{r_h}|^2},2,\frac{5 r}{{r_h}}-5\right) = \nonumber\\
& &  -\frac{{r_h}^3}{3^{2/3} \pi  {g_s} m^2 N (r-{r_h}) \Gamma \left(-\frac{3^{2/3} {g_s} m^2 N \pi }{5 {r_h}^2}\right)}+\frac{\psi
   ^{(0)}\left(1-\frac{3^{2/3} {g_s} m^2 N \pi }{5 {r_h}^2}\right)+\log \left(\frac{5 r}{{r_h}}-5\right)+2 \gamma -1}{\Gamma
   \left(-\frac{3^{2/3} {g_s} m^2 N \pi }{5 {r_h}^2}\right)}\nonumber\\
   & & +\frac{(r-{r_h}) \left(5 {r_h}^2-3^{2/3} \pi  {g_s} m^2 N\right) \left(2
   \psi ^{(0)}\left(2-\frac{3^{2/3} {g_s} m^2 N \pi }{5 {r_h}^2}\right)+2 \log (r-{r_h})+2 \log \left(\frac{5}{{r_h}}\right)+4 \gamma
   -5\right)}{4 {r_h}^3 \Gamma \left(-\frac{3^{2/3} {g_s} m^2 N \pi }{5 {r_h}^2}\right)}\nonumber\\
   & & +{\cal O}\left((r-{r_h})^2\right),
\end{eqnarray}
$r_h=T\sqrt{4\pi g_s N}$ - up to LO in $N$ - requires $c_2=0$ and
 \begin{eqnarray}
 \label{0-+-x}
 \frac{3^{2/3} m^2}{20 T^2}=n,
 \end{eqnarray}
 implying:
 \begin{eqnarray}
 \label{0-+-xi}
 m_n^{0^{-+}} = \frac{2 \sqrt{5} \sqrt{n} T}{\sqrt[3]{3}}.
 \end{eqnarray}
 One can show that imposing Dirichlet boundary condition $a_{\theta_2}(r=r_h)=0$, yields the same spectrum as (\ref{0-+-xi}).
If the temperature $T$ gets identified with $a$ of \cite{Czaki_et_al-0-+}, then the ground state, unlike \cite{Czaki_et_al-0-+}, is massless; the excited states for lower $n$'s are closer to $a=0$ and the higher excited states are closer to $a\rightarrow\infty$ in \cite{Czaki_et_al-0-+}.

\subsubsection{$r_h=0$ Limit of (\ref{0-+-vi})}

The $r_h=0$ limit of (\ref{0-+-vi}) gives:
\begin{equation}
\label{0-+_EOM_T=0}
\sqrt{3} r^4 {a_{\theta_2}}''(r)-4 \sqrt{3} r^3 {a_{\theta_2}}'(r)+3 \sqrt[6]{3} {\tilde{m}}^2 r_0^2 {a_{\theta_2}}(r) = 0,
\end{equation}
which near $r=r_0$ yields:
\begin{eqnarray}
\label{0-+_EOM_solution}
a_{\theta_2}(r) = (4 r-3 \tilde{m})^{5/4} \left(c_1 U\left(\frac{5}{4}-\frac{\tilde{m}^2}{4 \sqrt[3]{3}},\frac{9}{4},\frac{3 r}{\tilde{m}}-\frac{9}{4}\right)+c_2
   L_{\frac{1}{12} \left(3^{2/3} \tilde{m}^2-15\right)}^{\frac{5}{4}}\left(\frac{3 r}{\tilde{m}}-\frac{9}{4}\right)\right).
\end{eqnarray}
Imposing Neumann boundary condition on (\ref{0-+_EOM_solution}) yields:
\begin{eqnarray}
\label{0-+_T=0_spectrum}
& & m^{0^{-+}}(r_h=0) = 0\nonumber\\
& & m^{0^{-+*}}(r_h=0) \approx 3.4 \frac{r_0}{L^2}\nonumber\\
& & m^{0^{-+**}}(r_h=0) \approx 4.35 \frac{r_0}{L^2}.
\end{eqnarray}
One can similarly show that imposing Dirichlet boundary condition on (\ref{0-+_EOM_solution}) for $c_2=0$ yields:
\begin{eqnarray}
\label{0-+_T=0_spectrum-ii}
& & m^{0^{-+}}(r_h=0) = 0\nonumber\\
& & m^{0^{-+*}}(r_h=0) \approx 3.06 \frac{r_0}{L^2}\nonumber\\
& & m^{0^{-+**}}(r_h=0) \approx 4.81 \frac{r_0}{L^2}.
\end{eqnarray}

\subsubsection{WKB Quantization for $r_h\neq0$ }

The potential corresponding to the Schr\"{o}dinger-like equation a la \cite{Minahan}, substituting $m = \tilde{m} \frac{\sqrt{y_h}}{L^2}$, is given by:
\begin{eqnarray}
\label{V_0-+-Minahan-i}
& & V = \frac{e^z \left(4\ 3^{2/3} \tilde{m}^2 \left(3 e^z+e^{2 z}+2\right)-64 e^z-108 e^{2 z}-25 e^{3 z}+96\right)}{16
   \left(e^z+1\right)^2 \left(e^z+2\right)^2}.
\end{eqnarray}
Therefore, in the IR:
\begin{eqnarray}
\label{V_0-+Minahan-ii}
& & V(z\ll0)= \left(-\frac{3}{16} 3^{2/3} \tilde{m}^2-\frac{11}{2}\right) e^{2 z}+\left(\frac{1}{8} 3^{2/3}
   \tilde{m}^2+\frac{3}{2}\right) e^z + {\cal O}(e^{-3z}),
\end{eqnarray}
the turning points being given by $-\infty$ and $\log\left(\frac{2\ 3^{2/3} \tilde{m}^2+24}{3\ 3^{2/3} \tilde{m}^2+88}\right)\approx-0.405$. But only $z\in(-\infty,-2.526]$ corresponds to the IR in our calculations. So,
\begin{equation}
\label{WKB-Minahan-0-+_i}
\int_{-\infty}^{-2.526}\sqrt{V} = 0.283 \tilde{m} = \left(n + \frac{1}{2}\right)\pi,
\end{equation}
which obtains:
\begin{equation}
\label{WKB-Minahan-0-+_ii}
 m_n^{0^{-+}}(T, IR) = 5.56 (1 + 2n)\frac{r_h}{L^2}.
\end{equation}

Similarly, in the UV:
\begin{equation}
\label{V0-+_T_UV}
V(UV,T) = \left({\frac{1}{4} 3^{2/3} \tilde{m}^2+\frac{21}{8}}\right)e^{-z}+\left(\frac{9}{16}-\frac{3}{4} 3^{2/3}
   \tilde{m}^2\right) e^{-2 z}-\frac{25}{16} + {\cal O}(e^{-3z}),
\end{equation}
whose turning points are: $\log\left(\frac{1}{25} \left(2\ 3^{2/3} \tilde{m}^2\pm\sqrt{6} \sqrt{2 \sqrt[3]{3} \tilde{m}^4-36\ 3^{2/3}
   \tilde{m}^2+111}+21\right)\right) \\= \log(3 + {\cal O}\left(\frac{1}{\tilde{m}^2}\right)),\log \left(0.33 \tilde{m}^2 - 1.32 + {\cal O}\left(\frac{1}{\tilde{m}^2}\right)\right) $. Now:
   $\sqrt{V(UV,T)} = \frac{1}{2} \sqrt[3]{3} \tilde{m} e^{-z} \sqrt{e^z-3} + {\cal O}\left(\frac{1}{\tilde{m}}\right)$. Therefore,
   \begin{equation}
   \label{WKB0-+_UV_T}
   \int_{\log 3}^{\log (0.33 \tilde{m}^2 - 1.32)}\sqrt{V(UV,T)} = 0.654 \tilde{m} - 2.5 = \left(n + \frac{1}{2}\right)\pi,
   \end{equation}
   which obtains:
   \begin{equation}
   \label{mn0-+_UV_T}
   m_n^{0^{-+}}(UV,T) = \left(6.225 + 4.804 n\right)\pi T.
   \end{equation}

\subsubsection{WKB Quantization  at $r_h=0$}

The `potential' is given by:
\begin{eqnarray}
\label{V0-+_T=0}
V(0^{-+},r_h=0) = \frac{4 3^{2/3} \tilde{m}^2 e^{2 z}+\frac{e^{2 z} \left(e^z+1\right) \left(5 e^z+12\right)^2}{\left(e^z+2\right)^2}-2\left(e^z+1\right)
\left(14 e^z+25 e^{2 z}+4\right)}{16 \left(e^z+1\right)^3} + {\cal O}\left(\frac{g_sM^2}{N}\right).
\end{eqnarray}
Therefore, in the IR:
\begin{equation}
\label{V0-+_IR-T=0}
V(IR,r_h=0) = -\frac{1}{2} - \frac{3}{4}e^z + \left(\frac{9}{8} + \frac{3^{\frac{2}{3}}}{4}\tilde{m}^2\right)e^{2z} + {\cal O}(e^{3z}),
\end{equation}
and in the IR, the domain of integration becomes: $[\log\left(\frac{\sqrt{2}}{3^{\frac{1}{3}}\tilde{m}}\right),\log(\delta^2-1)]$:
$\int_{\log\left(\frac{\sqrt{2}}{3^{\frac{1}{3}}\tilde{m}}\right)}^{\log(\delta^2-1)}\sqrt{V(IR,r_h=0)} = \frac{3^{\frac{1}{3}}(\delta^2-1)}{2} \tilde{m} - 1.1126$, yielding:
\begin{equation}
\label{mn0-+_T=0_IR}
m_n^{0^{-+}}(IR,r_h=0) = \frac{\left(3.72 + 4.36 n\right)}{\left(\delta^2-1\right)}\frac{r_0}{L^2}.
\end{equation}
Also, in the UV:
\begin{equation}
\label{V0-+_UV-T=0}
V(UV,r_h=0) = \left(-\frac{3}{4} 3^{2/3} \tilde{m}^2-\frac{103}{16}\right) e^{-2 z}+\left(\frac{1}{4} 3^{2/3} \tilde{m}^2+\frac{21}{8}\right)
   e^{-z}-\frac{25}{16},
\end{equation}
whose turning points are: $\frac{1}{25} \left(2 3^{2/3} \tilde{m}^2\pm\sqrt{12 \sqrt[3]{3} \tilde{m}^4-216 3^{2/3} \tilde{m}^2-2134}+21\right)=\left(3 + {\cal O}\left(\frac{1}{\tilde{m}^2}\right),\frac{4}{25}3^{\frac{2}{3}}\tilde{m}^2 - \frac{33}{25}\right)$, yielding: $\int_{3}^{\frac{4}{25}3^{2/3}\tilde{m}^2 - \frac{33}{25}}\sqrt{V(UV,r_h=0)}
=\frac{\pi}{4 3^{\frac{1}{6}}} = \left(n + \frac{1}{2}\right)\pi$ which obtains:
\begin{equation}
\label{m0-+_T=0}
m_n^{0^{-+}}(UV,r_h=0) = 4.804\left(n + \frac{1}{2}\right)\frac{r_0}{L^2}.
\end{equation}

\section{Glueball ($0^{--(***...)}$) Masses}

\subsection{$r_h\neq0$ and Neumann/Dirichlet Boundary Conditions at $r=r_h$}

Given the Weiss-Zumino term $A^{\mu\nu}d^{abc}{\rm Tr}\left(F_{\mu\rho}^aF^{b\ \rho}_\lambda F^{c\ \lambda}_\nu\right)$ and the two-form potential $A_{\mu\nu}$ is dual to a pseudo-scalar, for $r_h\neq0$, corresponding to ${\rm QCD}_3$, one writes down the EOM for the fluctuation $\delta A^{23}$. The $B_{MN}, C_{MN}$ EOMs are:
\begin{eqnarray}
\label{B_C_EOMs}
& & D^M H_{MNP} = \frac{2}{3}F_{NPQRS}F^{QRS},\nonumber\\
& & D^M F_{MNP} = - \frac{2}{3}F_{NPQRS}H^{QRS},
\end{eqnarray}
or defining $A_{MN} = B_{MN} + i C_{MN}$, (\ref{B_C_EOMs}) can be rewritten as:
\begin{equation}
\label{A_EOM}
D^M\partial_{[M}A_{NP]} = - \frac{2i}{3}F_{NPQRS}\partial^{[Q}A^{RS]}.
\end{equation}
Now, $A_{MN}\rightarrow A^{(0)}_{MN} + \delta A_{MN}$ with $\delta A^{MN} = \delta^M_2 \delta^N_3 \delta A_{23}$, the EOM satisfied by $\delta A_{23}(x^{0,1,2,3},r) = \int \frac{d^4k}{\left(2\pi\right)^4}e^{i k\cdot x}g_{22} G(r)$ reduces to:
\begin{equation}
\label{A_EOM_ii}
\partial_\mu\left(\sqrt{-g}g^{22}g^{33}g^{\mu\nu}\partial_\nu \delta A_{23}\right) = 0.
\end{equation}
Assuming $a=\left(\alpha + \beta \frac{g_s M^2}{N} + \gamma \frac{g_s M^2}{N}\log r_h \right)r_h$ \cite{K. Dasgupta  et al [2012],EPJC-2} for $(\alpha,\beta,\gamma)=(0.6,4,4)$ \cite{EPJC-2} and $k^\mu=(\omega,k_1,0,0): k^2=-m^2$, and defining ${\cal G}(r)\equiv g_{22} G(r)$ the EOM for $G(r)$ is:
{
\begin{eqnarray}
\label{0--i}
& &  {\cal G}''(r)\nonumber\\
 & &  + {\cal G}'(r) \Biggl(\frac{3 {g_s} M^2 ({g_s} {N_f} ({\log  N}-6+\log (16))-24 {g_s} {N_f} \log (r)-8 \pi )}{64 \pi ^2 N r}\nonumber\\
 & & -\frac{75.
   {r_h}^2 \left(4. {g_s} M^2 \log ({r_h})+4. {g_s} M^2+0.6 N\right)^2}{N^2 r^3}+\frac{5 r^4-{r_h}^4}{r^5-r {r_h}^4}\Biggr)\nonumber\\
   & &  -\frac{{g_s}m^2 {\cal G}(r)}{4 \pi  r^2 \left(r^4-{r_h}^4\right)} \left(\frac{3 {r_h}^2 \left(4. {g_s} M^2 \log ({r_h})+4. {g_s} M^2+0.6 N\right)^2}{N^2}-r^2\right)\nonumber\\
   & & \times \biggl[36 {g_s}^2 M^2 {N_f}
   \log ^2(r)-3 {g_s} M^2 \log (r) ({g_s} {N_f} ({\log  N}-6+\log (16))-8 \pi )+16 \pi ^2 N\biggr] = 0.\nonumber\\
   & &
\end{eqnarray}}
The EOM (\ref{0--i}), near $r=r_h$ can be approximated as:
\begin{equation}
\label{0--ii}
 {\cal G}''(r)+\left({b_1} + \frac{1}{r-{r_h}}\right) {\cal G}'(r) + {\cal G}(r) \left(\frac{{a_2}}{r-{r_h}}+{b_2}\right)=0,
\end{equation}
where
{\footnotesize
\begin{eqnarray}
\label{0--iii}
& &\hskip -0.6in b_1=\frac{{g_s} M^2 ({g_s} (0.005 {\log  N}-0.015) {N_f}+(-0.114 {g_s} {N_f}-360.) \log ({r_h})-360.119)}{N {r_h}}-\frac{24.5}{{r_h}},\nonumber\\
& &\hskip -0.6in a_2=\frac{0.02 {g_s}^2 m^2 M^2 \left(\log ({r_h}) ({g_s} (0.24 {\log  N}-0.775) {N_f}-2279.99)-2.88 {g_s} {N_f} \log
   ^2({r_h})-2273.96\right)}{{r_h}^3}-\frac{0.251 {g_s} m^2 N}{{r_h}^3},\nonumber\\
   & &\hskip -0.6in b_2=\frac{0.04 {g_s}^2 m^2 M^2 \left(\log ({r_h}) ({g_s} {N_f} (8.158 -3.42 {\log  N})+4065.38)+{g_s} (0.12 {\log  N}-0.387) {N_f}+41.04
   {g_s} {N_f} \log ^2({r_h})+3976.41\right)}{{r_h}^4}\nonumber\\
   & & \hskip -0.6in +\frac{7.163 {g_s} m^2 N}{{r_h}^4}.
\end{eqnarray}}
The solution to (\ref{0--ii}) is given by:
\begin{eqnarray}
\label{0--iv}
& & {\cal G}(r) =  e^{\frac{1}{2} r \left(-\sqrt{{b_1}^2-4 {b_2}}-{b_1}\right)}\Biggl[ c_1 U\left(-\frac{2 {a_2}-{b_1}-\sqrt{{b_1}^2-4 {b_2}}}{2
   \sqrt{{b_1}^2-4 {b_2}}},1,\sqrt{{b_1}^2-4 {b_2}} r-\sqrt{{b_1}^2-4 {b_2}} {r_h}\right)\nonumber\\
  & &   +c_2  L_{\frac{2 {a_2}-\sqrt{{b_1}^2-4 {b_2}}-{b_1}}{2 \sqrt{{b_1}^2-4 {b_2}}}}\left(r
   \sqrt{{b_1}^2-4 {b_2}}-{r_h} \sqrt{{b_1}^2-4 {b_2}}\right)\Biggr],
\end{eqnarray}
implying:
{\footnotesize
\begin{eqnarray}
\label{0--v}
& & G^\prime(r) = \left(\frac{{\cal G}(r)}{g_{22}}\right)^\prime = \frac{1}{\Gamma \left(\frac{-2 {a_2}+{b_1}+\sqrt{{b_1}^2-4 {b_2}}}{2 \sqrt{{b_1}^2-4 {b_2}}}\right)}\sum_{n=-1}^\infty a_n(N,M,N_f,g_s,r_h)(r-r_h)^n.
\end{eqnarray}}
Assuming $c_2=0$, the Neumann boundary condition at $r=r_h$ can be satisfied by setting the argument of the gamma function to a negative integer $n$. It runs out setting $\frac{-2 {a_2}+{b_1}+\sqrt{{b_1}^2-4 {b_2}}}{2 \sqrt{{b_1}^2-4 {b_2}}}=-n\in\mathbb{Z}^-\cup\left\{0\right\}$ produces a negligible ground state $0^{--}$ mass. Hence, we consider $\frac{-2 {a_2}+{b_1}+3\sqrt{{b_1}^2-4 {b_2}}}{2 \sqrt{{b_1}^2-4 {b_2}}}=-n\in\mathbb{Z}^-\cup\left\{0\right\}$, which gives a finite ground state mass.  This condition up to LO in $N$ yields:
\begin{eqnarray}
\label{0--vii}
& & \frac{T^2 \left(1.5 \sqrt{\frac{1.82719\times 10^{12} T^2-1.66774\times 10^9 m^2}{T^2}}-675867.\right)+265.153 m^2}{T \sqrt{1.82719\times 10^{12} T^2-1.66774\times
   10^9 m^2}} = - n\in\mathbb{Z}^-\cup\{0\},
\end{eqnarray}
the solution to which are given below:
\begin{eqnarray}
\label{0--viii}
& & m_{0^{--}} = 32.461 T\nonumber\\
& & m_{0^{--}}^* = 32.88 T \nonumber\\
& & m_{0^{--}}^{**} = 32.989 T \nonumber\\
& & m_{0^{--}}^{***} = 33.033 T \nonumber\\
& & m_{0^{--}}^{****} = 33.055 T.
\end{eqnarray}
One can show that one obtains the same spectrum as in (\ref{0--viii}) after imposing Dirichlet boundary condition $G(r=r_h)=0$.

\subsection{$r_h=0$ limit of (\ref{0--i})}

\begin{eqnarray}
\label{0--T=0-i}
& & {\cal G}''(r)+\frac{{\cal G}'(r) \left(\frac{3 {g_s} M^2 ({g_s} {N_f} ({\log N}-6+\log (16))-24 {g_s} {N_f} \log (r)-8 \pi
   )}{\pi ^2 N}+320\right)}{64 r}\nonumber\\
   & & +\frac{{g_s} m^2  \left(36 {g_s}^2 M^2 {N_f} \log ^2(r)-3 {g_s} M^2 \log (r)
   ({g_s} {N_f} ({\log N}-6+\log (16))-8 \pi )+16 \pi ^2 N\right)}{4 \pi  r^4}{\cal G}(r) = 0.\nonumber\\
   & &
\end{eqnarray}
The EOM (\ref{0--T=0-i}) near $r=r_0$ - IR cut-off at $r_h=0$, reduces to:
\begin{equation}
\label{0--T=0-ii}
{\cal G}''(r) + (\alpha_1 + \beta_1(r - r_0)){\cal G}'(r) + (\alpha_2 + \beta_2(r - r_0)){\cal G}(r) = 0
\end{equation}
where
\begin{eqnarray}
\label{0--T=0-iii}
& & \alpha_1 = \frac{\frac{3 {g_s}^2 {\log N} M^2 {N_f}}{\pi ^2 N}+320}{64 {r_0}},\nonumber\\
& & \beta_1 = -\frac{3 {g_s}^2 M^2 {N_f}}{64 \pi ^2 N {r_0}^2}-\frac{5}{{r_0}^2},\nonumber\\
& & \alpha_2 = \frac{4 \pi  {g_s} m^2 N}{{r_0}^4}-\frac{3 {g_s}^3 m^2 M^2 {N_f} \log ({r_0})}{4 \pi  {r_0}^4},\nonumber\\
& & \beta_2 = \frac{3 {g_s}^3 {\log N} m^2 M^2 {N_f} (4 \log ({r_0})-1)}{4 \pi  {r_0}^5}-\frac{16 \pi  {g_s} m^2
   N}{{r_0}^5}.
\end{eqnarray}
The solution to (\ref{0--T=0-ii}) is given by:
\begin{eqnarray}
\label{0--T=0-iv}
& & {\cal G}(r) = e^{-{\alpha_1} r+\frac{{\beta_2} r}{{\beta_1}}-\frac{{\beta_1} r^2}{2}+{\beta_1} r {r_0}} \Biggl[c_2 \,
   _1F_1\left(\frac{{\beta_1}^3-{\alpha_2} {\beta_1}^2+{\alpha_1} {\beta_2} {\beta_1}-{\beta_2}^2}{2
   {\beta_1}^3};\frac{1}{2};\frac{\left((r-{r_0}) {\beta_1}^2+{\alpha_1} {\beta_1}-2 {\beta_2}\right)^2}{2
   {\beta_1}^3}\right)\nonumber\\
   & & +c_1 H_{\frac{-{\alpha_1} {\beta_1} {\beta_2}+{\alpha_2}
   {\beta_1}^2-{\beta_1}^3+{\beta_2}^2}{{\beta_1}^3}}\left(\frac{{\alpha_1} {\beta_1}+{\beta_1}^2 (r-{r_0})-2
   {\beta_2}}{\sqrt{2} {\beta_1}^{3/2}}\right)\Biggr].
\end{eqnarray}
One can then work out  $G'(r=r_0) = \left.\left(\frac{{\cal G}(r)}{g_{22}}\right)^\prime\right|_{r=r_0}$.
Now, setting $c_2=0$, defining $\tilde{m}$ via: $m = \tilde{m} \frac{r_0}{L^2}$, and using the large $\tilde{m}$-limit of Hermite functions:
\begin{eqnarray}
\label{0--T=0-vi}
& &   H_{\frac{-{\alpha_1} {\beta_1} {\beta_2}+{\alpha_2}
   {\beta_1}^2-(1\ {\rm or}\ 2){\beta_1}^3+{\beta_2}^2}{{\beta_1}^3}}\left(\frac{{\alpha_1} {\beta_1}-2 {\beta_2}}{\sqrt{2}
   {\beta_1}^{3/2}}\right)\longrightarrow H_{-\frac{16 \tilde{m}^4}{125}}\left(\frac{2^{\frac{5}{2}}}{5^{\frac{3}{2}}}\tilde{m}^2\right),
\end{eqnarray}
and
\begin{equation}
\label{Hn[x]_large-n}
H_n(x)\stackrel{n\gg1}{\longrightarrow}\frac{2^{\frac{n}{2}+\frac{1}{2}} e^{\frac{x^2}{2}} \left(\frac{n}{e}\right)^{n/2} \cos
   \left(\frac{\pi  n}{2}-x \sqrt{2 n-\frac{x^2}{3}+1}\right)}{\sqrt[4]{1-\frac{x^2}{2 n}}},
\end{equation}
one can show that the Neumann(/Dirichlet: $G(r=r_0)=0$) boundary condition at $r=r_0$ is equivalent to the condition:
\begin{equation}
\label{0--T=0-Neumann_+bc}
\frac{8}{375} \left(\sqrt{6} \tilde{m}^2 \sqrt{375-64 \tilde{m}^4}-6 i \pi  \tilde{m}^4\right)=i \pi  (2 n+1),
\end{equation}
yielding:
\begin{equation}
\label{0--T=0-ix}
m_n^{0^{--}}(r_h=0) = \frac{1}{2} 5^{3/4} \sqrt[4]{\frac{-2 \left(\sqrt{6} \sqrt{\pi ^2 \left(16 n^2+22 n+7\right)+6}+6\right)-3 \pi ^2 (2 n+1)}{3 \pi ^2-32}}\frac{r_0}{L^2}.
\end{equation}

\subsection{$0^{--}$ Glueball Spectrum from WKB Method  for $r_h\neq0$}

Using the variables of \cite{Minahan}, the potential in the IR is given as:
\begin{eqnarray}
\label{V0--_IR_i}
& & V(IR,T) = \left(6-0.01 \tilde{m}^2\right) e^z+\left(0.15 \tilde{m}^2-16.1875\right) e^{2 z} + {\cal O}\left(e^{3z}\right),
\end{eqnarray}
where in the `large' $\tilde{m}$-limit, $V(IR,T)>0$ for (i)$z: e^z>\frac{2.31799\times 10^{31}-3.86332\times 10^{28} \tilde{m}^2}{6.25374\times 10^{31}-5.79497\times 10^{29} \tilde{m}^2} = 0.067 + {\cal O}\left(\frac{1}{\tilde{m}^2}\right)$ if $\tilde{m}>24.495$ and (ii) $0<e^z<\frac{2.31799\times 10^{31}-3.86332\times 10^{28} \tilde{m}^2}{6.25374\times 10^{31}-5.79497\times 10^{29} \tilde{m}^2} = 0.067 + {\cal O}\left(\frac{1}{\tilde{m}^2}\right)$ if $10.388<\tilde{m}<24.495$. One can show that:
$\int_{\log(0.067)}^{-2.526}\sqrt{V(IR,T)} = \int_{-\infty}^{\log(0.067)}\sqrt{V(IR,T)}\approx 0$, implying there is no contribution to the WKB quantization condition in the IR.

Now, consider:
\begin{equation}
\label{V0--_UV}
V(UV,T) = \left(-1.02 \tilde{m}^2-22.5\right) e^{-2 z}+\left(0.25 \tilde{m}^2+8.25\right) e^{-z}-1 + {\cal O}\left(e^{-3z}\right).
\end{equation}
For $\tilde{m}>4.29$ the turning points of $V(UV,T)$ are $0.125 \tilde{m}^2-0.025 \sqrt{25. \tilde{m}^4+18. \tilde{m}^2-8775.}+4.125 = 4.08 + {\cal O}\left(\frac{1}{\tilde{m}^2}\right)<e^z<0.125 \tilde{m}^2+0.025 \sqrt{25.
   \tilde{m}^4+18. \tilde{m}^2-8775.}+4.125 = 0.25 \tilde{m}^2 + 4.17 + {\cal O}\left(\frac{1}{\tilde{m}^2}\right)$.
   Hence,
   \begin{eqnarray}
   \label{WKB-0--_UV}
   & & \int_{\log (4.08)}^{\log(0.25 \tilde{m}^2 + 4.17 )}\sqrt{V(UV,T)} =\int_{\log (4.08)}^{\log(0.25 \tilde{m}^2 + 4.17 )} e^{-z}\sqrt{0.25 e^z - 1.02} + {\cal O}\left(\frac{1}{\tilde{m}}\right)\nonumber\\
   & & = 0.389 \tilde{m} - 2 + {\cal O}\left(\frac{1}{\tilde{m}}\right) = \left(n + \frac{1}{2}\right)\pi.
   \end{eqnarray}
   Hence one obtains isospectrality with $0^{++}$; for large $n$, there is also isospectrality with $2^{++}$.

\subsection{WKB Method  at $r_h=0$}

In this section we will discuss obtaining the spectrum at $r_h=0$ using WKB quantization condition at LO in $N$ in {\bf 5.3.1} and up to NLO in $N$ in {\bf 5.3.2}.

\subsubsection{ LO in $N$ }

In the IR, the WKB `potential' can be shown to be given by:
\begin{equation}
\label{WKB-integ_T=0}
V(IR,r_h=0) = -\frac{1}{4} + \frac{1}{4}\left(-3 + \tilde{m}^2\right)e^{2z} + {\cal O}(e^{3z}),
\end{equation}
with turning points: $(\log\left(\frac{1}{m_0} + {\cal O}\left(\frac{1}{m_0^3}\right)\right)\approx-\log(m_0),\log \left(\delta^2-1\right))$.
Further dropping ${\cal O}\left(\frac{1}{m_0^3}\right)$ terms, $\int_{-\log m_0}^{-2.526}\sqrt{V(IR,r_h=0)} = \frac{\left(\delta^2-1\right)}{2} \tilde{m} - 0.785 = \left(n + \frac{1}{2}\right)\pi$ yielding:
\begin{equation}
\label{QKB0--_IR_T=0}
m_n^{0^{--}}(IR,r_h=0) = \frac{(3+4n)\pi}{2\left(\delta^2-1\right)}\frac{r_0}{L^2} = \left(\frac{4.71 + 6.28 n}{\delta^2-1}\right)\frac{r_0}{L^2}.
\end{equation}

In the UV,
\begin{eqnarray}
\label{V0--_UV_T=0}
& & V(UV,r_h=0) = -\frac{3}{4} \left({\tilde{m}}^2+3\right) e^{-2 z}+\frac{1}{4} \left({\tilde{m}}^2+6\right) e^{-z}-1,
\end{eqnarray}
with turning points: $\left(\log\left(3 + {\cal O}\left(\frac{1}{\tilde{m}^2}\right)\right),\log\left(\frac{\tilde{m}^2}{4} - \frac{3}{2} + {\cal O}\left(\frac{1}{\tilde{m}^2}\right)\right)\right)$, and $\sqrt{V(UV,r_h=0)} = \frac{e^{-z}}{2}\sqrt{e^z - 3}\tilde{m} + {\cal O}\left(\frac{1}{\tilde{m}}\right)$:
\begin{eqnarray}
\label{WKB_0--_UV_T=0}
\int_{\log 3}^{\log\left(\frac{\tilde{m}^2}{4} - \frac{3}{2}\right)}\frac{e^{-z}}{2}\sqrt{e^z - 3}\tilde{m} = \left(n + \frac{1}{2}\right)\pi,
\end{eqnarray}
implying:
\begin{equation}
\label{mn0--_UV}
m_n^{0^{--}}(UV,r_h=0) =  (7.87 + 6.93 n)\frac{r_0}{L^2}.
\end{equation}

\subsubsection{NLO (in $N$)/Non-Conformal Corrections}

Up to NLO in $N$, in the IR, the potential `$V(IR,r_h=0)$' is given by:
{\footnotesize
\begin{eqnarray}
\label{V2++_NLO_T=0}
& & V(IR,r_h=0) =  \frac{1}{256 \pi ^2 N}\Biggl\{e^{2 z} \Biggl(-{g_s}^2 M^2 {N_f} (6 {\log N}-72+\log (16777216))+36 {g_s}^2 M^2 \tilde{m}^2 {N_f} \log ^2({y_0})+{g_s}
   M^2 \log ({y_0})\nonumber\\
    & & \times\left({g_s} {N_f} \left(72-\tilde{m}^2 (6 {\log N}-36+\log (16777216))\right)+48 \pi  \tilde{m}^2\right) +48 \pi
    {g_s} M^2+64 \pi ^2 \left(\tilde{m}^2-3\right) N\Biggr)\Biggr\}-\frac{1}{4} + {\cal O}(e^{-3z}).\nonumber\\
    & &
\end{eqnarray}}
The turning points of (\ref{V2++_NLO_T=0}) up NLO in $N$ are given by:\\
$\Biggl[\log\left(\frac{1}{\tilde{m}}\left[1-\frac{{g_s} M^2 \log ({y_0}) (-{g_s} {N_f} (6 {\log N}-36+\log (16777216))+36 {g_s} {N_f} \log ({y_0})+48 \pi )}{128 \pi ^2
    N}\right]\right),\log(\delta^2-1)\Biggr]$. After evaluation of the integral of $\sqrt{V(IR,r_h=0)}$ between the aforementioned turning points, in the large-$\tilde{m}$-limit, one obtains the following quantization condition:
    \begin{eqnarray}
    \label{WKB_NLO_N+0++_T=0}
    & & \left(\frac{(\delta^2-1)}{2}-\frac{3(\delta^2-1)g_sM^2(g_sN_f)\log N\ \log r_0}{64\pi^2N}\right)\tilde{m} -\frac{\pi}{4} = \left(n + \frac{1}{2}\right)\pi,
    \end{eqnarray}
    which yields:
    \begin{eqnarray}
    \label{mn0--_NLON_T=0}
    & & m_n^{0^{--}}(r_h=0) = \frac{6.28319 n+4.71239}{\delta ^2-1}\left(1 + \frac{0.01 {g_s}^2 {\log N} M^2 {N_f} \log ({r_0})}{ N}\right).
    \end{eqnarray}

\section{ Glueball Masses from M theory}
The glueball spectrum for spin $0^{++}, 1^{++}$ and $2^{++}$ is calculated in this section from the M-theory perspective both by considering $r_h\neq0$ and an IR radial cut-off (for $r_h=0$) in the background.
The $11$ dimensional M-theory action is given as:
\begin{equation}\label{Mtheoryaction}
S_{M}=\frac{1}{16\pi}\int_{M}d^{11}x\sqrt{G}\ R-\frac{1}{4}\sqrt{G}\int_{M}d^{11}x G_4\wedge *_{11}G_4,
\end{equation}
where $G_4=d C_3 + A_1 \wedge d B_2 + dx_{10}\wedge dB_2$, and $C_{\mu \nu 10}^M = B_{\mu \nu}^{IIA}, C_{\mu \nu \rho}^M = C_{\mu \nu \rho}^{IIA}$. Now, as shown in \cite{MQGP}, no $F_4^{IIA}$ (to be obtained via a triple
T-dual of type IIB $F_{1,3,5}$ where $F_1\sim F_{x/y/z}, F_3\sim F_{xy r/\theta_1/\theta_2}, F_{xz r/\theta_1/\theta_2},
 F_{yz r/\theta_1/\theta_2}$ and $F_5\sim F_{xyz \beta_1\beta_2}$ where $\beta_i=r/\theta_i$) can be generated.
\footnote{Consider $T_x$ followed by $T_y$ followed by $T_z$ where $T_i$ means T-dualizing along i-th direction. As an example, $T_x F_x^{IIB}\rightarrow
{\rm non-dynamical\ 0-form\ field\ strength}^{IIA}\cite{kiritsis-book}, T_y T_x F_x^{IIB} \rightarrow F_y^{IIB}$,\\
 $T_z F_y^{IIB} \rightarrow F_{yz}^{IIA}$ implying one can never generate
$F_4^{IIA}$ from $F_1^{IIB}$.
    As also an example consider $T_x F_{xy\beta_i}^{IIB}\rightarrow F_{y \beta_i}^{IIA}, T_yF_{y \beta_i}^{IIA}\rightarrow F_{\beta_i}^{IIB}, T_z F_{\beta_i}^{IIB}
\rightarrow F_{\beta_i z}^{IIA}$ again not generating $F_4^{IIA}$;
   $ T_x F_{xyz \beta_1 \beta_2}^{IIB}\rightarrow F_{yz \beta_1 \beta_2}^{IIA}$,
    $T_y F_{yz \beta_1 \beta_2}^{IIA}\rightarrow F_{z \beta_1 \beta_2}^{IIB},
T_z F_{z \beta_1 \beta_2}^{IIB}\rightarrow F_{\beta_1 \beta_2}^{IIB}$; thus one can not generate $F_4^{IIA}$.}
Thus, the four-form flux $G_4=d\left(C_{\mu\nu10}dx^\mu\wedge dx^\nu\wedge dx_{10}\right)
 + \left(A^{F_1}_1 + A^{F_3}_1 + A^{F_5}_1\right)\wedge H_3=H_3\wedge dx_{10} + A\wedge H_3$, where $C_{\mu\nu10}\equiv B_{\mu\nu}$.

\begin{eqnarray}
\label{flux_action_D=11-ii}
& & \int G_4\wedge *_{11}G_4 = \int \left(H_3\wedge dx_{10} + A\wedge H_3\right)\wedge *_{11}\left(H_3\wedge dx_{10} + A\wedge H_3\right).
\end{eqnarray}
Now, $H_3\wedge dx_{10}\wedge *_{11}\left(H_3\wedge A\right)=0$ as neither $H_3$ nor $A$ has support along $x_{10}$. Hence,
\begin{eqnarray}\begin{split}
\label{flux_action_D=11-iii}
& H_3\wedge dx_{10}\wedge *_{11}\left(H_3\wedge dx_{10}\right)
\\&=\sqrt{G}H_{\mu\nu\rho10}G^{\mu\mu_1}G^{\nu\nu_1}G^{\rho\rho_1}G^{10\lambda_1}H_{\mu_1\nu_1\rho_1\lambda_1}
dt\wedge...dx_{10}\\
& = \sqrt{G} H_{\mu\nu\rho10}\left(-G^{\mu10}G^{\nu\nu_1}G^{\rho\rho_1}G^{10\lambda_1}H_{\nu_1\rho_1\lambda_1} + G^{\mu\mu_1}G^{\nu10}G^{\rho\rho_1}G^{10\lambda_1}H_{\mu_1\rho_1\lambda_1} \right.\\
& \left.- G^{\mu\mu_1}G^{\nu\nu_1}G^{\rho10}G^{10\lambda_1}H_{\mu_1\nu_1\lambda_1} + G^{\mu\mu_1}G^{\nu\nu_1}G^{\rho\rho_1}G^{10\ 10}H_{\mu_1\nu_1\rho_1}\right)dt\wedge...dx_{10},
\end{split}
\end{eqnarray}
where $H_{\mu\nu\rho10}=H_{\mu\nu\rho}$, and
\begin{eqnarray}
\label{flux_action_D=11-iv}
& & \left(H\wedge A\right)\wedge *_{11}\left(H\wedge A\right)=\sqrt{G} H_{[\mu\nu\rho}A_{\lambda]}G^{\mu\mu_1}G^{\nu\nu_1}G^{\lambda\lambda_1}H_{[\mu_1\nu_1\rho_1}A_{\lambda_1]},
\end{eqnarray}
with
$H_{[\mu_1\mu_2\mu_3}A_{\mu_4]}\equiv H_{\mu_1\mu_2\mu_3}A_{\mu_4} - \left(H_{\mu_2\mu_3\mu_4}A_{\mu_1} - H_{\mu_3\mu_4\mu_1}A_{\mu_2} + H_{\mu_4\mu_1\mu_2}A_{\mu_3}\right)$.

It was shown in \cite{MQGP} that in the MQGP limit (\ref{limits_Dasguptaetal-ii}), in equation (\ref{flux_action_D=11-iii}), contribution of $ H_3\wedge dx^{11}\wedge *_{11}(H_3\wedge dx^{11})$ is always dominated by $ \sqrt{G} H_{\mu\nu\omega} G^{\mu\sigma} G^{\nu\psi} G^{\omega\alpha} G^{10\ 10}H_{\sigma\psi\alpha}$ term  and in equation (\ref{flux_action_D=11-iv}), contribution of $(H\wedge A)\wedge *_{11}(H\wedge A)$ is dominated by $ \sqrt{G} H_{\theta_1\theta_2 y}A_\psi G^{\theta_1\alpha} G^{\theta_2\rho} G^{y\sigma} G^{\psi\beta}H_{\alpha\rho\beta}  $ term. Therefore, for simplicity in calculations, we assume that leading contribution in equations ($\ref{flux_action_D=11-iii}$) and ($\ref{flux_action_D=11-iv}$) are governed by aforementioned terms.

The relevant inverse components of the 11-dimensional metric, in the MQGP limit (\ref{limits_Dasguptaetal-ii}), as worked out in \cite{MQGP} using which the most dominant contribution near $\theta_{1,2}=0$ in the equations (\ref{flux_action_D=11-iii}) and (\ref{flux_action_D=11-iv}), as shown in \cite{MQGP}, are given by the following analytical expressions:
{\small
\begin{eqnarray}
\label{flux_action_D=11-v}
& &  \left.H_3\wedge dx^{11}\wedge *_{11}\left(H_3\wedge dx^{11}\right)\right|_{\theta_1\sim\frac{\alpha_{\theta_1}}{N^{\frac{1}{5}}},\theta_2\sim\frac{\alpha_{\theta_2}}{N^{\frac{3}{10}}}}  = \frac{\sqrt{3}\left(- 4 \alpha_{\theta_2}^2 + 27 \alpha_{\theta_1}^6\right)r^3 N^{\frac{9}{10}}}{2 \alpha_{\theta_1}^3\alpha_{\theta_2}^6 g_s^{\frac{11}{4}}\pi^{\frac{3}{4}}} + \frac{\alpha_{\theta_1}\left(82 r^4 - r_h^4\right)N^{\frac{1}{4}}}{2\sqrt{3}\alpha_{\theta_2}^4g_s^{\frac{11}{4}} \pi^{\frac{3}{4}} r}
\end{eqnarray}}
and
{\small
\begin{eqnarray}
\label{flux_action_D=11-vi}
& & \hskip -0.3in \left.\left(H\wedge A\right)\wedge *_{11}\left(H\wedge A\right)\right|_{\theta_1\sim\frac{\alpha_{\theta_1}}{N^{\frac{1}{5}}},\theta_2\sim\frac{\alpha_{\theta_2}}{N^{\frac{3}{10}}}}
= {\cal F}(\alpha_{\theta_1},\alpha_{\theta_2};a; g_s,M,N_f)N^{\frac{17}{20}},
\end{eqnarray}}
where  ${\cal F}(\alpha_{\theta_1},\alpha_{\theta_2};a; g_s,M,N_f)$ is a well-defined function of the parameters indicated.  One hence notes that $\lim_{r_\Lambda\rightarrow\infty}\int_0^{\frac{\sqrt{4\pi g_s N}}{r_h}}\int_{r_h}^{r_\Lambda}\left(H\wedge A\right)\wedge *_{11}\left(H\wedge A\right)\sim \frac{r_\Lambda}{r_h} N^{\frac{27}{20}}$, and is UV-divergent. Also, this yields a large cosmological constant in the IR because : $\frac{G_4\wedge * G_4}{\sqrt{G}}\ni \frac{|H\wedge A|^2}{\sqrt{G}}\sim \frac{N^{\frac{17}{20}}}{r^3 N^{\frac{17}{20}}}=\frac{1}{r^3}$. To take care of both these issues, from the discussion on holographic renormalizability of the $D=11$ supergravity action in \cite{MQGP}, one sees that this term can be cancelled by a boundary counter term: $\int_{r=r_\Lambda}\sqrt{G}|G_4|^2$.

Now, using:
\begin{equation}
\label{sqdetG0}
\left.\sqrt{G}\right|_{\theta_1\sim\frac{\alpha_{\theta_1}}{N^{\frac{1}{5}}},\theta_2\sim\frac{\alpha_{\theta_2}}{N^{\frac{3}{10}}}}
= \frac{2 N^{17/20} r^3}{27\ 3^{5/6} \sqrt[4]{\pi } \alpha_{\theta_1}^4 {\alpha_{\theta_2}} {g_s}^{35/12}},
\end{equation}
 and (\ref{flux_action_D=11-v}), one sees that one obtains a large-$N$ suppressed cosmological constant from the second term in $\frac{|H\wedge dx^{10}|^2}{\sqrt{G}}$ that remains small $\forall r>r_h$. To ensure one does not generate an $N$-enhanced cosmological constant from the first term in (\ref{flux_action_D=11-v}), one imposes the condition:
$- 4 \alpha_{\theta_2}^2 + 27 \alpha_{\theta_1}^6 = 0$, i.e., $\alpha_{\theta_2} = \frac{3^{\frac{3}{4}}\alpha_{\theta_1}^{\frac{3}{2}}}{\sqrt{2}}$.

One hence obtains the following flux-generated cosmological constant (with a slight abuse of notation):
\begin{equation}
\label{cc-G4squaredoversqrtGM}
\left. \frac{G_4\wedge *G_4}{\sqrt{G}}\right|_{\theta_1\sim\frac{1}{N^{\frac{1}{5}}},\theta_2\sim\frac{1}{N^{\frac{3}{10}}}} = \frac{3^{\frac{13}{12}}\sqrt{\alpha_{\theta_1}}g_s^{\frac{1}{6}}\left(82 r^4 - r_h^4\right)}{N^{\frac{3}{5}}\sqrt{2\pi}r^4}.
\end{equation}
\paragraph{Metric Fluctuations:}
The background metric $g^{(0)}_{\mu\nu}$ is linearly perturbed as $g_{\mu\nu}=g^{(0)}_{\mu\nu}+h_{\mu\nu}$. With this perturbation the equation of motion follows from the action (\ref{Mtheoryaction}) as:
\begin{equation}\label{metricEOM}
R^{(1)}_{\mu\nu}=\frac{-1}{12}\frac{G_{4}^{2}}{\sqrt{G}} h_{\mu\nu},
\end{equation}
Now we assume the perturbation to have the following form: $h_{\mu\nu}=\epsilon_{\mu\nu}(r)e^{ikx_1}$. Clearly there is a $SO(2)$ rotational symmetry in the $x_2-x_3$ plane which allow us to classify different perturbations into three categories, namely tensor, vector and scalar type of metric perturbations.

The mass spectrum was obtained by (i) solving equation (\ref{metricEOM}) and applying Neumann/Diriclet boundary condition near $r_h/r_0$,  (ii) following the redefinition of variables in \cite{Minahan} and then considering the WKB quantization condition.
\subsection{$0^{++}$ Glueball spectrum}

The $0^{++}$ glueball in M-theory corresponds to scalar metric perturbations \cite{Mathur_et_al-0++-Mtheory}:
\begin{eqnarray}
\label{0++_M_i}
& & h_{tt} = g_{tt} e^{i q x_1} q_1(r),\nonumber\\
& & h_{x_1r} = h_{rx_1} = i q~ g_{x_1x_1} e^{i q x_1} q_3(r),\nonumber\\
& & h_{rr} = g_{rr} e^{i q x_1} q_2(r),
\end{eqnarray}
where $g_{tt}$, $g_{x_ix_i}$ and $g_{rr}$ are the metric components of the M-theory and is given in equation (\ref{Mtheory met}).
\subsubsection{M-theory background with $r_h\neq0$}
Considering these components at leading order in $N$, and taking into account the above perturbation, we get the following differential equation for $q_3(r)$ with $q^2=-m^2$ from (\ref{metricEOM}):
\begin{eqnarray}
\label{q3-EOM}
& &  {q_3}''(r) +  \frac{{q_3}'(r)}{12 \pi  {g_s}
   N r^3 \left(r^4-{r_h}^4\right)}\Biggl\{ \Biggl[r^2 \left(16 \pi ^2 {g_s}^2 m^2 N^2 r^2+12 \pi  {g_s} N \left(9 r^4-{r_h}^4\right)-3 r \left(r^4-{r_h}^4\right)^2\right)\nonumber\\
   & & -3
   a^2 \left(16 \pi ^2 {g_s}^2 m^2 N^2 r^2+36 \pi  {g_s} N \left(r^4-{r_h}^4\right)+3 r \left(r^4-{r_h}^4\right)^2\right)\Biggr]\Biggr\}\nonumber\\
   & &  + \frac{{q_3}(r)}{12 \pi  {g_s} N r^4 \left(r^4-{r_h}^4\right)}\Biggl\{ \Biggl[r^2 \left(32 \pi ^2 {g_s}^2 m^2 N^2 r^2+12 \pi  {g_s} N \left(15 r^4+{r_h}^4\right)-3 r
   \left(5 r^8-6 r^4 {r_h}^4+{r_h}^8\right)\right)\nonumber\\
   & & -36 a^2 \left(4 \pi ^2 {g_s}^2 m^2 N^2 r^2+\pi  {g_s} N \left(11 r^4+{r_h}^4\right)+r^9-r^5
   {r_h}^4\right)\Biggr]\Biggr\}=0.
\end{eqnarray}
\paragraph{(a) Spectrum from Neumann/Dirichlet Boundary Condition:}

Equation (\ref{q3-EOM}), for  $a = 0.6 r_h$ near $r=r_h$ (writing $m = \tilde{m} \frac{r_h}{L^2}$) simplifies to:
\begin{equation}
\label{EOM_r=rh-i}
{q_3}''(r) + \frac{\left(2 -0.00666667 \tilde{m}^2\right) {q_3}'(r)}{r-{r_h}}+\frac{\left(0.76 -0.103333 \tilde{m}^2\right)
   {q_3}(r)}{{r_h} (r-{r_h})}=0.
\end{equation}
Lets write the above equation of the following form,
\begin{equation}
\label{EOM_r=rh-ii}
h''(r) +  \frac{p}{r-{r_h}}h'(r) + \frac{s}{r-{r_h}}h(r) =0,
\end{equation}
where we have, $p=\left(2 -0.00666667 \tilde{m}^2\right)$,~~$s=\frac{\left(0.76 -0.103333 \tilde{m}^2\right)
   }{{r_h}}$.

The solution to (\ref{EOM_r=rh-ii}) is given by:
\begin{eqnarray}
\label{EOM_r=rh-iii}
& & h(r) = c_1 (2 r-2 {r_h})^{p/2} (r-{r_h})^{-p/2} (-s (r-{r_h}))^{\frac{1}{2}-\frac{p}{2}} I_{p-1}\left(2 \sqrt{-s (r-{r_h})}\right)\nonumber\\
& & + (-1)^{1-p} c_2
   (2 r-2 {r_h})^{p/2} (r-{r_h})^{-p/2} (-s (r-{r_h}))^{\frac{1}{2}-\frac{p}{2}} K_{p-1}\left(2 \sqrt{-s (r-{r_h})}\right).
\end{eqnarray}
Setting $c_2=0$, one can verify that one satisfy the Neumann boundary condition: $h^\prime(r=r_h)=0$ provided:
\begin{equation}
\label{EOM_r=rh-iv}
p = - n\in\mathbb{Z}^-\cup\{0\},
\end{equation}
implying:
\begin{equation}
\label{EOM_r=rh-v}
\tilde{m} = 12.25\sqrt{2+n}.
\end{equation}
One can similarly show that by imposing Dirichlet boundary condition: $h(r=r_h)=0$:
\begin{equation}
\label{EOM_r=rh-vi}
\tilde{m} = 12.25\sqrt{1+n}.
\end{equation}
\paragraph{(b) Spectrum using WKB method:}

Following the redefinition of (\cite{Minahan}), equation (\ref{q3-EOM}) can be rewritten as a Schrodinger like form, where for $ a = 0.6 r_h$ and setting $g_s=0.9, N\sim(g_s)^{-39}\sim100$ - in the MQGP limit of \cite{MQGP} - the `potential'  in the IR can be shown to be given by:
\begin{equation}
\label{V0++-WKB-IR}
V_{IR}(z) = {e^z} \left(-0.00576389 \tilde{m}^4-0.0708333 \tilde{m}^2+0.\right)+0.00201389 \tilde{m}^4+0.00333333 \tilde{m}^2-0.25 + {\cal O}\left(e^{2z}\right)
\end{equation}
The potential (\ref{V0++-WKB-IR}) is positive for $z\in(-\infty,\log(0.349)]$, but to remain within the IR, one truncates this domain to $z\in(-\infty,-2.526]$ and the same yields:
\begin{equation}
\label{WKB0++-IR}
\int_{-\infty}^{-2.626}\sqrt{V_{IR}(z)} = 0.0898 \tilde{m}^2 \log (\tilde{m})-0.0439 \tilde{m}^2=\pi  \left(n+\frac{1}{2}\right) = \left(n + \frac{1}{2}\right)\pi,
\end{equation}
or
\begin{equation}
\label{mn0++-WKB}
m_n^{0^{++}}= \frac{\sqrt{70.0055 n+35.0027}}{\sqrt{{\cal PL}(26.3065 n+13.1533)}}.
\end{equation}

In the UV, one can show that:
\begin{equation}
\label{VUV}
V_{UV}(z) = \frac{-0.00694444 \tilde{m}^4+0.295 \tilde{m}^2+1.62}{{e^{2z}}}+\frac{-0.0833333 \tilde{m}^2-0.54}{{e^z}}-0.25<0,
\end{equation}
implying no turning points in the UV.
\subsubsection{M-theory background with IR Cut-Off $r_0$}
In this case, equation (\ref{q3-EOM}) is modified only by the limit $(r_h,a)\rightarrow 0$. The equation in this limit is given as,
\begin{equation}
\label{0++EOM_T=0}
q_3''(r) + q_3'(r) \left(\frac{4 \pi  {g_s} m^2 N}{3 r^3}-\frac{r^4}{4 \pi  {g_s} N}+\frac{9}{r}\right)+q_3(r) \left(\frac{8 \pi  {g_s} m^2 N}{3
   r^4}-\frac{5 r^3}{4 \pi  {g_s} N}+\frac{15}{r^2}\right)=0.
\end{equation}
\paragraph{(a) Spectrum from Neumann/Dirichlet Boundary Condition:} Near the cut-off at $r=r_0$, equation (\ref{0++EOM_T=0}) is given by:
\begin{eqnarray}
\label{EOM0++_r=r0}
& & q_3''(r) + \left(\frac{4 \tilde{m}^2+108}{12 {r_0}}-\frac{\left(\tilde{m}^2+9\right) (r-{r_0})}{{r_0}^2}\right) q_3'(r)+q_3(r)
   \left(\frac{8 \tilde{m}^2+180}{12 {r_0}^2}-\frac{\left(32 \tilde{m}^2+360\right) (r-{r_0})}{12 {r_0}^3}\right)=0,\nonumber\\
   & &
\end{eqnarray}
whose solution is given by:
\begin{eqnarray}
\label{0++EOM-solution}
& & q_3(r) = e^{-\frac{2 \left(4 \tilde{m}^2+45\right) r}{3 \left(\tilde{m}^2+9\right) {r_0}}} \Biggl[c_1 H_{-\frac{2 \tilde{m}^6+71 \tilde{m}^4+828
   \tilde{m}^2+2835}{9 \left(\tilde{m}^2+9\right)^3}}\left(\frac{3 \left(\tilde{m}^2+9\right)^2 r-2 \left(2 \tilde{m}^4+37
   \tilde{m}^2+153\right) {r_0}}{3 \sqrt{2} \left(\tilde{m}^2+9\right)^{3/2} {r_0}}\right)\nonumber\\
   & & +c_2 \, _1F_1\left(\frac{2 \tilde{m}^6+71
   \tilde{m}^4+828 \tilde{m}^2+2835}{18 \left(\tilde{m}^2+9\right)^3};\frac{1}{2};\frac{\left(3 \left(\tilde{m}^2+9\right)^2 r-2 \left(2
   \tilde{m}^4+37 \tilde{m}^2+153\right) {r_0}\right)^2}{18 \left(\tilde{m}^2+9\right)^3 {r_0}^2}\right)\Biggr].\nonumber\\
   & &
\end{eqnarray}
Thus:
\begin{eqnarray}
\label{dq30++_T=0}
& & q_3^\prime(r=r_0) = \frac{1}{27 {r_0}^2}\Biggl\{e^{-\frac{2 \left(4 \tilde{m}^2+45\right)}{3 \left(\tilde{m}^2+9\right)}} \Biggl(\left[2 \tilde{m}^6+71 \tilde{m}^4+828
   \tilde{m}^2+2835\right]\nonumber\\
    & & \times\Biggl[-\frac{3 \sqrt{2} c_1 {r_0} H_{-\frac{11 \tilde{m}^6+314 \tilde{m}^4+3015 \tilde{m}^2+9396}{9
   \left(\tilde{m}^2+9\right)^3}}\left(-\frac{\tilde{m}^4+20 \tilde{m}^2+63}{3 \sqrt{2}
   \left(\tilde{m}^2+9\right)^{3/2}}\right)}{\left(\tilde{m}^2+9\right)^{5/2}}\nonumber\\
   & & -\frac{c_2 \left(\tilde{m}^4+20 \tilde{m}^2+63\right)
   {r_0} \, _1F_1\left(\frac{2 \tilde{m}^6+71 \tilde{m}^4+828 \tilde{m}^2+2835}{18
   \left(\tilde{m}^2+9\right)^3}+1;\frac{3}{2};\frac{\left(\tilde{m}^4+20 \tilde{m}^2+63\right)^2}{18
   \left(\tilde{m}^2+9\right)^3}\right)}{\left(\tilde{m}^2+9\right)^4}\Biggr]\nonumber\\
   & & -\frac{1}{\tilde{m}^2+9}\Biggl\{18 \left(4 \tilde{m}^2+45\right) {r_0} \Biggl(c_1
   H_{-\frac{2 \tilde{m}^6+71 \tilde{m}^4+828 \tilde{m}^2+2835}{9 \left(\tilde{m}^2+9\right)^3}}\left(-\frac{\tilde{m}^4+20
   \tilde{m}^2+63}{3 \sqrt{2} \left(\tilde{m}^2+9\right)^{3/2}}\right)\nonumber\\
   & & +c_2 \, _1F_1\left(\frac{2 \tilde{m}^6+71 \tilde{m}^4+828
   \tilde{m}^2+2835}{18 \left(\tilde{m}^2+9\right)^3};\frac{1}{2};\frac{\left(\tilde{m}^4+20 \tilde{m}^2+63\right)^2}{18
   \left(\tilde{m}^2+9\right)^3}\right)\Biggr)\Biggr\}\Biggr)\Biggr\}.
\end{eqnarray}
Numerically/graphically we see that for $c_1 = -0.509 c_2$, one gets $q_3(r=r_0,\tilde{m}\approx 4.1)=0$.
We hence estimate the ground state of $0^{++}$ from metric fluctuations in M-theory to be
$4.1 \frac{r_0}{L^2}$.
\paragraph{(b) Spectrum using WKB method:} Defining $m = \tilde{m} \frac{\sqrt{y_0}}{L^2}$ and following \cite{Minahan}, the potential term in the Schr\"{o}dinger-like equation from (\ref{0++EOM_T=0}) is given as,
{
\begin{eqnarray}
\label{V0++_WKB_T=0}
& & V(z) = \nonumber\\
 & & \frac{1}{2304 \pi ^2 {g_s}^2 N^2 \left(e^z+1\right)^4}\Biggl\{-16 \pi ^2 {g_s}^2 N^2 \left(12 \left(\tilde{m}^2+12\right) e^{3 z}+\left(\tilde{m}^4+12 \tilde{m}^2+216\right) e^{2 z}+144 e^z+36 e^{4
   z}+36\right)\nonumber\\
   & & +24 \pi  {g_s} N {y_0}^{5/2} e^{2 z} \left(e^z+1\right)^{7/2} \left(\tilde{m}^2+9 e^z+9\right)-9 {y_0}^5 e^{2 z}
   \left(e^z+1\right)^7\Biggr\}.\nonumber\\
   & &
\end{eqnarray}}
We hence see that in the large-$N$ limit, $V(z)<0$ and hence has no turning points. The WKB method a la \cite{Minahan} does not work in this case.

\subsection{$2^{++}$ Glueball spectrum}

To study the spectrum of spin $2^{++}$ glueball, we consider the tensor type of metric perturbations where the non-zero perturbations are given as:
\begin{equation}
\begin{split}
&h_{x_2x_3}=h_{x_3x_2}=g_{x_1x_1}H(r)e^{ikx_1} \\&h_{x_2x_2}=h_{x_3x_3}=g_{x_1x_1}H(r)e^{ikx_1},
\end{split}
\end{equation}
where $g_{x_1x_1}$ is as given in (\ref{Mtheory met}).
\subsubsection{M-theory background with $r_h\neq0$}
Considering the tensor modes of metric perturbations and the M-theory metric components corrected upto NLO in $N$, as given in (\ref{Mtheory met}), we obtain a second order differential equation in $H(r)$ from (\ref{metricEOM}),
\begin{eqnarray}
\label{2++-EOM}
& & H''(r) + H'(r) \Biggl(-\frac{3 a^2}{r^3}+\frac{15 {g_s} M^2 ({g_s} {N_f} \log (N)-24 {g_s} {N_f} \log (r)-6 {g_s}
   {N_f}+{g_s} {N_f} \log (16)-8 \pi )}{64 \pi ^2 N r}\nonumber\\
   & & +\frac{5 r^4-{r_h}^4}{r^5-r {r_h}^4}\Biggr) +\Biggl(\frac{1}{4\pi r^{4}\left( r^{4}-rh^{4}\right) } \Biggl[ 8\pi \Biggl\{3 a^2 \left(-2 \pi  {g_s} N m^2 r^2-r^4+{r_h}^4\right)+2 \pi  {g_s} N m^2 r^4+4 r^6\Biggr\}\nonumber\\
   & & -
3 {g_s}^2 M^2 m^2 r^2 \left(r^2-3 a^2\right) \log (r) \Biggl\{{g_s} {N_f} \log (N)+{g_s} {N_f} (\log (16)-6)-8 \pi \Biggr\}\nonumber\\
& & +36 {g_s}^3 M^2 {N_f} m^2 r^2 \left(r^2-3 a^2\right)  \log ^2(r)\Biggr]\Biggr)H(r)=0,
\end{eqnarray}
where we assume $k^2=-m^2$ with $m$ being the mass of the corresponding glueball.
\paragraph{(a) Spectrum from Neumann/Dirichlet Boundary Condition:}
Near $r=r_h$, the solution to the above equation will be given on the same lines as {\bf 5.1} for $0^{--}$ glueballs, and the
analog of (\ref{0--vii}) is:
\begin{eqnarray}
\label{2++-solutioni}
& & \frac{T^2 \left(\frac{1.5 \sqrt{0.0536698 T^2-0.00186263 m^2}}{T}-0.115834\right)+0.00018196 m^2}{T \sqrt{0.0536698 T^2-0.00186263 m^2}}=-n,
\end{eqnarray}
the solutions to which are given as:
\begin{eqnarray}
\label{2++M-spectrum}
& & m_{2^{++}} = 5.086 T \nonumber\\
& & m_{2^{++}}^* = 5.269 T \nonumber\\
& & m_{2^{++}}^{**} = 5.318 T \nonumber\\
& & m_{2^{++}}^{***} = 5.338 T \nonumber\\
& & m_{2^{++}}^{****} = 5.348 T \nonumber\\
\end{eqnarray}
One can impose Dirichlet boundary condition: $H(r=r_h)=0$, and show that,

~~~~~~~~~~~~~~~~~~~~~~$m_n^{2^{++}\ {(\rm Neumann)}} = m_{n+1}^{2^{++}\ {(\rm Dirichlet)}}$, for $n=0,1,2,..$.
\paragraph{(b) Spectrum from WKB method:}

Using the variables of \cite{Minahan}, The potential term in the schrodinger like equation for $2^{++}$ glueball can be obtained from (\ref{2++-EOM}) and in  the IR region written in terms of $ m = \tilde{m}\frac{r_h}{L^2}=\tilde{m}\frac{\sqrt{y_h}}{L^2}$, can be shown to given by:
\begin{eqnarray}
\label{WKB_2++_ii}
& & V_{IR}(z) = e^z \left(0.52 -0.01 \tilde{m}^2\right)+\left(0.15 \tilde{m}^2-1.0275\right) e^{2 z} + {\cal O}\left(\frac{g_s M^2}{N},e^{3z}\right).
\end{eqnarray}
Now, negative $z$ implies being closer to the IR and the boundary would correspond to $r\leq \sqrt{3}a=0.6\sqrt{3}r_h$ corresponding to $z=-2.53$. It can be shown that for $\tilde{m}>7.211$  $V(z\in[-2.71,-2.53])>0$. Now $\int_{z=-2.71}^{z=-2.53}\sqrt{V_{IR}(z)}\approx 0$. Hence, the IR does not contribute to the $2^{++}$ glueball spectrum.

In the UV we must consider the limit $(z\rightarrow\infty)$. Moreover, in the UV $N_f=M=0$.
\begin{eqnarray}
\label{WKB_2++_iii}
& & V_{UV}(z) = e^{-2 z} \left(6.56 -1.02 \tilde{m}^2\right)+\left(0.25 \tilde{m}^2-2.77\right) e^{-z}+1. + {\cal O}\left(\frac{1}{\tilde{m}},e^{-3z}\right),\nonumber\\
& &
\end{eqnarray}
whose turning points are: $\{\log\left(4.08 + {\cal O}\left(\frac{1}{\tilde{m}^2}\right)\right),\infty\}$, giving the WKB quantization as:
\begin{eqnarray}
\label{WKB_2++_v}
\int_{\log 4.08}^{\infty}\sqrt{V_{UV}(z)} = 0.39 \tilde{m}  = \left(n + \frac{1}{2}\right)\pi,
\end{eqnarray}
implying:
\begin{equation}
\label{mn2++_Minahan_T}
m_n^{2^{++}}(T) = 8.08\left(n + \frac{1}{2}\right)\frac{r_h}{L^2}.
\end{equation}
\subsubsection{M-theory background with an IR Cut-Off $r_0$}
Considering the limit $(r_h,a\rightarrow0)$ equation (\ref{2++-EOM}) is given by,
\begin{equation}
\begin{split}
&H''(r)+\Biggl(\frac{5 \left(3 M^2 g_s \left(-24 N_f g_s \log (r)-6 N_f g_s+N_f g_s \log (N)+\log (16) N_f g_s-8 \pi \right)+64 \pi ^2 N\right)}{64 \pi ^2 N r}\Biggr)H'(r)\\&+\frac{1}{4 \pi  r^4}\Biggl(36 m^2 M^2 N_f g_s^3 \log ^2(r)-3 m^2 M^2 g_s^2 \log (r) \left(N_f g_s \log (N)+(\log (16)-6) N_f g_s-8 \pi \right)\\&+16 \pi  \left(\pi  m^2 N g_s+2
   r^2\right)\Biggr)H(r)=0
\end{split}
\end{equation}
\paragraph{(a) Neumann/Dirichlet Boundary Condition at $r=r_0$:}

Up to LO in $N$ near $r=r_0$, the above equation is given by:
\begin{eqnarray}
\label{2++-EOM_T=0}
& & H''(r)+\left(\frac{5}{{r_0}}-\frac{5 (r-{r_0})}{{r_0}^2}\right) H'(r)+H(r)
   \left(\frac{\tilde{m}^2+8}{{r_0}^2}-\frac{4 \left(\tilde{m}^2+4\right)
   (r-{r_0})}{{r_0}^3}\right)=0.
\end{eqnarray}
The solution of (\ref{2++-EOM_T=0}) is given by:
\begin{eqnarray}
\label{solution_2++_T=0}
& &  H(r) = e^{-\frac{4 \left(\tilde{m}^2+4\right) r}{5 {r_0}}} \Biggl(c_1 H_{\frac{1}{125} \left(16 \tilde{m}^4+53
   \tilde{m}^2+56\right)}\left(\frac{2 \left(4 \tilde{m}^2-9\right) {r_0}+25 r}{5 \sqrt{10}
   {r_0}}\right)\nonumber\\
   & & +c_2 \, _1F_1\left(\frac{1}{250} \left(-16 \tilde{m}^4-53
   \tilde{m}^2-56\right);\frac{1}{2};\frac{\left(25 r+2 \left(4 \tilde{m}^2-9\right) {r_0}\right)^2}{250
   {r_0}^2}\right)\Biggr).
\end{eqnarray}
The Neumann boundary condition $H'(r=r_0)=0$, numerically yields that for $c_1=-0.509 c_2$, the lightest $2^{++}$ glueball has a mass $1.137 \frac{r_0}{L^2}$. Similarly, by imposing Dirichlet boundary condition: $H(r=r_h)=0$, for $c_1=-0.509 c_2$, the lightest $2^{++}$ glueball has a mass $0.665 \frac{r_0}{L^2}$.

\paragraph{(b) Spectrum using WKB method:}

Following \cite{Minahan}, the `potential' term, in the IR region,  up to leading order in $N$  with $ m = \tilde{m}\frac{r_0}{L^2}=\tilde{m}\frac{\sqrt{y_0}}{L^2}$ is given as:
\begin{eqnarray}
\label{V2++IR_T=0}
& & V_{IR}(z) = \frac{1}{4} \left({\tilde{m}}^2+5\right) e^{2 z}-\frac{1}{4} + {\cal O}\left(e^{3z}\right),
\end{eqnarray}
and $V_{IR}(z)>0$ for $z\in[\log \left(\frac{1}{\tilde{m}} + \frac{1}{\tilde{m}^3}\right),\log(\delta^2-1)]\approx [-\log \tilde{m},\log(\delta^2-1)]$. Hence WKB quantization condition gives,
\begin{eqnarray}
\label{WKB-integral-2++_T=0}
\int_{-\log \tilde{m}}^{-2.526}\sqrt{\frac{1}{4} \left({\tilde{m}}^2+5\right) e^{2 z}-\frac{1}{4} } = \frac{(\delta^2-1)}{2}\tilde{m}  - 0.785 + {\cal O}\left(\frac{1}{\tilde{m}^3}\right) = \left(n + \frac{1}{2}\right)\pi,
\end{eqnarray}
 implying:
\begin{equation}
\label{WKB_IR}
m_n^{2^{++}}(IR,r_h=0) =m_n^{0^{--}}(IR,r_h=0).
\end{equation}
In the UV, we have:
\begin{eqnarray}
\label{V2++_UV_T=0}
& & V_{UV}(z) = \frac{1}{4} \left({\tilde{m}}^2-10\right) e^{-z}-\frac{3}{4} \left({\tilde{m}}^2-5\right) e^{-2 z}+1 + {\cal O}(e^{-3z})\nonumber\\
& & = \frac{e^{-z}}{2}\sqrt{e^z - 3}\tilde{m} + {\cal O}\left(e^{-3z},\frac{1}{\tilde{m}}\right),
\end{eqnarray}
and $V_{UV}(z)>0$ for $z>\log\left(\frac{1}{8} \left(-{\tilde{m}}^2+\sqrt{{\tilde{m}}^4+28 {m_0}^2-140}+10\right)\right) =
\log\left(3 + {\cal O}\left(\frac{1}{\tilde{m}^2}\right)\right)$. Further, $\int_{\log 3}^\infty\sqrt{V_{UV}(z)} = \frac{\pi \tilde{m}}{8\sqrt{3}}$, implying:
\begin{equation}
\label{WKB2++_IR+T=0}
m_n^{2^{++}}(UV) = \left(3.46 + 6.93\right)\frac{r_0}{L^2}.
\end{equation}

\paragraph{(c) NLO-in-$N$/Non-Conformal Corrections using WKB method:}
The `potential' inclusive of NLO-in-$N$ terms, in the IR region in the $r_h=0$ limit, is given by:
{\
\begin{eqnarray}
\label{V-NLO_IR_2++_T=0}
& & \hskip -0.6in V(IR,r_h=0) = \frac{1}{512 \pi ^2 N}\Biggl\{e^{2 z} \Biggl(-60 {g_s}^2 M^2 {N_f} ({\log N}-12+\log (16))+72 {g_s}^2 M^2 {m_0}^2 {N_f} \log ^2({y_0})+{g_s} M^2 \log
   ({y_0})\nonumber\\
    & & \hskip -0.7in\times\left({g_s} {N_f} \left({m_0}^2 (-12 {\log N}+72+\log (4096)-15 \log (16))+720\right)+96 \pi  {m_0}^2\right)+480 \pi  {g_s}
   M^2+128 \pi ^2 \left({m_0}^2+5\right) N\Biggr)\Biggr\}\nonumber\\
   & & \hskip -0.7in -\frac{1}{4} + {\cal O}(e^{-3z}),
\end{eqnarray}}
whose turning points are given by:\\
$\left[\log\left\{\frac{1-\frac{{g_s} M^2 \log ({y_0}) ({g_s} {N_f} (-12 {\log N}+72+\log (4096)-15 \log (16))+72 {g_s} {N_f} \log ({y_0})+96 \pi
   )}{256 \pi ^2 N}}{{m_0}}\right\},\log\left(\delta^2-1\right)\right]$. The integral of \\ $\sqrt{V(IR,r_h=0)}$ between these turning points, in the large-$\tilde{m}$ limit, yields the same spectrum as $0^{--}$ up to NLO in $N$.

\subsection{Spin-$1^{++}$ Glueball spectrum}
Here we need to consider the vector type of metric perturbation with the non-zero components given as: $h_{ti}=h_{it}=g_{x_1x_1}G(r)e^{ikx_1}$, $i={x_2,x_3}$.
\subsubsection{M theory background with $r_h\neq0$}
Substituting the above ansatz for the perturbation in (\ref{metricEOM}), the differential equation in $G(r)$ is given with $k^2=-m^2$ as,
{\footnotesize
\begin{eqnarray}
\label{1++_i}
\nonumber\\
   & & \hskip -0.6in G''(r) +
   G'(r) \left(-\frac{3 a^2}{r^3}-\frac{15 {g_s} M^2 (-{g_s} {N_f} \log (N)+24
   {g_s} {N_f} \log (r)+6 {g_s} {N_f}-2 {g_s} {N_f} \log (4)+8 \pi )}{64 \pi ^2 N r}+\frac{5}{r}\right)
   \nonumber\\
   & & \hskip -0.6in + \frac{G(r)}{4 \pi  r^4 \left(r^4-{r_h}^4\right)}\Biggl\{ \Biggl(36 {g_s}^3 M^2 {N_f} m^2 r^2 \left(r^2-3 a^2\right) \log ^2(r)-3 {g_s}^2 M^2 m^2 r^2 \left(r^2-3 a^2\right) \log (r) ({g_s}
   {N_f} \log (N)+{g_s} {N_f} (\log (16)-6)-8 \pi )\nonumber\\
   & & \hskip -0.6in +8 \pi  \left(3 a^2 \left(-2 \pi  {g_s} N m^2 r^2-r^4+{r_h}^4\right)+2 \pi  {g_s} N m^2
   r^4+4 r^6\right)\Biggr)\Biggr\} = 0.
\end{eqnarray}}
\paragraph{(a) Neumann/Diriclet boundary condition at $r=r_h$:}
Near $r=r_h$, equation (\ref{1++_i}) up to LO in $N$, is given by:
\begin{equation}
\label{Neumann_1++_T=0}
G''(r) + \left(\frac{3.92}{r_h}\right)G'(r) + \left(\frac{2 - 0.02 \tilde{m}^2}{r_h(r-r_h)} + \frac{-1.16 + 0.57 \tilde{m}^2}{r_h^2}\right)G'(r) = 0,
\end{equation}
whose solution is given by:
{\footnotesize
\begin{eqnarray}
\label{Neumann1++_T=0-solution}
& & G(r) =  e^{\left(\frac{0.5 r \left(-2.
   \sqrt{5.0016 -0.57 \tilde{m}^2}-3.92\right)+ {r_h} \log (r-{r_h})}{{r_h}}\right)}\nonumber\\
   & & \times\Biggl[ c_1 U\left(-\frac{-0.01 \tilde{m}^2-\sqrt{5.0016 -0.57 \tilde{m}^2}+1}{\sqrt{5.0016 -0.57 \tilde{m}^2}},2,\frac{2.
   \sqrt{5.0016 -0.57 \tilde{m}^2} r}{{r_h}}-2. \sqrt{5.0016 -0.57 \tilde{m}^2}\right)\nonumber\\
    & & +c_2 L_{\frac{-1. \sqrt{5.0016
   -0.57 \tilde{m}^2}-0.01 \tilde{m}^2+1}{\sqrt{5.0016 -0.57 \tilde{m}^2}}}^{1}\left(\frac{2 r \sqrt{5.0016 -0.57
   \tilde{m}^2}}{{r_h}}-2. \sqrt{5.0016 -0.57 \tilde{m}^2}\right) \Biggr].
\end{eqnarray}}
Imposing Neumann boundary condition at $r=r_h$ yields:
\begin{eqnarray}
\label{Neumann_1++_T=0}
& & \lim_{z\rightarrow0}\frac{1}{r_h}\Biggl\{e^{-\sqrt{5.0016 -0.57 \tilde{m}^2}} \Biggl[0.140858 c_1 {r_h} U\left(\frac{0.01 \tilde{m}^2}{\sqrt{5.0016 -0.57
   \tilde{m}^2}}-\frac{1}{\sqrt{5.0016 -0.57 \tilde{m}^2}}+1,2,z\right)\nonumber\\
   & & +0.140858 c_2 {r_h} L_{-\frac{0.01
   \tilde{m}^2}{\sqrt{5.0016 -0.57 \tilde{m}^2}}+\frac{1}{\sqrt{5.0016 -0.57 \tilde{m}^2}}-1}^{1}(z)\Biggr]\Biggr\}.
\end{eqnarray}
Considering $p=\frac{0.01 \tilde{m}^2}{\sqrt{5.0016 -0.57
   \tilde{m}^2}}-\frac{1}{\sqrt{5.0016 -0.57 \tilde{m}^2}}+1$ and setting $c_2=0$ in (\ref{Neumann_1++_T=0}), and then using $\lim_{z\rightarrow0} U(p,2,z\sim0)\sim\frac{z^{-1}
\ _1F_1(p-1;0;z)}{\Gamma(p)}$, one notes that one can satisfy the Neumann boundary condition at $r=r_h$ provided $\lim_{z\rightarrow0}\ _1F_1(p-1;0;z) = \lim_{b\rightarrow0}\lim_{z\rightarrow0}\ _1F_1(p-1;b;z)$ (i.e. first set $z$ to 0 and then $b$), $p=-n\in\mathbb{Z}^-$. Hence:
\begin{eqnarray}
\label{1++_spectrum_Neumann_T}
& & m^{1^{++}}(T) = 2.6956 \pi T\nonumber\\
& & m^{1^{++*}}(T) = 2.8995 \pi T\nonumber\\
& & m^{1^{++**}}(T) = 2.9346 \pi T\nonumber\\
& & m^{1^{++***}}(T) = 2.9467 \pi T.
\end{eqnarray}
One can show that one obtains the same spectrum as (\ref{1++_spectrum_Neumann_T}) even upon imposing Dirichlet boundary condition: $G(r=r_h)=0$.

\paragraph{(b) Spectrum using WKB method:}

Following \cite{Minahan}, the `potential' $V$ in the Schr\"{o}dinger-like equation  working with the dimensionless mass variable $\tilde{m}$ defined via: $ m = \tilde{m}\frac{r_h}{L^2}=\tilde{m}\frac{\sqrt{y_h}}{L^2}$, in the IR, can be shown to be given by:
\begin{equation}
\label{1++_iii}
V_{IR}(z)= {e^{2z}} \left(0.15 \tilde{m}^2-1.52\right)+{e^z} \left(1-0.01 \tilde{m}^2\right)-\frac{1}{4} + {\cal O}\left(e^{3z},\frac{1}{N}\right).
\end{equation}
The zeros of the potential as function of $e^z$, in (\ref{1++_iii}) are given by:\\ $\frac{-1.8765\times 10^{14} \tilde{m}^2\pm 6.022232598554301\times 10^{-7} \sqrt{9.70917\times 10^{40} \tilde{m}^4+1.26219\times
   10^{44} \tilde{m}^2-5.04877\times 10^{44}}+1.8765\times 10^{16}}{5.70456\times 10^{16}-5.6295\times 10^{15} \tilde{m}^2} = - \frac{25}{\tilde{m}^2} + {\cal O}\left(\frac{1}{\tilde{m}^2}\right),$ $0.07 + {\cal O}\left(\frac{1}{\tilde{m}^2}\right)$; the first not being permissible. Now, in the IR $r\in[r_h,\sqrt{3}a\approx0.6\sqrt{3}r_h]$ or in terms of $z:(-\infty,-2.526]$. Therefore the allowed domain of integration is: $[\log (0.07),-2.526]$. Thus, in the IR:
\begin{equation}
\label{1++_iv}
\int_{z=-2.704}^{z=-2.526} \sqrt{{e^{2z}} \left(0.15 \tilde{m}^2-1.52\right)+{e^z} \left(1-0.01 \tilde{m}^2\right)-\frac{1}{4}}\approx 0,
\end{equation}
implying a null contribution to the WKB quantization condition in the IR.

In the UV, the potential is given by:
\begin{eqnarray}
\label{1++_v}
& & \hskip -0.4in V_{UV}(z) = \nonumber\\
 & & \hskip -0.4in \frac{e^{3z} \left(-3 b^2+\tilde{m}^2+10\right)+ e^{2z} \left(-3 b^2 \left(\tilde{m}^2+2\right)+2 \tilde{m}^2+9\right)+ e^z
   \left(\left(1-3 b^2\right) \tilde{m}^2+1\right)+4 e^{4z}-2}{4 (e^z+1)^3 (e^z+2)},
\end{eqnarray}
which for $b=0.6$ obtains:
\begin{equation}
\label{1++_vi}
\hskip 0.35in V(N_f=M=0,UV) = e^{-2 z} \left(6.56 -1.02 \tilde{m}^2\right)+\left(0.25 \tilde{m}^2-2.77\right) e^{-z}+1 + {\cal O}(e^{-3z}).
\end{equation}
The zeros of the potential in the UV, as a function of $e^z$,in (\ref{1++_vi}) are given by:\\ $-0.125 \tilde{m}^2\pm0.005 \sqrt{625. \tilde{m}^4+26950. \tilde{m}^2-185671.}+1.385$ $=(-0.25 \tilde{m}^2 - 1.31, 4.08 + {\cal O}\left(\frac{1}{\tilde{m}^2}\right))$; the former not being permissible. Hence, the allowed domain of integration over which the potential is positive, is: $([\log (4.08),\infty)$. Performing a large-$\tilde{m}$-expansion, one obtains:
\begin{eqnarray}
\label{1++_vii}
& & \int_{\log (4.08)}^\infty\sqrt{e^{-2 z} \left(6.56 -1.02 \tilde{m}^2\right)+\left(0.25 \tilde{m}^2-2.77\right) e^{-z}+1}=\int_{z=\log (4.08)}^\infty e^{-z}\sqrt{0.25 e^z - 1.02} + {\cal O}\left(\frac{1}{\tilde{m}}\right)\nonumber\\
& &  = 0.389 \tilde{m} = \left(n + \frac{1}{2}\right)\pi,
\end{eqnarray}
yielding:
\begin{equation}
\label{1++_viii}
m_n^{1^{++}}(T) = 4.04\left(1 + 2n\right)\frac{r_h}{L^2}.
\end{equation}
\subsubsection{M theory background with an IR Cut-Off $r_0$}

\paragraph{(a) Neumann/Diriclet boundary condition at $r=r_0$:}
Considering the limit of $(r_h,a)\rightarrow 0$ in equation (\ref{1++_i}) up to LO in $N$ and imposing Neumann boundary condition at the IR cut-off $r=r_0$, yields isospectrality with $2^{++}$ glueball spectrum at $r_h=0$.
\paragraph{(b) Spectrum using WKB method:}

Using the redefinition of \cite{Minahan}, the `potential' up to leading order in $N$ is given by:
\begin{equation}
\label{VLON_1++_T=0}
V(z) = \frac{\left(\tilde{m}^2+2\right) e^{2 z}-3 e^z+4 e^{3 z}-1}{4 \left(e^z+1\right)^3} +
{\cal O}\left(\frac{g_s M^2}{N}\right).
\end{equation}
In the IR region we get the potential as:
\begin{equation}
\label{V1++_T=0_IR}
V_{IR}(z) = - \frac{1}{4} + \frac{1}{4}(5 + \tilde{m}^2)e^{2z} + {\cal O}(e^{3z}),
\end{equation}
giving the turning points as, $z\in[\log\left(\frac{1}{\tilde{m}} + {\cal O}\left(\frac{1}{\tilde{m}^3}\right)\right)\approx-\log \tilde{m},\log(\delta^2-1)]$ and the WKB quantization condition becomes:
\begin{equation}
\label{WKB_1++_T=0}
\int_{-\log \tilde{m}}^{\log(\delta^2-1)}\sqrt{V_{IR}(z)} = \frac{(\delta^2-1)}{2} \tilde{m} - \frac{\pi}{4} = \left(n + \frac{1}{2}\right)\pi.
\end{equation}
Therefore:
\begin{equation}
\label{m1++_T=0_IR}
m_n^{1^{++}}(IR,r_h=0) = m_n^{2^{++}}(IR,r_h=0) = m_n^{0^{--}}(IR,r_h=0).
\end{equation}

Further, in the UV:
\begin{equation}
\label{V1++_T=0_UV}
V_{UV}(z) = \frac{1}{4} \left(\tilde{m}^2-10\right) e^{-z}-\frac{3}{4} \left(\tilde{m}^2-5\right) e^{-2 z}+1 + {\cal O}\left(e^{-3z}\right),
\end{equation}
implying that $V_{UV}(z)>0$ for $z\in[\log\left(\frac{1}{8} \left(-\tilde{m}^2+\sqrt{\tilde{m}^4+28 \tilde{m}^2-140}+10\right)=\log\left(3 + {\cal O}\left(\frac{1}{\tilde{m}^2}\right)\right),\infty\right)$. This yields the following WKB quantization condition:
\begin{equation}
\label{WKB1++_T=0}
\int_{\log 3}^\infty\sqrt{V_{UV}(z)} = \frac{\pi\tilde{m}}{4\sqrt{3}} = \left(n + \frac{1}{2}\right)\pi,
\end{equation}
which obtains:
\begin{equation}
\label{mn1++_T=0_UV}
m_n^{1^{++}}(UV,r_h=0) = \left(3.46 + 6.93 n\right)\frac{r_0}{L^2}.
\end{equation}
\paragraph{(c) NLO in $N$/Non-Conformal Corrections using WKB method:}

In the IR region, the `potential' including NLO-in-$N$ corrections in the $r_h=0$ limit, is given by:
{\footnotesize
\begin{eqnarray}
\label{V2++_NLN_T=0}
& & V(IR,r_h=0) = \frac{1}{4} e^{2z} \left(\tilde{m}^2+5\right)\nonumber\\
& & +\frac{1}{512 \pi ^2 N}\Biggl\{{g_s} M^2 e^{2 z} \Biggl(\log ({y_0}) \left({g_s} {N_f}
   \left(\tilde{m}^2 (-12 {\log N}+72+\log (4096)-15 \log (16))+720\right)+96 \pi  \tilde{m}^2\right)\nonumber\\
   & & +60 (8 \pi -{g_s} {N_f}
   ({\log N}-12+\log (16)))+72 {g_s} \tilde{m}^2 {N_f} \log ^2({y_0})\Biggr)\Biggr\}-\frac{1}{4},
\end{eqnarray}}
whose turning points are:\\
$\left[\log\left(\frac{1}{\tilde{m}}\left[1-\frac{{g_s} M^2 \log ({y_0}) ({g_s} {N_f} (-12 {\log N}+72+\log (4096)-15 \log (16))+72 {g_s} {N_f} \log ({y_0})+96
   \pi )}{256 \pi ^2  N}\right]\right),\log(\delta^2-1)\right]$, and the integral of $\sqrt{V(IR,r_h=0)}$ between these turning points yields an isospectrality with the $0^{--}$ and $2^{++}$ NLO-in-$N$ spectrum.

\section{$2^{++}$ Glueball Masses from Type IIB}

\subsection{$r_h\neq0$}

The $10$-dimensional type IIB supergravity action in the low energy limit is given by,
\begin{equation}\label{action}
\frac{1}{2k_{10}^2}\left\{\int d^{10}x~ e^{-2\phi}\sqrt{-G}\left(R-\frac{1}{2}H_3^2\right)-\frac{1}{2}\int d^{10}x
~\sqrt{-G}\left(F_1^2+\widetilde{F_3^2}+\frac{1}{2}\widetilde{F_5^2}\right)\right\},
\end{equation}
where $\phi$ is the dilaton, $G_{MN}$ is the $10$-d metric and $F_1$, $H_3$, $\widetilde{F_3}$, $\widetilde{F_5}$ are different fluxes.

The five form flux $\widetilde{F_5}$ and the three form flux $\widetilde{F_3}$ are defined as,
\begin{equation}\label{F5F3tilde}
\widetilde{F_5}=F_5+\frac{1}{2}B_2\wedge F_3 ,~~~~~~~~~~~~~~~~~~~~            \widetilde{F_3}=F_3-C_0\wedge H_3,
\end{equation}
where $F_5$ and $F_3$ are sourced by the $D_3$ and $D_5$ branes respectively. $B_2$ is the NS-NS two form and $C_0$ is the axion. The three form fluxes $\widetilde{F_3}$, $H_3$, the two form $B_2$ and the axion $C_0$ are given as \cite{metrics} - see (\ref{three-form-fluxes}). Now varying the action in (\ref{action}) with respect to the metric $g_{\mu\nu}$ one get the following equation of motion,
\begin{equation}\label{ricci scalar}
\begin{split}
R_{\mu\nu} & =\left(\frac{5}{4}\right)e^{2\phi}\widetilde{F}_{\mu p_2p_3p_4p_5}\widetilde{F}_{\nu}^{p_2p_3p_4p_5}-\left(\frac{g_{\mu\nu}}{8}\right)e^{2\phi}\widetilde{F}_{5}^2+\left(\frac{3}{2}\right)H_{\mu\alpha_2\alpha_3}
H^{\alpha_2\alpha_3}_{\nu}\\ & -\left(\frac{g_{\mu\nu}}{8}\right)H_3^2
 +\left(\frac{3}{2}\right)e^{2\phi}\widetilde{F}_{\mu\alpha_2\alpha_3}
\widetilde{F}^{\alpha_2\alpha_3}_{\nu}
-\left(\frac{g_{\mu\nu}}{8}\right)e^{2\phi}\widetilde{F}_{3}^2+\left(\frac{1}{2}\right)e^{2\phi}F_{\mu}F_{\nu}.
\end{split}
\end{equation}
 we consider the following linear perturbation of the metric,
\begin{equation}
g_{\mu\nu}=g_{\mu\nu}^{(0)}+h_{\mu\nu},
\end{equation}
where as before $\mu,\nu=\{t,x_1,x_2,x_3,r,\theta_1,\theta_2,\phi_1,\phi_2,\psi\}$. Here the only non zero component according to the tensor mode of metric fluctuation is $h_{x_2x_3}$. Since the non zero components of $F_1$, $\widetilde{F_3}$ and $H_3$ has no indices as $\{x_2,x_3\}$, the final equation of motion gets simplified and can be shown to be given as:
\begin{equation}\label{final EOM}
\begin{split}
R^{(1)}_{x_2x_3} & =\left(\frac{5}{4}\right)e^{2\phi}\left(4\widetilde{F}_{x_2 x_3p_3p_4p_5}\widetilde{F}_{x_2x_3 q_3q_4q_5}g^{p_3q_3}g^{p_4q_4}g^{p_5q_5}h^{x_2x_3}\right)-\left(\frac{h_{x_2x_3}}{8}\right)e^{2\phi}\widetilde{F}_{5}^2
\\&-\left(\frac{h_{yz}}{8}\right)H_3^2
-\left(\frac{h_{x_2x_3}}{8}\right)e^{2\phi}\widetilde{F}_{3}^2.
\end{split}
\end{equation}
Working at a particular value of $\theta_1$ and $\theta_2$ given as: $\theta_1=N^{-1/5}$ and $\theta_2=N^{-3/10}$, the square of different fluxes figuring in (\ref{final EOM}) at the lowest-order in $N$, are given in (\ref{fluxes-squared}).
Writing the perturbation $h_{x_2x_3}$ as $h_{x_2x_3}=\frac{r^2}{2 (g_{s}\pi N)^{1/2}} H(r)e^{i k x_1}$, (\ref{final EOM}) reduces to the following second order differential equation in $H(r)$:
{\scriptsize
\begin{eqnarray}
\label{EOM}
& & \hskip -0.4in H^{\prime\prime}(r)+\Biggl(\frac{5 r^4-{r_h}^4}{r \left(r^4-{r_h}^4\right)}-\frac{9 a^2}{r^3}+\Biggl\{\frac{3}{256 \pi ^2 N^{2/5} r^3}\Biggl[-54 a^2 {g_s}^2 M^2 {N_f}-72 \pi  a^2 {g_s} M^2+768 \pi ^2 a^2+12 {g_s}^2 M^2 {N_f} r^2+\nonumber\\
& & \hskip -0.4in  9 a^2 {g_s}^2 M^2 {N_f} \log (16)-2 {g_s}^2 M^2 {N_f} r^2 \log (16)+16 \pi  {g_s} M^2 r^2+{g_s}^2 M^2 {N_f} \left(9 a^2-2 r^2\right) \log (N)-24 {g_s}^2 M^2 {N_f}\nonumber\\
& & \hskip -0.4in \left(9 a^2-2 r^2\right) \log (r)\Biggr]\Biggr\}\Biggr)H^{\prime}(r)+\Biggl(\frac{1}{4 \pi  r^4 \left(r^4-{r_h}^4\right)}\Biggr\{8 \pi  \left(a^2 \left(6 \pi  {g_s} N q^2 r^2-9 r^4+9 {r_h}^4\right)-2 \pi  {g_s} N q^2 r^4+4 r^6\right)\nonumber\\
& &\hskip -0.4in  +3 {g_s}^2 M^2 q^2 r^2 \left(r^2-3 a^2\right) \log (r) ({g_s} {N_f} \log (16 N)-6 {g_s} {N_f}-8 \pi )-36 {g_s}^3 M^2 {N_f} q^2 r^2 \left(r^2-3 a^2\right) \log ^2(r)\Bigg\}-\nonumber\\
& &\hskip -0.4in  \frac{{g_s}^{2}}{512\pi^{3}}\Biggl\{\frac{34992 a^2 {g_s} M^2 \left(\sqrt[5]{N}+3\right) {N_f}^2 \log (r)}{r^3}+9 a^2 {g_s} {N_f} \left(\frac{7831552 \pi ^5}{\left(r^4-{r_h}^4\right) ({g_s} {N_f} \log (16 N)-3 {g_s} {N_f} \log (r)+4 \pi )^3}-\frac{81 M^2
  \left(7 \sqrt[5]{N}-1\right) {N_f}}{r^4}\right)+\nonumber\\
& & \hskip -0.4in  \frac{2 \left(243 {g_s} M^2 \left(\sqrt[5]{N}+1\right) {N_f}^2+\frac{3915776 \pi ^5 r^4}{\left(r^4-{r_h}^4\right) ({g_s} {N_f} \log (16 N)-3 {g_s}
   {N_f} \log (r)+4 \pi )^2}\right)}{r^2}\Biggr\}\Biggr)H(r)=0.
\end{eqnarray}}

\subsubsection{Mass Spectrum from Neumann Boundary Condition at $r=r_h$}

To get a sensible answer, one has to perform a small-$T$ expansion when one rewrites and solves (\ref{EOM}) around $r=r_h$. This time around the analog of (\ref{0--vii}) becomes:
\begin{equation}
\label{2++-IIB-i}
0.5 -\frac{0.174071 m^2 \left(1-\frac{3. \left({g_s} M^2 (2. \log (N)+11.8963)+4. {g_s} M^2 \log (T)+0.6
   N\right)^2}{N^2}\right)}{T \sqrt{-m^2 \left(4-7 \left(1-\frac{3. \left({g_s} M^2 (2. \log (N)+11.8963)+4. {g_s} M^2 \log (T)+0.6
   N\right)^2}{N^2}\right)\right)}}=-n,
\end{equation}
which for $g_s=0.8,N=g_s^{-39}\sim6000, M=3$ yields:
\begin{eqnarray}
\label{2++IIB-spectrum}
& & m_{2^{++}} = 4.975 T \nonumber\\
& & m_{2^{++}}^* = 14.925 T \nonumber\\
& & m_{2^{++}}^{**} = 24.876 T \nonumber\\
& & m_{2^{++}}^{***} = 34.826 T \nonumber\\
& & m_{2^{++}}^{****} = 44.776 T \nonumber\\
\end{eqnarray}
So, one obtains an approximate match between the ground state $2^{++}$ mass from M theory and type IIB string theory.

\subsubsection{WKB Quantization Method}

Using the variables of \cite{Minahan}, the `potential',  defining $m = \tilde{m}\frac{\sqrt{y_h}}{L^2}$, yields:
\begin{eqnarray}
\label{V_2++_IIB-i}
& & V(2^{++}, IIB, r_h\neq0) = \frac{1}{4 \left(e^z+1\right)^3 \left(e^z+2\right)^2}\Biggl\{e^z \Biggl(3 b^2 \left(e^z+2\right) \left(-\left(\tilde{m}^2-6\right) e^z-\tilde{m}^2+3 e^{2 z}+6\right)\nonumber\\
& & +\left(-e^z-1\right)
   \left(\left(25-3 \tilde{m}^2\right) e^z-\left(\tilde{m}^2-18\right) e^{2 z}-2 \left(\tilde{m}^2-6\right)+4 e^{3
   z}\right)\Biggr)\Biggr\} + {\cal O}\left(\frac{g_s M^2}{N}\right).
\end{eqnarray}
In the IR, the potential is given by:
\begin{eqnarray}
\label{V-IR}
& & V(IR,T) = e^z \left(\left(0.15 \tilde{m}^2-1.3375\right) e^z-0.01 \tilde{m}^2+0.06\right) + {\cal O}(e^{3z}),
\end{eqnarray}
and in the IR, for $\tilde{m}>2.986$, $V(IR<T)>0$ for $z\in[\log\left(0.067 + {\cal O}\left(\frac{1}{\tilde{m}^2}\right)\right)\approx -2.708,-2.526]$ and: $\int_{2.708}^{-2.526}\sqrt{V(IR,T)}\approx0$ - hence the IR provides no contribution to the WKB quantization.

In the UV:
\begin{eqnarray}
\label{V_UV_T}
& & V(UV,T) = \frac{1}{4} \left(\tilde{m}^2+9.24\right) e^{-z}-\frac{3}{4} \left(\tilde{m}^2+0.36 \left(\tilde{m}^2+9\right)+3\right) + {\cal O}\left(e^{-3z}\right),
   e^{-2 z}-1
\end{eqnarray}
the turning points for $\tilde{m}>7.141$ are $\left(\log(4.08 + {\cal O}\left(\frac{1}{\tilde{m}^2}\right),0.25\tilde{m}^2 - 1.77)\right)$. Hence,
\begin{eqnarray}
\label{WKB-2++_T_IIB}
& & \int_{1.406}^{\log(0.25\tilde{m}^2 - 1.77)}\sqrt{V(UV,T)} = \int_{1.406}^{\log(0.25\tilde{m}^2 - 1.77)} \frac{e^{-z}}{2}\sqrt{e^z - 4.08} \tilde{m} + {\cal O}\left(\frac{1}{\tilde{m}}\right)\nonumber\\
& & = 0.389 \tilde{m} - 2 = \left(n + \frac{1}{2}\right)\pi,
\end{eqnarray}
which obtains:
\begin{equation}
\label{mn2++_T_IIB}
m_n^{2^{++}}(T) = (9.18 + 8.08 n)\frac{r_h}{L^2}.
\end{equation}
Hence, the string theory $2^{++}$ glueball is isospectral with $0^{++}$; in the large $n$-limit of the spectrum, the M-theory and type IIB spectra coincide.

\subsection{$r_h=0$}

\subsubsection{WKB Method}

Once again defining $r=\sqrt{y}, r_h = \sqrt{y_0}, y = y \left(1 + e^z\right)$ analogous to their $r_h\neq0$ redefinitions of \cite{Minahan}, the `potential' in the IR, is given by:
\begin{equation}
\label{V1++_ii}
 V(r_h=0) = -\frac{1}{4} + \frac{1}{4}\left(1 + \tilde{m}^2\right)e^{2z} + {\cal O}(e^{-3z}).
\end{equation}
The domain in the IR over which $V(r_h=0)>$ is: $[-\frac{1}{2}\log(5 + \tilde{m}^2),\log(\delta^2-1)]$ and:
\begin{equation}
\label{WKB_1++_T=0}
\int_{-\frac{1}{2}\log(5 + \tilde{m}^2)}^{-2.526}\sqrt{-\frac{1}{4} + \frac{1}{4}\left(1 + \tilde{m}^2\right)e^{2z}} = \frac{(\delta^2-1)}{2}\tilde{m} - \frac{\pi}{4} = \left(n + \frac{1}{2}\right)\pi,
\end{equation}
yielding:
\begin{equation}
\label{mn1++-IR}
m_n^{2^{++}}(IR,IIB,r_h=0) = m_n^{2^{++}}(IR,M\ {\rm theory},r_h=0).
\end{equation}

In the UV
\begin{equation}
\label{V1++-UV}
V(UV,r_h=0) = \frac{1}{4} \left(\tilde{m}^2-10\right) e^{-z}-\frac{3}{4} \left(\tilde{m}^2-5\right) e^{-2 z}+1 + {\cal O}\left(e^{-3z}\right),
\end{equation}
the zeros of which, as functions of $e^z$, are at $(\frac{1}{8} \left(-\tilde{m}^2\pm\sqrt{\tilde{m}^4+28 \tilde{m}^2-140}+10\right)$ $=(-\frac{\tilde{m}^2}{4} - \frac{1}{2}, 3 + {\cal O}\left(\frac{1}{\tilde{m}^2}\right))$. Hence the domain of integration over which $V(UV,r_h=0)>0$ is: $[\log 3,\infty)$. Therefore:
\begin{eqnarray}
\label{WKB1++-i}
& & \int_{\log 3}^\infty\sqrt{V(UV, r_h=0)} = \frac{1}{2}\int_{\log 3}^\infty e^{-z}\sqrt{e^z - 3}\tilde{m} + {\cal O}\left(\frac{1}{\tilde{m}}\right) = \frac{\tilde{m}\pi}{4\sqrt{3}} = \left(n + \frac{1}{2}\right)\pi,
\end{eqnarray}
yielding:
\begin{equation}
\label{mn1++-ii}
m_n^{2^{++}}(r_h=0) = (3.464 + 6.928 n)\frac{r_0}{L^2}.
\end{equation}

\subsubsection{NLO-in-$N$/Non-Conformal Corrections in the IR  in the $r_h=0$ Limit}

The `potential' inclusive of the NLO-in-$N$ corrections in the IR in the $r_h=0$ limit, reads as:
\begin{eqnarray}
\label{IIB_2++_IR_T=0}
& & V(IR,r_h=0) = e^{2 z} \Biggl(\frac{{g_s} M^2 \left(\frac{1}{N}\right)^{2/5} ({g_s} {N_f} (6 {\log N}-72+\log (16777216))-72 {g_s} {N_f} \log ({y_0})-48
   \pi )}{512 \pi ^2}\nonumber\\
   & & +\frac{1}{4} \left(\tilde{m}^2-3\right)\Biggr)-\frac{1}{4} + {\cal O}(e^{-3z}),
\end{eqnarray}
whose turning points, in the large-$\tilde{m}$ limit are: $[\log\left\{\frac{1}{\tilde{m}} + {\cal O}\left(\frac{1}{\tilde{m}^3}\right)\right\},\log(\delta^2-1)]\approx[-\log\tilde{m},\log\left(\delta^2-1\right)]$.
Now,
\begin{eqnarray}
\label{WKB_NLO_IIB_2++_T=0}
& & \int_{-\log\tilde{m}}^{\log(\delta^2-1)}\sqrt{V(IR,r_h=0)} = \frac{\left(\delta^2-1\right)}{2}\tilde{m} - \frac{\pi}{4} + {\cal O}\left(\frac{1}{\tilde{m}}\right) = \left(n + \frac{1}{2}\right),
\end{eqnarray}
which yields the same {\it LO} spectrum as $0^{--}, 1^{++}$ and the $2^{++}$ spectrum obtained from M theory. Hence, the type IIB at $r_h=0$ is unable to capture the non-conforamal NLO-in-$N$ corrections in the $2^{++}$ corrections, as the same either precisely cancel out or are $\frac{1}{\tilde{m}}$-suppressed in the large-$\tilde{m}$ limit, in a IIB computation.

\section{Summary and Discussion}

Supergravity calculations of glueball spectra in top-down holographic duals of large-$N$ thermal QCD at {\it finite string/gauge coupling} and not just $g_sN\gg1$, have thus far, been missing in the literature.  Such a limit is particularly relevant to sQGP \cite{Natsuume}. This work fills in this gap by working out the spectra of $0^{++},0^{-+}, 0^{--}, 1^{++}, 2^{++}$ glueballs in a type IIB/delocalized SYZ IIA mirror/(its) M-theory (uplift) model corresponding to the top-down holographic dual of \cite{metrics} in the MQGP limit introduced in \cite{MQGP}. As discussed in {\bf 2.1}, towards the end of a Seiberg duality cascade in the IR, despite setting $M(=N_c$ in the IR) to three in the MQGP limit, due a flavor-color enhancement of the length scale as compared to the KS model, one can trust supergravity calculations without worrying about stringy corrections. Further, in the MQGP limit, all physical quantities as seen in all calculations in \cite{EPJC-2}, receive  non-conformal corrections that appear at the NLO in $N$ and display a universal $\frac{g_sM^2(g_sN_f)}{N}$-suppression.

It should be noted that a numerical computation like the `shooting computation' used in a lot of holographic glueball spectrum computations will not be feasible to use for the following reason. In the `shooting method', like \cite{Ooguri_et_al}, one can first solve the EOMs in the UV using the infinite series/Frobenius method and then numerically (via Euler's method, etc.) obtain the solution at the horizon where one imposes a Neumann boundary condition. By matching the value obtained by numerically `shooting' from the UV to the horizon in the IR and matching the radial derivative of the solution so obtained to zero, one can obtain quantized values of the glueball masses. The caveat is that one should have at hand the exact radial profile of the effective number of fractional $D3$-branes ($D5$-branes wrapping the small two-cycle) and the number of flavor branes which would correctly interpolate between $(M,N_f)=(0,0)$ in the UV and ($M=3, N_f =2$) in the IR. But, we do not have this information - we know the values in the IR and the UV but not for the interpolating region. Hence, numerical methods such as the `shooting method' could at best be used, to obtain only the LO-in-$N$ results, not the NLO-in-$N$ results which is one of the main objectives of our computations.

 The summary of all calculations is given in  tables 1 (and Fig. 2) and 3 - the former  table/graph having to do with a WKB quantization calculation using the coordinate/field redefinitions of \cite{Minahan} and the latter table having to do with obtaining the mass spectrum by imposing Neumann/Dirichlet boundary condition at $r_h$/IR cut-off $r_0$. Some of the salient features of the results are given as separate bullets.

 It should be noted that the last two columns in Tables 1 and 3 have been prepared in the same spirit as the last columns in Table 2 of \cite{Boschi+Braga_AdS_BH_AdS_slice}.

 \newpage
\begin{table}[h]
\begin{tabular}{|c|c|c|c|}\hline
S. No. & {\scriptsize Glueball} & {\scriptsize $\tilde{m}$  using WKB $r_h\neq0$} & {\scriptsize
$\tilde{m}$ using WKB  $r_h=0$}   \\
&& {\scriptsize (units of $\pi T$, up to LO in $N$)} & {\scriptsize (units of $\frac{r_0}{L^2}$, up to NLO in $N$)}\\
&& {\scriptsize (large-$\tilde{m}$ limit)} & {\scriptsize (large-$\tilde{m}$ limit)}\\ \hline
1 & $0^{++}$ & (M theory) & (M theory)  \\
& {\scriptsize (Fluctuations: $h_{00,rx_1,rr}$}& $\frac{\sqrt{35 + 70 n}}{\sqrt{{\cal PL}(13.15 + 26.31 n)}}$ & {\scriptsize\rm No\ turning\ points} \\
& {\scriptsize in M-theory metric)} &  \mytriangle{red} & \\ \cline{2-4}
& $0^{++}$ & (Type IIB) & (Type IIB) \\
& {\scriptsize (Dilaton Fluctuations)}& $9.18 + 8.08 n$ \mytriangle{green} & {\scriptsize$ \frac{(4.64 + 6.28 n)}{(\delta^2-1)}\left[1 - 0.01 \frac{g_s M^2 }{N}(g_s N_f)\log N\log r_0\right]$} \\  \hline
2 & $0^{-+}$ & (Type IIA) & (Type IIA) \\
& {\scriptsize(1-form fluctuation $a_{\theta_2}$) }&  {\scriptsize  $11.12\left(n + \frac{1}{2}\right), n=0$} \mysquare{blue}& {\scriptsize $\frac{3.72 + 4.36 n}{(\delta^2-1)}, n=0$}  \\
& & {\scriptsize  $(6.22 + 4.80 n), n\in\mathbb{Z}^+$} \mytriangle{blue} & {\scriptsize $4.8\left(n + \frac{1}{2}\right), n\in\mathbb{Z}^+$} \\ \hline
3 & $0^{--}$ & (Type IIB) & (Type IIB) \\
&{\scriptsize 2-form fluctuation $A_{23}$}& {\scriptsize $= m_n^{0^{++}}({\rm dilaton},T)$} & {\scriptsize  $ \frac{6.28 n+4.71}{(\delta ^2-1)}\left(1 + \frac{0.01 {g_s}^2 {\log N} M^2 {N_f} \log ({r_0})}{ N}\right), n=0$} \\
&& \mytriangle{green} & {\scriptsize  $(7.87 + 6.93 n), n\in\mathbb{Z}^+$} \\ \hline
4 & $1^{++}$ & (M theory) & (M theory) \\
&{\scriptsize (Fluctuations: $h_{it}=h_{ti},i=x_{2,3}$}& {\scriptsize $8.08\left(n + \frac{1}{2}\right)$} & {\scriptsize $m_n^{1^{++}}(n=0,r_h=0) = m_n^{0^{--}}(n=0,r_h=0)$} \\
&{\scriptsize in M-theory metric)}& \mytriangle{purple} &  {\scriptsize $(3.46 + 6.93 n),n\in\mathbb{Z}^+$} \\ \hline
5 & $2^{++}$ & (M theory) & (M theory) \\
&{\scriptsize (Fluctuations: $h_{x_2x_3}=h_{x_3x_2}, $}& {\scriptsize $8.08\left(n + \frac{1}{2}\right) = m_n^{1^{++}}(T)$} & {\scriptsize $=m_n^{1^{++}}(r_h=0)$} \\
& {\scriptsize $h_{x_2x_2}=h_{x_3x_3}$ in M-theory metric)} &\mytriangle{purple}&  \\ \cline{2-4}
& $2^{++}$& (Type IIB) & (Type IIB) \\
& {\scriptsize (Fluctuation $h_{x_2x_3}=h_{x_3x_2}$}& {\scriptsize $9.18 + 8.08 n = m_n^{0^{++}}(IIB,T)$} & {\scriptsize $=m_n^{1^{++}}(r_h=0)$} \\
& {\scriptsize in type IIB metric)}& \mytriangle{green} &
\\ \hline
   \end{tabular}
   \caption{Summary of Glueball Spectra: $m = \tilde{m}\frac{r_h}{L^2}$  from Type IIB, IIA and M Theory using WKB quantization condition for $r_h\neq0$, and $m = \tilde{m} \frac{r_0}{L^2}$ for $r_h=0$ (equalities in the $r_h=0$ column, are valid up to NLO in $N$); the colored triangles/square in the third column correspond to the colored triangles/square that appear  in Fig. 2 in  the combined plot of $r_h\neq0$ supergravity calculations of glueballs}
   \end{table}

The $r_h\neq0$ glueball spectra is plotted in Figure 2.
\newpage
\begin{figure}
 \begin{center}
 \includegraphics[scale=0.8]{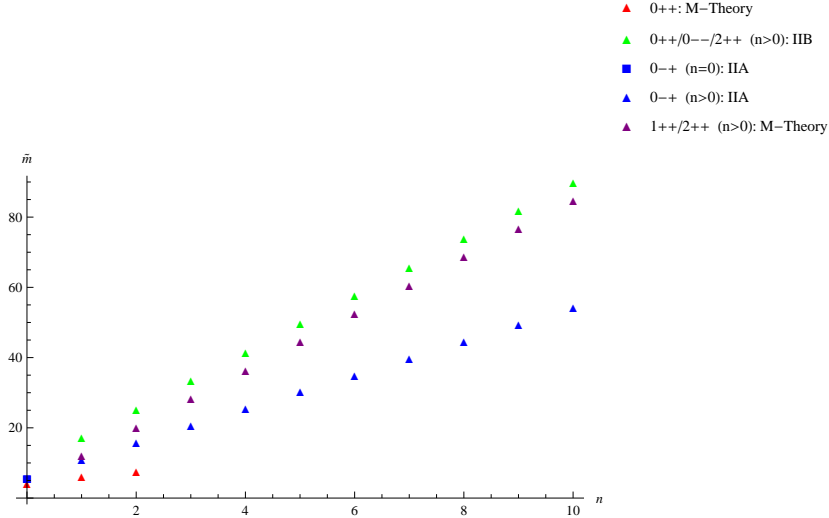}
 \end{center}
\caption{Plots of Supergravity $0^{++}, 0^{-+}, 0^{--}, 1^{++}$ Glueball Spectra for $r_h\neq0$}
\end{figure}

Some of the salient features of Table 1 and Figure 2 are presented below:
\begin{enumerate}
\item
Interestingly, via a WKB quantization condition using coordinate/field redefinitions of \cite{Minahan}, the lightest $0^{++}$ glueball spectrum for $r_h\neq0$ coming from scalar metric fluctuations in M theory compares rather well with the $N\rightarrow\infty$ lattice results of \cite{Teper98} - refer to Table 2. Also, similar to \cite{Brower}, the $0^{++}$ coming from the scalar fluctuations of the M theory metric  is lighter than the $0^{++}$ coming from type IIB dilaton fluctuations. Further, interestingly, one can show that by using the coordinate and field redefinitions of \cite{WKB-i} when applied to the EOM for dilaton fluctuation to yield a WKB quantization condition, for $a=0.6 r_h$ - as in \cite{EPJC-2} - one obtains a match with the UV limit of the $0^{++}$ glueball spectrum as obtained in \cite{Minahan}. For our purpose, the method based on coordinate/field redefinitions of \cite{WKB-i}, is no good for obtaining the $0^{++}$ glueball ground state and was not used for any other glueball later on in subsequent calculations in this paper.
\begin{table}[h]
\begin{tabular}{|c|c|c|c|}\hline
{\scriptsize State} & $N\rightarrow\infty$ {\scriptsize Entry in Table 34} of \cite{Teper98} & {\scriptsize M-theory scalar metric perturbations}  & {\scriptsize Type IIB Dilaton  fluctuations of} \cite{Ooguri_et_al}\\
& {\scriptsize in units of square root of} & ({\bf 6.1.2} - {\scriptsize in units of} & {\scriptsize in units of reciprocal of} \\
& {\scriptsize string tension} & $\frac{r_h}{L^2}$) & {\scriptsize temporal circle's diameter} \\ \hline
$0^{++}$ & $4.065\pm0.055$ & 4.267 & 4.07 ({\scriptsize normalized to match lattice}) \\ \hline
$0^{++*}$ & $6.18\pm0.13$ & 6.251 & 7.02 \\ \hline
$0^{++**}$ & $7.99\pm0.22$ & 7.555 & 9.92 \\ \hline
$0^{++**}$ & - & 8.588 & 12.80 \\ \hline
$0^{++***}$ & - & 9.464 & 15.67 \\ \hline
\end{tabular}
\caption{Comparison of \cite{Teper98}'s $N\rightarrow\infty$ lattice results for $0^{++}$ glueball with our supergravity results obtained  using WKB quantization condition and redefinitions of \cite{Minahan} for M theory scalar metric fluctuations}
\end{table}

\item
Also, from Table 1/Figure 2, $m_{n>0}^{2^{++}}>m_{n>0}^{0^{++}}$(scalar metric perturbations), similar to
\cite{Brower}.

\item
The higher excited states of the type IIA $0^{-+}$ glueball, for both $r_h\neq0$ and $r_h=0$, are isospectral. This is desirable because large-$n$ corresponds to the UV and that takes one away from the BH geometry, i.e., towards $r_h=0$.

\item
The non-conformal corrections up to NLO in $N$, have a semi-universal behavior of $\frac{(g_sM^2)(g_sN_f)\log r_0}{N}$ and turn out to be multiplied by a numerical pre-factor of ${\cal O}(10^{-2})$; we could disregard the same in the MQGP limit.

\item
As per a more recent lattice calculation \cite{Chen_et_al_latest_lattice_2006}\footnote{We thank P.Majumdar for bringing this reference to our attention.}, the $0^{++}$-glueball has a mass $4.16\pm0.11\pm0.04$ (in units of the reciprocal of the `hadronic scale parameter' of \cite{Sommer-r0}), which compares rather well with $m_{n=0}^{0^{++}}=4.267$ (in units of $\frac{r_h}{L^2}$) of Table 2 coming from scalar fluctuations of the M theory metric. Similarly, the $0^{-+}$-glueball in \cite{Chen_et_al_latest_lattice_2006} has a mass $6.25\pm0.06\pm0.06$ and from Table 1, which matches rather nicely with  $m_{n=0}^{0^{-+}}(\delta=1.26)=6.25$ (in units of $\frac{r_0}{L^2}$) of Table 1 coming from type IIA one-form fluctuation.

\item
The ground state and the $n\gg1$ excited states of $1^{++}$ and $0^{--}$ glueballs are isospectral.

\item
The higher excited $r_h\neq0$ $2^{++}$ glueball states corresponding to metric fluctuations of the M-theory metric and the ones corresponding to fluctuations of the type IIB metric, are isospectral. The $r_h=0$ $2^{++}$ glueball states corresponding to metric fluctuations of the M-theory/type IIB string theory, are isospectral. Further, it turns out that due to internal cancellation of terms and $\frac{1}{\tilde{m}}$-suppression, a type IIB $r_h=0$ $2^{++}$ glueball spectrum, unlike an M-theoretic computation, is unable to capture the NLO-in-$N$ corrections to the LO-in-$N$ type IIB $2^{++}$ glueball spectrum.

\item
$m^{2^{++}}_n({\rm NLO},r_h=0) = m_n^{1^{++}}({\rm NLO},r_h=0)\stackrel{n\gg1}{\longrightarrow}m_n^{0^{--}}({\rm NLO},r_h=0)$, where the `NLO' implies equality with the inclusion of NLO-in-$N$ corrections.

\end{enumerate}
\newpage
\begin{table}[h]
\begin{tabular}{|c|c|c|c|}\hline
S. No. & Glueball & Spectrum Using   & Spectrum Using     \\
&& N(eumann)/D(irichlet) & N(eumann)/D(irichlet) \\
&& b.c., $r=r_h$(units of $\pi T$) & b.c., $r=r_0$(units of $\frac{r_0}{L^2}$)\\ \hline
1 & $0^{++}$ & (M theory) & (M theory)  \\
&& (N) {\scriptsize $12.25\sqrt{2+n}$} & (N) 4.1 \\
&& (D) {\scriptsize $12.25\sqrt{1+n}$} & \\ \hline
2 & $0^{-+}$ & (Type IIA) & (Type IIA) \\
& & (N/D) {\scriptsize $\frac{3.1}{\pi}\sqrt{n}$} & (N) {\scriptsize $m_{n=0}^{0^{-+}}=0, m_{n=1}^{0^{-+}}\approx 3.4, m_{n=2}^{0^{-+}}\approx 4.35$} \\
&&& (D) {\scriptsize $m_{n=0}^{0^{-+}}=0, m_{n=1}^{0^{-+}}\approx 3.06, m_{n=2}^{0^{-+}}\approx 4.81$} \\ \hline
3 & $0^{--}$ & (Type IIB) & (Type IIB) \\
&& (N/D) {\scriptsize $m_{n=0}^{0^{--}}(T)=0, m_{n=1}^{0^{--}}(T)=\frac{32.46}{\pi},$} & (large $n$)  \\
&& {\scriptsize $m_{n=2}^{0^{--}}(T)=\frac{32.88}{\pi}$} & (N/D)  \\
&&& {\scriptsize ${\scriptsize \frac{1}{2} 5^{3/4} \sqrt[4]{\frac{2 \left(\sqrt{6} \sqrt{\pi ^2 \left(16 n^2+22 n+7\right)+6}+6\right)+3 \pi ^2 (2 n+1)}{32 - 3 \pi ^2}}}$} \\ \hline
4 & $1^{++}$ & (M theory) & (M theory) \\
& & (N/D) {\scriptsize $m_{n=0}^{1^{++}}(T) = 2.6956, m_{n=1}^{1^{++}}(T)=2.8995$}  & (N) {\scriptsize $m_{n=0}^{1^{++}}(r_h=0)\approx1.137$} \\
& & {\scriptsize $m_{n=2}^{1^{++}}(T) = 2.9346$} & (D) {\scriptsize $m_{n=0}^{1^{++}}(r_h=0)\approx0.665$} \\ \hline
5 & $2^{++}$ & (M theory) & (M theory) \\
&& (N) {\scriptsize $m_{n=0}^{2^{++}}(T)=\frac{5.086}{\pi}, m_{n=1}^{2^{++}}(T)=\frac{5.269}{\pi}$} & $=m_n^{1^{++}}(r_h=0)$ \\
&& {\scriptsize $m_{n=2}^{2^{++}}(T)=\frac{5.318}{\pi}$} &  \\
&& {\scriptsize $m_{n=0}^{2^{++}}(D,T)=0, m_{n+1}^{2^{++}}(D,T)=m_n^{2^{++}}(N,T)$} & \\ \hline
   \end{tabular}
   \caption{Summary of Glueball Spectra from Type IIB, IIA and M Theory for $r_h\neq0/r_h=0$ using Neumann/Dirichlet boundary conditions at the horizon $r_h$/IR cut-off $r_0$}
   \end{table}
\newpage
Some salient features of Table 3 are presented below:
\begin{itemize}

\item The following is the comparison of ratios of $0^{--}$ glueball masses obtained in this work from Neumann/Dirichlet boundary conditions at the horizon,  with \cite{Ooguri_et_al}:

\begin{table}[h]
\begin{center}
\begin{tabular}{|c|c|c|}\hline
Ratio & Our Results & \cite{Ooguri_et_al}'s Results\\  \hline
$\frac{m_{0^{--}}^*}{m_{0^{--}}}$ & 1.0129 & 1.5311\\ \hline
$\frac{m_{0^{--}}^{**}}{m_{0^{--}}^*}$ & 1.0033 & 1.3244\\ \hline
$\frac{m_{0^{--}}^{***}}{m_{0^{--}}^{**}}$ & 1.0013 & 1.2393\\ \hline
$\frac{m_{0^{--}}^{****}}{m_{0^{--}}^{***}}$ & 1.0007 & 1.1588\\ \hline
 \end{tabular}
   \caption{Comparison of ratios of $0^{--}$ glueball masses obtained  from Neumann/Dirichlet boundary conditions at the horizon,  with \cite{Ooguri_et_al}}
   \end{center}
   \end{table}

Hence, for higher excited states, the ratio of masses of successive excited states approaches unity faster as per our results as compared to \cite{Ooguri_et_al}.

 \item From a comparison of results in Tables 1/2 or Figures 2 with $N\rightarrow\infty$ lattice results, it appears that WKB quantization-based spectra are closer to $N\rightarrow\infty$ lattice results than the computations involving imposing Neumann/Dirichlet boundary conditions at the horizon/IR cut-off. In particular, it is pleasantly surprising that the WKB quantization method applied to the $0^{++}, 0^{-+}$ glueball spectra, is able to provide a good agreement (in fact for the lightest $0^{++}$ glueball spectrum, better than the classic computations of \cite{Ooguri_et_al}) with lattice results even for the ground and the lower excited states.

     \end{itemize}

\section*{Acknowledgements}

KS is supported by is supported by a senior research fellowship (SRF) from the Ministry of Human Re-source and Development (MHRD), Govt. of India. VY is supported by a junior research fellowship (JRF) from the University Grants Commission, Govt. of India. We would like to thank P. Majumdar and S. Gupta  for useful communications, H. B. Filho for a useful clarification, and R. Das for participating in the material in {\bf 3.2.1} as part of his Masters' project work. One of us (AM) would like to dedicate this work to the memory of his father, the late R. K. Misra who was and (whose memory) still remains a true driving force and inspiration for him.
\newpage
\appendix

\section{$\widetilde{F^2_5}, \widetilde{F^2_3}, H_3^2$}
\setcounter{equation}{0}\seceqaa

The expressions of squares of various fluxes that figure in the EOM (\ref{final EOM}) are given below for ready reference:
{\
\begin{eqnarray}
\label{fluxes-squared}
& & \widetilde{F^2_5}=-\frac{8}{\sqrt{\pi } \sqrt{N} \sqrt{g_s}}-\frac{1}{254803968 \pi ^{13/2} N^{7/10}}
\nonumber\\
&&\times\Biggl[M^4 r^6 N_f^4 g_s^{11/2} \left(r^4-r_h^4\right) \left(\phi _1+\phi_2-\psi \right){}^2\Biggl(N_f g_s \log (N) (2 (r+1) \log (r)+1)
+ 2 \Biggl\{-9 (r+1) N_f g_s \log ^2(r)\nonumber\\
&&-2 (r+1) \log (r) \left(2 \pi -\log (4) N_f g_s\right)+\log (4) N_f g_s\biggr\}\Biggr){}^2\Biggr]
\nonumber\\
&&+a^2\biggl[+\frac{\pi ^{3/2} r^{10} g_s^{3/2}}{956593800 N^{7/10}}-\frac{24}{\sqrt{\pi } \sqrt{N} \sqrt{g_s}r^2}-\frac{1}{84934656 \pi ^{13/2} N^{7/10}} \nonumber\\
&&\times\Biggl\{M^4 r^6 N_f^4 g_s^{11/2} \left(r^4-r_h^4\right) \left(\phi _1+\phi_2-\psi \right){}^2(24 r\log{r}-1)\Biggl(N_f g_s \log (N) (2 (r+1) \log (r)+1)\nonumber\\
&&+ 2 \Biggl\{-9 (r+1) N_f g_s \log ^2(r)-2 (r+1) \log (r) \left(2 \pi -\log (4) N_f g_s\right)+\log (4) N_f g_s\biggr\}\Biggr){}^2\Biggr\}\Biggr];\nonumber \\
&&\widetilde{F^2_3}=\frac{729 M^2 N_f^2 \sqrt{g_s} \left(r^4-r_h^4\right) \left(72 a^2 N^{2/5} \log (r)+a^2 \left(-\left(3 N^{2/5}+4\right)\right)+2 N^{2/5} r^2\right)}{128
   \pi ^{7/2} N^{11/10} r^6};\nonumber\\
&&   H_3^2 =\frac{243 M^2 N_f^2 g_s^{5/2} \left(r^4-r_h^4\right) \left(144 a^2 \left(\sqrt[5]{N}+3\right) r \log (r)+a^2 \left(9-15 \sqrt[5]{N}\right)+2
   \left(\sqrt[5]{N}+1\right) r^2\right)}{256 \pi ^{7/2} \sqrt{N} r^6};\nonumber\\
&&\widetilde{F}_{x_2 x_3p_3p_4p_5}\widetilde{F}_{x_2x_3 q_3q_4q_5}g^{p_3q_3}g^{p_4q_4}g^{p_5q_5}h^{x_2x_3}=\frac{60 r^4}{\pi ^{3/2} N^{3/2} g_s^{3/2}}-\frac{1}{169869312 \pi ^{13/2} N^{7/10}}\nonumber\\
&&\times\Biggl[5 M^4 r^{10} N_f^4 g_s^{9/2} \left(r^4-r_h^4\right) \left(\phi _1+\phi _2-\psi \right){}^2\Biggl(N_f g_s \log (N) (2 (r+1) \log (r)+1)
+ 2 \Biggl\{-9 (r+1) N_f g_s \log ^2(r)\nonumber\\
& &-2 (r+1) \log (r) \left(2 \pi -\log (4) N_f g_s\right)+\log (4) N_f g_s\biggr\}\Biggr){}^2\Biggr]+a^2\Biggl[\frac{180 r^2}{\pi ^{3/2} N^{3/2} g_s^{3/2}}+\frac{1}{56623104 \pi ^{15/2} N^{17/10}}\nonumber\\
&&\times\Biggl\{5 M^4 r^{10} N_f^4 g_s^{9/2} \left(r^4-r_h^4\right) \left(\phi _1+\phi _2-\psi \right){}^2(24 r\log{r}-1)\Biggl(N_f g_s \log (N) (2 (r+1) \log (r)+1)
\nonumber \\
&& + 2 \Biggl\{-9 (r+1) N_f g_s \log ^2(r)-2 (r+1) \log (r) \left(2 \pi -\log (4) N_f g_s\right)+\log (4) N_f g_s\biggr\}\Biggr){}^2\Biggr\}\Biggr].
\end{eqnarray}}


\begin{thebibliography}{99}
\bibitem{metrics}  M.~Mia, K.~Dasgupta, C.~Gale and S.~Jeon, {\it Five Easy Pieces: The Dynamics of Quarks in Strongly Coupled Plasmas}, Nucl.\ Phys.\ B {\bf 839}, 187 (2010) [arXiv:hep-th/0902.1540].
\bibitem{MQGP}M.~Dhuria and A.~Misra, {\it Towards MQGP}, JHEP 1311 (2013) 001, [arXiv:hep-th/1306.4339].
\bibitem{NPB}K.~Sil and A.~Misra, {\it On Aspects of Holographic Thermal QCD at Finite Coupling},
  Nucl.\ Phys.\ B {\bf 910}, 754 (2016) [arXiv:1507.02692 [hep-th]].
 \bibitem{Natsuume}M.~Natsuume, {\it String theory and quark-gluon plasma}, hep-ph/0701201.
 \bibitem{EPJC-2}K.~Sil and A.~Misra,
  {\it New Insights into Properties of Large-N Holographic Thermal QCD at Finite Gauge Coupling at (the Non-Conformal/Next-to) Leading Order in N}, Eur.\ Phys.\ J.\ C {\bf 76}, no. 11, 618 (2016)
  doi:10.1140/epjc/s10052-016-4444-7
  [arXiv:1606.04949 [hep-th]].
  \bibitem{maldacena} Juan~ M.~Maldacena, {\it The Large N Limit of Superconformal Field Theories and Supergravity}, Adv.Theor.Math.Phys.2:231-252,(1998), doi:10.1023/A:1026654312961 [	 arXiv:hep-th/9711200]
      \bibitem{lattice result} G. ~Boyd, J.~ Engels, F.~ Karsch, E.~ Laermann, C.~ Legeland, M.~ Luetgemeier, B.~ Petersson {\it Thermodynamics of SU(3) Lattice Gauge Theory}, Nucl.Phys. B469 (1996) 419-444, [	 arXiv:hep-lat/9602007]
  \bibitem{Colangelo}
  P.~Colangelo, F.~Giannuzzi and S.~Nicotri,
  {\it Holographic Approach to Finite Temperature QCD: The Case of Scalar Glueballs and Scalar and Scalar Mesons},
  Phys.\ rev.\ D {\bf 80}, 094019 (2009)
  doi:10.1103/PhysRevD.80.094019
  [arXiv:0909.1534 [hep-ph]].
   \bibitem{Nicotri}
  S.~Nicotri,
  {\it Scalar glueball in a holographic model of QCD},
  Nuovo Cim.\ B {bf 123},796 (2008)
  [Nuovo Cim.\ B {\bf 123}, 851 (2008)]
  doi:10.1393/ncb/i2008-10579-5, 10.1393/ncb/i2008-10580-0
  [arXiv:0807.4377 [hep-ph]].
   \bibitem{Forkel}
  H.~Forkel,
  {\it Glueball correlators as holograms},
  arXiv:0808.0304 [hep-ph]].
   \bibitem{ForkelStructure}
  H.~Forkel,
  {\it Holograhic glueball structure},
  Phys.\ Rev.\ D {\bf 78}, 025001 (2008)
  doi:10.1103/PhysRevD.78.025001
  [arXiv:071101179 [hep-ph]].
  \bibitem{Li}
  D.~Li and M.~Huang,
  {\it Dynamical holographic QCD model for glueball and light meson spectra},
  JHEP {\bf 1311}, 088 (2013)
  doi:10.1007/JHEP11(2013)088
  [arXiv:1303.6929 [hep-ph]].
  \bibitem{FolcoCapossoli} E.~Folco Capossoli and H.~Boschi-Filho, {\it  Renormalized AdS$_5$ Mass for even Spin Glueball and Pomeron Trajectory from a Holographic softwall model}, arXiv:1611.09817[hep-ph].
  \bibitem{Jugeau}
  F.~Jugeau,
  {\it Holographic description of glueballs in a deformed AdS-dilaton background},
  AIP Conf.\ Proc.\ {\bf 964} (2007) 151
  doi:10.1063/1.2823842
  [arXiv:0709.1093 [hep-ph]].
   \bibitem{Colangelolight}
  P.~Colangelo, F.~De Fazio, F.~jugeau and S.~Nicotri,
  {\it On the light glueball spectrum in a holographic description of QCD},
  Phys.\ Lett.\ B {\bf 652}, 73 (2007)
  doi:10.1016/j.physletb.2007.06.072
  [hep-th/0703316].
  \bibitem{Huang}
  W.~H.~Huang,
  {\it Holographic Description of Glueball and Baryon in Noncommulative Dipole Gauge Theory},
  JHEP {\bf 0806}, 006 (20008)
  doi:10.1088/1126-6708/2008/06/006
  [arXiv:0805.0985 [hep-ph]].
   \bibitem{Chen} Y.~Chen and M.~Huang, {\it Two-gluon and trigluon glueballs from dynamical holography QCD}, Chin.\ Phys.\ C {\bf 40}, no. 12, 123101 (2016) doi:10.1088/1674-1137/40/12/123101 [arXiv:1511.07018 [hep-ph]].
    \bibitem{Gordeli}
  I.~Gordeli and D.~Melnikov,
  {\it Calculation of glueball spectra in supersymmetric theories via holography},
  arXiv:1311.6537 [hep-ph]].
   \bibitem{Hashimoto}
  K.~Hashimoto, C.~I.~tan and S.~Terashima,
  {\it Glueball decay in holographic QCD},
  Phys.\ Rev.\ D {\bf 77}, 086001 (2008)
  doi:10.1103/PhysRevD.77.086001
  [arXiv:0709.2208 [hep-th]].
   \bibitem{Brunner0} F.~Brunner and A.~Rebhan, {\it Holographic QCD predictions for production and decay of pseudoscalar glueballs}, arXiv:1610.10034 [hep-ph].
   \bibitem{Brunner1} A.~Rebhan, F.~Brunner and D.~Parganlija,
 {\it Glueball decay patterns in top-down holographic QCD}
 PoS EPS {\bf -HEP2015}, 421 (2015)
 [arXiv:1511.01391 [hep-ph]].
  \bibitem{ParganlijaScalar}
 D.~Parganlija
 {\it Scalar Glueball in a Top-Down Holographic Approach to QCD},
 Acta Phys.\ Polon.\ Supp.\ {\bf 8}, no. 1, 219(2015)
 doi:10.5506/APhysPolBSupp.8.219
 [arXiv:1503.00550 [hep-ph]].
  \bibitem{Brunnerdecay}
 F.~Brunner, D.~Parganlija and A.~Rebhan,
 {\it Top-down Holographic Glueball Decay Rates},
 AIP Conf.\ Proc.\ {\bf 1701}, 090007 (2016)
 doi:10.1063/1.4938709
 [arXiv:1502.00456 [hep-ph]].
 \bibitem{Brunner2}
  F.~Brunner, D.~Parganlija and A.~Rebhan,
  {\it Holographic Glueball Decay},
  Acta Phys.\ Polon.\ Supp.\ {bf 7}, no. 3, 533 (2014)
  doi:10.5506/APhysPolBSupp.7.533
  [arXiv:1407.6914 [hep-ph]].
   \bibitem{Parganlija} D.~Paragnlija,
 {\it Tensor Glueball in a Top-Down Holographic Approach to QCD},
 Acta Phys.\ Polon.\ Supp.\ {\bf 8}, no. 2, 289(2015)
 doi:10.5506/APhysPolBSupp.8.289
 [arXiv:1506.03000 [hep-ph]].
 \bibitem{Capossoli} E.~Folco Capossoli and H.~Boschi-Filho,
 {\it Glueball spectra and Regge trajectories from a modified holographic softwall model},
 Phys.\ Lette.\ B {\bf 753}, 419 (2016)
 doi:10.1016/j.physletb.2015.12.034
 [arXiv:1510.03372 [hep-ph]].
  \bibitem{witten} E. Witten, Anti-de Sitter space, {\it thermal phase transition and confinement in gauge theories},
Adv. Theor. Math. Phys. 2 (1998) 505 [hep-th/9803131].
\bibitem{polchinski} J. Polchinski and M.J. Strassler,{\it Hard scattering and gauge/string duality}, Phys. Rev. Lett.
88 (2002) 031601 [hep-th/0109174].
\bibitem{transport-coefficients}M.~Dhuria and A.~Misra, {\it Transport Coefficients of Black MQGP M3-Branes},
  Eur.\ Phys.\ J.\ C {\bf 75}, no. 1, 16 (2015)  [arXiv:1406.6076 [hep-th]].
  \bibitem{KW} Igor R. Klebanov and Edward Witten, {\it Superconformal Field Theory on Threebranes at a Calabi-Yau Singularity},  Nucl. Phys. B 536, 199 (1998)[arXiv:hep-th/9807080].
  \bibitem{Gubser-Tpq}  S.~S.~Gubser, {\it Einstein manifolds and conformal field theories},
  Phys.\ Rev.\ D {\bf 59}, 025006 (1999)
  doi:10.1103/PhysRevD.59.025006
  [hep-th/9807164].
  \bibitem{glueball-QCD-instantons}N.~Kochelev,  {\it Ultralight glueballs in Quark-Gluon Plasma},
  Phys.\ Part.\ Nucl.\ Lett.\  {\bf 13}, no. 2, 149 (2016)  doi:10.1134/S1547477116020138
 [arXiv:1501.07002 [hep-ph]].
\bibitem{WKB-i} A.~S.~Miranda, C.~A.~Ballon Bayona, H.~Boschi-Filho and N.~R.~F.~Braga,
  {\it Black-hole quasinormal modes and scalar glueballs in a finite-temperature AdS/QCD model},
  JHEP {\bf 0911}, 119 (2009) [arXiv:0909.1790 [hep-th]].4
  \bibitem{Minahan}
  J.~A.~Minahan,
  {\it Glueball mass spectra and other issues for supergravity duals of QCD models}
  JHEP {\bf 9901}, 020 (1999)
  doi:10.1088/1126-6708/1999/01/020
  [hep-th/9811156].
\bibitem{KT} I.R. Klebanov and A. Tseytlin, {\it Gravity Duals of Supersymmetric $SU(N)\times SU(M+N)$ Gauge Theories}, [hep-th/0002159].
\bibitem{KS} I.~R.~Klebanov and  M.~J.~Strassler, {\it Supergravity and a Confining Gauge Theory: Duality Cascades and $X$SB-Resolution of Naked Singularities}, JHEP 0008:052,2000 [arXiv:hep-th/0007191].
    \bibitem{ouyang} P.~Ouyang, {\it Holomorphic D7-Branes and Flavored N=1 Gauge Theories}, Nucl.Phys.B 699:207-225 (2004), [arXiv:hep-th/0311084].
    \bibitem{Buchel} A.~Buchel, {\it Finite temperature resolution of the Klebanov-Tseytlin singularity}, Nucl.\ Phys.\ B {\bf 600}, 219 (2001) [hep-th/0011146].
\bibitem{Gubser-et-al-finitetemp}  S.~S.~Gubser, C.~P.~Herzog, I.~R.~Klebanov and A.~A.~Tseytlin, {\it Restoration of chiral symmetry: A Supergravity perspective},  JHEP {\bf 0105}, 028 (2001) [hep-th/0102172]; A.~Buchel, C.~P.~Herzog, I.~R.~Klebanov, L.~A.~Pando Zayas and A.~A.~Tseytlin, {\it Nonextremal gravity duals for fractional D-3 branes on the conifold},   JHEP {\bf 0104}, 033 (2001)[hep-th/0102105].
\bibitem{Leo-ii} M.~Mahato, L.~A.~Pando Zayas and C.~A.~Terrero-Escalante, {\it Black Holes in Cascading Theories: Confinement/Deconfinement Transition and other Thermal Properties}, JHEP {\bf 0709}, 083 (2007)  [arXiv:0707.2737 [hep-th]].
  \bibitem{K. Dasgupta  et al [2012]} M.~Mia, F.~Chen, K.~Dasgupta, P.~Franche and S.~Vaidya, {\it Non-Extremality, Chemical Potential and the Infrared limit of Large N Thermal QCD}, Phys.\ Rev.\ D {\bf 86}, 086002 (2012)[arXiv:1202.5321 [hep-th]].
    \bibitem{IR-UV-desc_Dasgupta_etal} M.~Mia, K.~Dasgupta, C.~Gale and S.~Jeon, {\it Toward Large N Thermal QCD from Dual Gravity: The Heavy Quarkonium Potential},  Phys.\ Rev.\ D {\bf 82}, 026004 (2010) [arXiv:1004.0387 [hep-th]].
\bibitem{syz} A.~Strominger, S.~T.~Yau and E.~Zaslow, {\it Mirror symmetry is T duality},  Nucl.\ Phys.\ B {\bf 479}, 243 (1996)  [hep-th/9606040].
    \bibitem{M.Ionel and M.Min-OO (2008)}M.~Ionel and M.~Min-OO, {\it Cohomogeneity One Special Lagrangian 3-Folds in the Deformed and the Resolved Conifolds},  Illinois Journal of Mathematics, Vol. 52, Number 3 (2008).
\bibitem{M. Becker et al [2004]}S.~Alexander, K.~Becker, M.~Becker, K.~Dasgupta, A.~Knauf and R.~Tatar, {\it In the realm of the geometric transitions,}  Nucl.\ Phys.\ B {\bf 704}, 231 (2005) [hep-th/0408192].
\bibitem{F. Chen et al [2010]}F.~Chen, K.~Dasgupta, P.~Franche, S.~Katz and R.~Tatar, {\it Supersymmetric Configurations, Geometric Transitions and New Non-Kahler Manifolds}, Nucl.\ Phys.\ B {\bf 852}, 553 (2011) [arXiv:hep-th/1007.5316].
\bibitem{T-dual-NS5-Taub-NUT-Tong} D.~Tong, {\it NS5-branes, T duality and world sheet instantons}, JHEP {\bf 0207}, 013 (2002)  [hep-th/0204186].
\bibitem{KK-monopoles-A-Sen} A.~Sen, {\it Dynamics of multiple Kaluza-Klein monopoles in M and string theory}, Adv.\ Theor.\ Math.\ Phys.\  {\bf 1}, 115 (1998)[hep-th/9707042].
\bibitem{Dasguptaetal_G2_structure}F.~Chen, K.~Dasgupta, P.~Franche, S.~Katz and R.~Tatar, {\it Supersymmetric Configurations, Geometric Transitions and New Non-Kahler Manifolds},  Nucl.\ Phys.\ B {\bf 852}, 553 (2011) [arXiv:1007.5316 [hep-th]].
     \bibitem{torsion}G.~Lopes Cardoso, G.~Curio, G.~Dall'Agata, D.~Lust, P.~Manousselis and G.~Zoupanos,
  {\it NonKahler string backgrounds and their five torsion classes},
  Nucl.\ Phys.\ B {\bf 652}, 5 (2003)
  [hep-th/0211118].
        \bibitem{Butti et al [2004]}A.~Butti, M.~Grana, R.~Minasian, M.~Petrini and A.~Zaffaroni, {\it The baryonic branch of Klebanov-Strassler solution: A supersymmetric family of SU(3) structure backgrounds}, JHEP 0503, 069
(2005) [arXiv:hep-th/0412187].
\bibitem{Grigorian}S.~Grigorian, {\it Deformations of G2-Structures with Torsion}, arXiv:1108.2465.
\bibitem{Bryant} R.~L.~Bryant, {\it Some remarks on G(2)-structures},  math/0305124 [math-dg].
\bibitem{karigiannis-2007}S.~Karigiannis, {\it Geometric {F}lows on {M}anifolds with ${G_2}$ {S}tructure},[arXiv:math/0702077].
\bibitem{G2-Structure} P.~Kaste, R.~Minasian, M.~Petrini and A.~Tomasiello, {\it Kaluza-Klein bundles and manifolds of exceptional holonomy},  JHEP {\bf 0209}, 033 (2002) [hep-th/0206213].
\bibitem{Czaki_et_al-0-+}C.~Csaki, Y.~Oz, J.~Russo and J.~Terning,
{\it Large N QCD from rotating branes},
  Phys.\ Rev.\ D {\bf 59}, 065012 (1999)
  doi:10.1103/PhysRevD.59.065012
  [hep-th/9810186].
     \bibitem{Teper98} M.~J.~Teper, {\it SU(N) gauge theories in (2+1)-dimensions},
  Phys.\ Rev.\ D {\bf 59}, 014512 (1999)
  doi:10.1103/PhysRevD.59.014512
  [hep-lat/9804008].
   \bibitem{Brower}
   R.~C.~Brower, S.~D.~Mathur and C.~I.~Tan,
   {\it Discrete spectrum of the graviton in the AdS$_5$ black hole background}
   Nucl.\ Phys.\ B  {\bf 574}, 219 (2000)
   doi:10.1016/S0550-3213(99)00802-0
   [hep-th/9908196].
  \bibitem{Chen_et_al_latest_lattice_2006}Y.~Chen {\it et al.},
  {\it Glueball spectrum and matrix elements on anisotropic lattices},
  Phys.\ Rev.\ D {\bf 73}, 014516 (2006)
  doi:10.1103/PhysRevD.73.014516
  [hep-lat/0510074].
  \bibitem{Sommer-r0}R.~Sommer,
{\it A New way to set the energy scale in lattice gauge theories and its applications to the static force and alpha-s in SU(2) Yang-Mills theory,}
  Nucl.\ Phys.\ B {\bf 411}, 839 (1994)
  doi:10.1016/0550-3213(94)90473-1
  [hep-lat/9310022].
  \bibitem{Ooguri_et_al}  C.~Csaki, H.~Ooguri, Y.~Oz and J.~Terning,
  {\it Glueball mass spectrum from supergravity},
  JHEP {\bf 9901}, 017 (1999)
  doi:10.1088/1126-6708/1999/01/017
  [hep-th/9806021].
\bibitem{decay} Frederic.~Brunner, Denis.~Parganlija and Anton.~Rebhan, {\it Glueball decay patterns in top-down holographic QCD},[arXiv:1511.01391].
   \bibitem{deformed conifold} Elena.~Caceres and Rafel.~Hernandez, {\it Glueball Masses for the Deformed Conifold Theory},[arXiv:hep-th/0011204].
    \bibitem{deformed conifold} Elena.~Caceres and Xavier.~Amador, {\it Spin two glueball mass and glueball regge trajectory from supergravity },JHEP{\bf 11}(2004) 022.
  \bibitem{BoschiFilho}
  H.~Boschi-Filho and N.~R.~F.~Braga,
  {\it QCD/string holographic mapping and glueball mass spectrum},
  Eur.\ Phys.\ J.\ C {\bf 32}, 529 (2004)
  doi:10.1140/epjc/s2003-01526-4
  [hep-th/0209080].
  \bibitem{Myers}
  N.~R.~Constable and R.~C.~Myers,
  {\it Spin two glueballs, positive energy theorems and the AdS/CFT correspondence},
  JHEP {\bf 9910}, 037 (1999)
  doi:10.1088/1126-6708/1999/10/037
  [hep-th/9908175].
  \bibitem{Mathur et al}
  R.~C.~Brower, S.~D.~Mathur and C.~I.~Tan,
  {\it Glueball spectrum for QCD from AdS supergravity Duality},
  Nucl.\ Phys.\ B {\bf 587}, 249 (2000)
   doi:10.1016/S0550-3213(00)00435-1
   [hep-th/0003115].
   \bibitem{Ouguri et al}
   C.~Csaki, H.~Ouguri, Y.~Oz, J.~Terning,
   {\it Glueball mass spectrum from supergravity},
   JHEP {\bf 9901}, 017 (1999)
   doi:10.1088/1126-6708/1999/01/017
   [hep-th/9806021].
\bibitem{KN}, I.R. Klebanov and N. Nekrasov, {\it Gravity Duals of Fractional Branes and Logarithmic RG Flow}, [hep-th/9911096].
\bibitem{Leo-i} B.~A.~Burrington, J.~T.~Liu, L.~A.~Pando Zayas and D.~Vaman, {\it Holographic duals of flavored N=1 super Yang-mills: Beyond the probe approximation}, JHEP {\bf 0502}, 022 (2005)[hep-th/0406207].
  \bibitem{Mathur_et_al}  R.~C.~Brower, S.~D.~Mathur and C.~I.~Tan, {\it Glueball spectrum for QCD from AdS supergravity duality},  Nucl.\ Phys.\ B {\bf 587}, 249 (2000)
  doi:10.1016/S0550-3213(00)00435-1
  [hep-th/0003115].
   \bibitem{kiritsis-book}{\it String Theory in a Nutshell}, E.~Kiritsis, Princeton University Press (2007).
   \bibitem{Mathur_et_al-0++-Mtheory}R.~C.~Brower, S.~D.~Mathur and C.~I.~Tan,
  {\it Discrete spectrum of the graviton in the AdS(5) black hole background},
  Nucl.\ Phys.\ B {\bf 574}, 219 (2000)
  doi:10.1016/S0550-3213(99)00802-0
  [hep-th/9908196].
 \bibitem{Chiossi+Salamon}S.~ Chiossi, S.~ Salamon, {\it The IntrinsicTorsion of SU(3) and G2 Structures}, arXiv:math/0202282 [math.DG].
     \bibitem{Boschi+Braga_AdS_BH_AdS_slice}{\it Gauge/string duality and scalar glueball mass ratios}, Henrique Boschi-Filho, Nelson R. F. Braga, JHEP0305:009,2003 [hep-th/0212207].
\end{thebibliography}
\end{document}